\documentclass[11pt,a4paper]{article}
\pdfoutput=1
\usepackage[utf8]{inputenc}
\usepackage[a4paper, total={16.5cm, 22.5cm}]{geometry}
\usepackage{graphicx}
\usepackage{cancel}
\usepackage{textcomp}
\usepackage{cite}
\usepackage{bm}
\usepackage{times}
\usepackage{color}
\usepackage{hyperref}
\usepackage{setspace}
\usepackage{paralist}
\usepackage{multirow}
\usepackage{soul}
\usepackage{bbm}
\usepackage{enumitem}

\usepackage{xcolor}

\usepackage{slashed}
\usepackage{comment}
\usepackage{epstopdf}
\usepackage{array}
\usepackage{tikz}

\usepackage{rotating}
\usepackage{framed}

\usepackage{placeins}
\usepackage{amsmath} 
\usepackage{amsthm} 
\usepackage{amssymb}	

\definecolor{Darkgreen}{RGB}{30,120,30}

\newcommand{\abs}[1]{\left| #1 \right|} 
 
\newcommand{\ket}[1]{\left| #1 \right>} 
\newcommand{\bra}[1]{\left< #1 \right|} 

\definecolor{Darkgreen}{RGB}{30,120,30}

\let\baraccent=\= 
\renewcommand{\=}[1]{\stackrel{#1}{=}} 

\theoremstyle{definition}

\theoremstyle{remark}

\def\beqn{\begin{eqnarray}}
\def\eeqn{\end{eqnarray}}
\def\no{\nonumber}
\hypersetup{
    pdfnewwindow=true,      
    colorlinks=true,       
    linkcolor=black,          
    citecolor=blue,        
    filecolor=blue,      
    urlcolor=blue           
}

\makeatletter
\newcommand\xleftrightarrow[2][]{%
  \ext@arrow 9999{\longleftrightarrowfill@}{#1}{#2}}
\newcommand\longleftrightarrowfill@{%
  \arrowfill@\leftarrow\relbar\rightarrow}
\makeatother

\begin{document}

\thispagestyle{empty}
\hfill\mbox{IFIC/19-03, FTUV/19-0419}
\vspace{3cm}

\begin{center}

{\huge\sc 
Global fit to  $b \to c \tau \nu$ transitions
}

\vspace{1cm}

{\sc
Clara Murgui,${}^a$ Ana Pe\~nuelas,${}^a$ Martin Jung${}^{b,c}$ and Antonio Pich${}^a$
}

\vspace*{.7cm}

{\sl
${}^a$ Departament de F\'\i sica Te\`orica, IFIC, Universitat de Val\`encia -- CSIC\\
Apt. Correus 22085, E-46071 Val\`encia, Spain,

${}^b$ Excellence Cluster Universe, Technische  Universit\"at  M\"unchen\\
Boltzmannstr. 2,  D-85748  Garching,  Germany,

${}^c$ Dipartimento di Fisica, Universit\`a di Torino and INFN Sezione di Torino,\\
Via P. Giuria 1, I-10125 Torino, Italy.
}

\end{center}

\date{\vspace{-5ex}}

\begin{abstract}
We perform a general model-independent analysis of $b \to c \tau \bar{\nu}_\tau $ transitions, including measurements of $\mathcal{R}_D$, $\mathcal{R}_{D^*}$, their $q^2$ differential distributions, the recently measured longitudinal $D^*$ polarization $F_L^{D^*}$, and constraints from the $B_c \to \tau \bar{\nu}_\tau$ lifetime, each of which has significant impact on the fit. A global fit to a general set of Wilson coefficients of an effective low-energy Hamiltonian is presented, the solutions of which are interpreted in terms of hypothetical new-physics mediators. From the obtained results we predict selected $b \to c\tau\bar\nu_\tau$ observables, such as the baryonic transition $\Lambda_b \to \Lambda_c \tau \bar{\nu}_\tau$, the ratio $\mathcal{R}_{J/\psi}$, the forward-backward asymmetries ${\cal A}_\text{FB}^{D^{(*)}}$, the $\tau$ polarization asymmetries $\mathcal{P}_\tau^{D^{(*)}}$, and the longitudinal $D^*$ polarization fraction $F_L^{D^*}$. The latter shows presently a slight tension with any new-physics model, such that an improved measurement could have an important impact.
We also discuss the potential change due the very recently announced preliminary $\mathcal{R}_{D^{(*)}}$ measurement by the Belle collaboration.
\end{abstract}

%

\section{Introduction}
\label{sec:Introduction}
The success of the Standard Model (SM) has reached its climax with the discovery of the Brout-Englert-Higgs boson \cite{PhysRevLett.13.508, PhysRevLett.13.321, PhysRevLett.13.585}, which seems to suggest the simplest scenario where the electroweak spontaneous symmetry breaking is linearly realized.
In spite of its success as a low-energy effective field theory (EFT), there are both experimental signals and conceptual issues that cannot be accommodated in the SM framework and, therefore, motivate the search of New Physics (NP) beyond the SM. In this context, the series of anomalies in semi-leptonic $B$-meson decays, recently reported by several experiments, have caught a great attention in the scientific community. The unexpected deviations seem to appear in both $b \to c$ and $b \to s$ semi-leptonic decay transitions when different generations of leptons are involved, see Ref.~\cite{Bifani:2018zmi} for a recent review.

The $b \to c$ transitions are of particular interest, because the necessary NP effect would be comparable with the tree-level contribution of the SM, which in turn would require NP to be either rather light or strongly coupled to the SM particles. Deviations from the SM predictions in those modes have been recently observed by the BaBar \cite{Lees:2012xj, Lees:2013uzd}, Belle  \cite{Huschle:2015rga, Sato:2016svk, Hirose:2016wfn} and LHCb  \cite{Aaij:2015yra, Aaij:2017uff} collaborations in the ratios
\begin{equation}
\mathcal{R}_{D^{(*)}} \,\equiv\, \frac{\mathcal{B} (B \to D^{(*)}\tau \bar{\nu}_\tau)}{\mathcal{B} (B \to D^{(*)}\ell \bar{\nu}_\ell)} \, ,
\label{eq:RD}
\end{equation}
where $\mathcal{B}$ represents the branching ratio of the decay and $\ell$ denotes the light leptons, i.e., $\ell= e, \mu$.
The combination of these measurements performed  by the Heavy Flavour Averaging Group (HFLAV)~\cite{Amhis:2016xyh} reads
\beqn \label{eq:RDavg}
\mathcal{R}_{D}^{\text{avg}} &= 0.407 \pm 0.039 \pm 0.024  \qquad\quad \text{  and  } \qquad\quad
\mathcal{R}_{D^*}^{\text{avg}} &= 0.306 \pm 0.013 \,  \pm 0.007\, ,  
\eeqn
with a correlation of $-20\%$, which shows a tension of $4.4\sigma$ with our SM predictions (see also
\cite{Aoki:2016frl,Fajfer:2012vx,Bigi:2016mdz,Bernlochner:2017jka,Bigi:2017jbd,Jaiswal:2017rve,Amhis:2016xyh}),
\beqn \label{eq:RDSM}
\mathcal{R}_{D}^{\text{SM}} &= 0.300^{+0.005}_{-0.004} \qquad\quad  \text{  and  } \qquad\quad
\mathcal{R}_{D^*}^{\text{SM}} &= 0.251^{+0.004}_{-0.003}\,,
\eeqn
to be discussed below.\footnote{Note that this prediction does not rely on experimental inputs, but includes only part of the $1/m_q^2$ corrections in heavy quark effective theory.}
 Apart from the above observables, also the recent LHCb measurement  \cite{Aaij:2017tyk} of the $B_c \to J/\Psi$ ratio, 
\begin{equation}
\mathcal{R}_{J/\psi} \,\equiv\, \frac{\mathcal{B}(B_c \to J/\psi \tau \bar{\nu}_\tau)}{\mathcal{B}(B_c \to J/\psi \mu \bar{\nu}_\mu)} \, =\, 0.71 \pm 0.17 \pm 0.18\, ,
\end{equation}
deviates from the SM predictions  $\mathcal{R}_{J/\psi}^{SM} \approx 0.25$--$0.28$~\cite{Anisimov:1998uk,Kiselev:2002vz,Ivanov:2006ni,Hernandez:2006gt,Huang:2007kb,Wang:2008xt,Issadykov:2018myx,Wen-Fei:2013uea,Hu:2019qcn,Leljak:2019eyw,Azizi:2019aaf,Tran:2018kuv}.
This points naively into the same direction, although the central value is in fact so large that it cannot be accommodated with NP contributions either.

These deviations could be interpreted as hints at lepton flavour universality violation (LFUV), which cannot be accommodated in the SM and therefore suggest the existence of NP. The lack of evidence of similar discrepancies in $K$ and $\pi$ semi-leptonic and purely leptonic decays, or in electroweak precision observables, favours a scenario in which the potential NP contribution responsible for LFUV is only coupled to the third generation of leptons and quarks. The fact that in universality ratios large parts of the hadronic uncertainties cancel, renders underestimated theory uncertainties as an explanation extremely unlikely. This remains true considering recent discussions of radiative corrections \cite{deBoer:2018ipi,Cali:2019nwp}, see also, \emph{e.g.}, Refs.~\cite{Becirevic:2009fy,Atwood:1989em} for earlier discussions. The correct inclusion
of radiative corrections is, however, very important for the forthcoming precision analyses.

However, recent measurements of $\mathcal{R}_{D^*}$ by LHCb~\cite{Aaij:2017uff} and Belle~\cite{Hirose:2016wfn}, which identify the final $\tau$ through its hadronic decays, result in values more compatible with the SM and yield a downward shift in the average that might suggest that the anomaly is smaller than indicated by the above numbers.~\footnote{The average of these measurements, only, agrees with the SM at the level of $1\text{-}1.5\sigma$.}

Our work aims at a better understanding of the nature of these anomalies, assuming in the following that they are indeed due to NP contributions and not due to underestimated systematic uncertainties or statistical fluctuations. 
Instead of considering any specific NP model, we follow a bottom-up approach, in which the available experimental input is used to constrain any possible higher-scale effect and in this way infer information on NP without prejudice. We do exploit, however, the consequences of the apparent absence of NP close to the electroweak scale. Only afterwards we investigate which indications for more specific NP scenarios can be inferred.
Numerous discussions can be found in the literature~\cite{Celis:2012dk,Datta:2012qk,Duraisamy:2013kcw, Dutta:2013qaa,Sakaki:2013bfa,Duraisamy:2014sna, Freytsis:2015qca,Alonso:2015sja,Boubaa:2016mgn, Bardhan:2016uhr, Bhattacharya:2016zcw, Alonso:2016oyd, Choudhury:2016ulr, Celis:2016azn, Alok:2016qyh,Alok:2017qsi,Bernlochner:2017jka, Capdevila:2017iqn,Altmannshofer:2017poe,Buttazzo:2017ixm,Cai:2017wry,  Crivellin:2017zlb,Jung:2018lfu, Biswas:2018jun,Azatov:2018knx,Hu:2018veh,Angelescu:2018tyl,Aebischer:2018iyb,Blanke:2018yud, Bhattacharya:2018kig,Dutta:2017xmj,Dutta:2015ueb,Huang:2018nnq,Azizi:2018axf}, where the $b \to c \tau \bar{\nu}_\tau$ transitions are studied from a model-independent point of view. However, most of these works restrict their analyses to either effects from a single NP operator or a single heavy particle mediating the interaction. We will adopt the most general possible scenario under a set of well-motivated assumptions instead. 

In addition to the
ratios defined in Eq.~\eqref{eq:RD} we consider the normalized experimental distributions of $\Gamma(B \to D^{(*)} \tau \bar{\nu}_\tau)$ measured by BaBar~\cite{Lees:2013uzd} and Belle~\cite{Huschle:2015rga}. Although this shape information was shown to provide quite stringent constraints in Ref.~\cite{Lees:2013uzd,Sakaki:2014sea,Freytsis:2015qca,Celis:2016azn, Bhattacharya:2016zcw}, it has been so far ignored in most phenomenological analyses. 
We also analyze the effect of including the recently announced value for $F_L^{D^{*}}$ by the Belle collaboration~\cite{Abdesselam:2019wbt}, 
\begin{equation}
F_L^{D^*}=0.60 \pm 0.08 \text{ (stat) }\pm 0.04 \text{ (syst)} ,
\end{equation}
which differs from its SM prediction by $1.6\sigma$, and discuss its consequences in detail. Other related observables, such as $\mathcal{P}_{\tau}^{D^*}$~\cite{Hirose:2016wfn} and $\mathcal{R}_{J/\psi}$~\cite{Aaij:2017tyk}, are not included due to their large experimental uncertainties, but are predicted from our fits. 
Very recently, the Belle collaboration has announced a new preliminary measurement of $\mathcal{R}_D$ and $\mathcal{R}_{D^*}$~\cite{RDmoriondBelle,Abdesselam:2019XXX}:
\begin{equation}
{\cal R}_D^\text{Belle} = 0.307 \pm 0.037 \pm 0.016 \qquad \text{ and } \qquad
{\cal R}_{D^*}^\text{Belle} = 0.283 \pm 0.018 \pm 0.014\,,
\end{equation}
with a correlation of $-54\%$. This result is compatible with the SM at the $1.2\sigma$ level. Including this measurement in the global average yields
\beqn 
\mathcal{R}_{D}^{\text{avg,new}} &= 0.337\pm 0.030 \qquad\quad  \text{  and  } \qquad\quad
\mathcal{R}_{D^*}^{\text{avg,new}} &= 0.299\pm0.013\,,
\eeqn
which reduces the significance of the anomaly slightly; however, it still amounts to  $4\sigma$ relative to the above SM prediction.\footnote{We find a milder reduction in the overall significance of the anomaly than what was stated in the Belle presentation. We obtain $\chi^2=20.0$ for 2 d.o.f. for the new average. Relative to the SM prediction quoted by HFLAV \cite{Amhis:2016xyh} the significance is reduced to $3.4\sigma$. 
}

We present at the end of Sec.~\ref{sec:FitResults} an updated analysis, including in the fit these preliminary data, and discuss their implications.

Our paper is organized as follows: In Sec.~\ref{sec:TheoricalFramework}, the theoretical framework used in this work is presented, and the physical observables and experimental inputs are defined. In Sec.~\ref{sec:FitResults}, we discuss our global $\chi^2$ fit and detail the resulting values of the fitted parameters. The interpretation of these results and their relation to NP are given in Sec.~\ref{sec:interpretation}, 
where we complete our discussion with several additional fits, relaxing some of the assumptions. A set of predictions for relevant observables, for which measurements will be published or improved soon, is presented in Sec.~\ref{sec:Predictions}.  Finally, we draw conclusions in Sec.~\ref{sec:Conclusions}. Some technical details are relegated to the appendices.

\section{Theoretical 
framework}
\label{sec:TheoricalFramework}
\subsection{Effective Hamiltonian}
\label{sec:Heff}

We adopt the most general $SU(3)_C \otimes U(1)_Q$-invariant effective Hamiltonian  describing $b \to c \ell \bar{\nu}_\ell$ transitions at the bottom quark scale, not considering the possibility of light right-handed neutrinos:
\begin{equation}
{\cal H}_{\text{eff}}^{b\to c \ell \nu} =  \frac{4G_F}{\sqrt{2}}V_{cb}\big [ \left(  1 + C_{V_L} \right){\cal O}_{V_L} +C_{V_R} {\cal O}_{V_R}
+C_{S_R} {\cal O}_{S_R} +C_{S_L} {\cal O}_{S_L}+C_{T} {\cal O}_T \big ]  + \text{h.c.} .
\label{eq:effH}
\end{equation}
The above fermionic operators are given by\footnote{Note that in full generality, the neutrino flavour can be different from the charged-lepton one. However, phenomenologically these situations are very similar, apart from the fact that no interference with the SM occurs when the two flavours are different. We do not consider this possibility in the following.}
\begin{equation}
    {\cal O}_{V_{L,R}} = \left( \bar{c}\, \gamma^{\mu} b_{L,R} \right)\left( \bar{\ell}_L \gamma_{\mu} \nu_{\ell L} \right),\quad  {\cal O}_{S_{L,R}} = \left( \bar{c}\,  b_{L,R} \right)\left( \bar{\ell}_R \nu_{\ell L} \right),\quad {\cal O}_{T} = \left( \bar{c}\, \sigma^{\mu \nu} b_L \right)\left( \bar{\ell}_R \sigma_{\mu \nu} \nu_{\ell L} \right)\,,\label{eq:effop}
\end{equation}
and are weighted by the corresponding Wilson coefficients $C_i$, which are, in general, lepton and flavour dependent, and parametrize any possible deviation from the SM, i.e., $C_{i}^{\text{SM}} \equiv 0$.
This effective Hamiltonian forms the basis of our analysis, restricted only by a minimal set of well-motivated assumptions:
\begin{itemize}
\item Possible NP contributions are assumed to be present only in the third generation of leptons. 
This is motivated by the absence of experimental evidence of deviations from the SM in tree-level transitions involving light leptons; specifically, precision measurements like the ratio $\mathcal{B}(\tau \to \mu \nu_{\tau} \bar{\nu}_{\mu})/\mathcal{B}(\tau \to e \nu_{\tau} \bar{\nu}_{e}) = 0.9762\pm 0.0028$ \cite{Tanabashi:2018oca} and the analysis of $b\to c(e,\mu)\bar{\nu}_{(e,\mu)}$ transitions in Ref.~\cite{Jung:2018lfu} constrain potential effects to be negligible in the present context.

\item  
The coefficient $C_{V_R}$ is assumed to be lepton-flavour universal in our main fit. This statement can be derived~\cite{Cirigliano:2009wk,Alonso:2014csa, Cata:2015lta} in the context of the Standard Model Effective Field Theory (SMEFT) \cite{Buchmuller:1985jz,Grzadkowski:2010es}, which is the appropriate effective theory in the presence of a sizeable energy gap above the electroweak scale if the electroweak symmetry breaking is linearly realized. The experimental facts that no new states beyond the SM have been found so far up to an energy scale of approximately 1~TeV and that measurements of the Higgs couplings are all consistent with the SM expectations support this scenario. In this case, $C_{V_R}$ is strongly constrained from $b\to c(e,\mu)\bar{\nu}_{(e,\mu)}$ data \cite{Jung:2018lfu}, and we set it to zero for convenience. If the assumption of linearity is relaxed, a non-universal $C_{V_R}$ coefficient can be generated \cite{Cata:2015lta}; we will consider this case separately.

\item The CP-conserving limit is taken, so all Wilson coefficients $C_i$ are assumed to be real. This is mostly done for convenience; however, none of the measurements related to the $B$ anomalies
refers to a CP-violating observable. Possible CP-violating contributions have been analyzed before in, e.g., Ref.~\cite{Celis:2016azn,Becirevic:2018afm,Bhattacharya:2018kig,Blanke:2018yud, Bhattacharya:2019olg}. Note that in the presence of such couplings other observables can become relevant, like electric dipole moments, see, e.g., \cite{Jung:2013hka,Dekens:2018bci}.
This assumption will be briefly commented in Section~\ref{sec:FitResults}.

\end{itemize}

\subsection{Form Factors}
\label{sec:FFs}

The relevance of hadronic uncertainties in the determination of $|V_{cb}|$ has opened an intense debate about the most adequate way to parametrise the relevant hadronic form factors \cite{Lattice:2015rga, Bigi:2016mdz,Bigi:2017njr,Bigi:2017jbd, Bernlochner:2017jka,Bernlochner:2017xyx,Grinstein:2017nlq,Jaiswal:2017rve}. 
It has been suggested that the accuracy of the usually adopted Caprini-Lellouch-Neubert (CLN) parametrisation \cite{Caprini:1997mu} has been probably overestimated and the current experimental precision requires to use more generic functional forms such as the one advocated by Boyd, Grinstein and Lebed (BGL) \cite{Boyd:1994tt,Boyd:1995sq,Boyd:1997kz}.
However, we note that the observables considered here are mostly ratios, reducing the overall form-factor sensitivity. 
We consider a heavy quark effective theory (HQET) \cite{Neubert:1993mb,Manohar:2000dt} parametrization, including corrections of order $\alpha_s$, $\Lambda_{\text{QCD}}/m_{b,c}$ and partly $\Lambda_{\text{QCD}}^2/m_c^2$, mostly following \cite{Bernlochner:2017jka,Jung:2018lfu}. In the heavy-quark limit all form factors either vanish or reduce to a common functional form, the Isgur-Wise function $\xi(q^2)$~\cite{Isgur:1989ed}.  Thus, it is convenient to factor out $\xi(q^2)$ by defining \cite{Bernlochner:2017jka}
\begin{equation}
\hat{h}(q^2)\, =\, h(q^2) / \xi(q^2)\, .
\label{eq:iwrel}
\end{equation}

The leading Isgur-Wise function can be more conveniently expressed in terms of the kinematical parameters
\begin{equation}
\omega(q^2)\, =\, \frac{m_B^2 + m_{D^{(*)}}^2-q^2}{2 m_B m_{D^{(*)}} } \,  
\qquad \text{ and } \qquad 
z(q^2)\, =\, \frac{\sqrt{\omega(q^2)+1}-\sqrt{2}}{\sqrt{\omega(q^2)+1}+\sqrt{2}} \, .
\label{eq:wrelat}
\end{equation}
The variable $\omega(q^2)$ is the inner product of the $B$ and $D^{(*)}$ velocities, so that $\omega=1$ corresponds to the zero-recoil point, $q^2_{\text{max}} = (m_B - m_{D^{(*)}})^2$, where $\xi(q^2_{\text{max}})=1$. The conformal mapping $z(q^2)$ encodes in a very efficient way the analyticity properties of the form factors, transforming the cut $q^2$ plane into the circle $|z|<1$  \cite{Bourrely:1980gp}, so that a perturbative expansion in powers of $z(q^2)$ has an optimized convergence. Up to $\mathcal{O}(z^4)$ corrections, $\xi(q^2)$ can be written as\footnote{The phenomenological necessity to include orders higher than $z^2$ in this expansion  has first been found in \cite{Bordone:2019vic}.}
\begin{eqnarray}
\xi(q^2)& =&
1 - \rho^2\, [\omega(q^2)-1] + c\, [\omega(q^2)-1]^2+ d\, [\omega(q^2)-1]^3 + \mathcal{O}([\omega-1]^4)
\nonumber\\ & = &
1 - 8\rho^2 z(q^2) +   (64 c - 16 \rho^2)\, z^2(q^2) + (256 c - 24 \rho^2 + 512 d)\, z^3(q^2) + \mathcal{O}(z^4)\, , 
\end{eqnarray}
and it is characterized through the parameters $\rho^2$, $c$ and $d$.

The functions $\hat h(q^2)$ introduce corrections of order $\Lambda_{\text{QCD}}/m_{b,c}$ and $\Lambda_{\text{QCD}}^2/m_c^2$ via the subleading Isgur-Wise functions $\chi_{2,3}(\omega),\eta(\omega)$ at order $1/m_{c,b}$ and $l_{1,2}(\omega)$ at order $1/m_{c}^2$, parametrized by the parameters $\{\chi_2(1), \chi_2'(1), \chi_3'(1), \eta(1), \eta'(1)\}$ and $\{l_1(1),l_2(1)\}$, respectively. They also include the corrections of order $\alpha_s$. 
The detailed parametrization of the different form factors can be found in Ref.~\cite{Bernlochner:2017jka,Jung:2018lfu}. The main difference to the latter article is the introduction of the $z^3$ term in the leading Isgur-Wise function, that renders the fit compatible with the extrapolation of the recent lattice data \cite{Na:2015kha,Lattice:2015rga} to large recoil.

We updated the corresponding fit to the inputs from lattice quantum chromodynamics (QCD) \cite{Na:2015kha,Lattice:2015rga,Bailey:2014tva,Harrison:2017fmw}, light-cone sum rules \cite{Faller:2008tr} and QCD sum rules \cite{Neubert:1992wq,Neubert:1992pn,Ligeti:1993hw} (see \cite{Jung:2018lfu} for details); note that this fit does not make use of experimental data, thereby rendering the form factors independent of the NP scenario considered. 
The results obtained for the 10 form-factor parameters are given in Table \ref{table:inputFF}, while the corresponding correlation matrix can be found in Table~\ref{table:FFinputcorrelation} of Appendix~\ref{app:addinput}.

\begin{table}[ht]
\centering
\begin{tabular}{| c | c |}
\hline
Parameter & Value \\
\hline
$\rho^2$ & $\phantom{-}1.32 \pm 0.06 $\\
$c$ & $\phantom{-}1.20 \pm 0.12 $\\
$d$ & $ -0.84 \pm 0.17$\\
$\chi_2(1)$ & $-0.058 \pm 0.020$ \\
$\chi_2^\prime(1)$ & $\phantom{-}0.001 \pm 0.020$\\
$\chi_3^\prime(1)$ & $\phantom{-}0.036 \pm 0.020$\\
$\eta(1)$  &  $\phantom{-}0.355 \pm 0.040$\\
$\eta^\prime(1)$ &$-0.03 \pm 0.11$\\
$l_1(1)$ & $\phantom{-}0.14 \pm 0.23$ \\
$l_2(1)$ & $\phantom{-}2.00 \pm  0.30$\\
\hline
\end{tabular}
\caption{Inputs used to determine the form factors in the HQET parametrization as in \cite{Bernlochner:2017jka}. The first three parameters determine the leading Isgur-Wise function, while the last seven enter in the $1/m_{c,b}$ and $1/m_c^2$ corrections. The correlations between these parameters can be found in  Table~\ref{table:FFinputcorrelation} of Appendix~\ref{app:addinput}\label{table:inputFF}.}
\end{table}
%
\subsection{Observables and experimental input}
\label{sec:obs}

We collect the formulae for the main observables entering our analysis. Starting with $B\to D^{(*)}\tau \bar{\nu}_\tau$ decays, we obtain from the effective Hamiltonian of Eq.~\eqref{eq:effH} their differential decay rates as a function of the general set of Wilson coefficients \cite{Tanaka:2012nw, Sakaki:2013bfa}:
 \beqn 
 \label{eq:GammaD}
 \no
 \frac{d \Gamma (\bar{B} \to D \tau \bar{\nu}_\tau)}{d q^2} & = &\frac{G_F^2 \abs{V_{cb}}^2}{192 \pi^3 m_B^3}\; q^2 \sqrt{\lambda_D(q^2)}\left( 1- \frac{m_{\tau}^2}{q^2} \right)^2 \\ \no 
& \times & \left\lbrace \abs{1+C_{V_L} + C_{V_R}}^2 \left[ \left( 1 + \frac{m_{\tau}^2}{2 q^2} \right) H_{V,0}^{s,2} + \frac{3}{2} \frac{m_{\tau}^2}{q^2}\, 
H_{V, t}^{s 2} \right] \right. \\ \no
& + & \frac{3}{2}\, \abs{C_{S_R} +C_{S_L} }^2 H_S^{s} + 8\, \abs{C_T}^2 \left( 1+ \frac{2 m_{\tau}^2}{q^2}\right) H_T^{s 2}  \\
& + & 3\, \mathcal{R}\mathrm{e} \left[ \left( 1 + C_{V_L} + C_{V_R} \right) \left(C_{S_R}^* + C_{S_L}^* \right) \right] \frac{m_{\tau}}{\sqrt{q^2}}\, H_S^s H_{V,t}^s \\ \no
  & - &   \left. 12\, \mathcal{R}\mathrm{e} \left[ \left( 1 + C_{V_L} + C_{V_R} \right) C_T^* \right] \frac{m_{\tau}}{\sqrt{q^2}}\, H_T^s H_{V,0}^s \right\rbrace \, , \no
 \eeqn
 and
 \beqn 
  \label{eq:GammaDstar}
 \no
 \frac{d \Gamma (\bar{B} \to D^* \tau \bar{\nu}_\tau)}{d q^2}  &=& \frac{G_F^2 \abs{V_{cb}}^2}{192 \pi^3 m_B^3}\; q^2 \sqrt{\lambda_{D^*} (q^2)}\left( 1- \frac{m_{\tau}^2}{q^2} \right)^2 \\ \no 
& \times & \left\lbrace \left( \abs{1+C_{V_L}}^2 + \abs{C_{V_R}}^2 \right) \left[ \left( 1 + \frac{m_{\tau}^2}{2q^2} \right) \left(H_{V,+}^2+H_{V,-}^2 + H_{V,0}^2\right) + \frac{3}{2}\frac{m_{\tau}^2}{q^2}\, H_{V,t}^2 \right] \right. \\  \no 
&-& 2\, \mathcal{R}\mathrm{e} \left[ \left( 1 + C_{V_L} \right) C_{V_R}^* \right] \left[ \left( 1 + \frac{m_{\tau}^2}{2q^2} \right) \left( H_{V,0}^2 + 2 H_{V, +} H_{V, -} \right) + \frac{3}{2} \frac{m_{\tau}^2}{q^2}\, H_{V, t}^2 \right] \\
&+& \frac{3}{2}\, \abs{C_{S_R} - C_{S_L}}^2 H_S^2 + 8\, \abs{C_T}^2 \left( 1 + \frac{2 m_{\tau}^2}{q^2}  \right) \left( H_{T,+}^2  + H_{T, -}^2 + H_{T, 0}^2 \right) \\ \no 
&+& 3\, \mathcal{R}\mathrm{e} \left[ \left( 1+ C_{V_L} - C_{V_R} \right) \left(C_{S_R}^* - C_{S_L}^* \right) \right] \frac{m_{\tau}}{\sqrt{q^2}}\, H_S H_{V,t} \\ \no 
&-& 12\, \mathcal{R}\mathrm{e} \left[ \left( 1+ C_{V_L} \right) C_T^* \right] \frac{m_{\tau}}{\sqrt{q^2}} \left( H_{T,0} H_{V,0} + H_{T, +} H_{V, +} - H_{T, -} H_{V, -} \right) \\ \no 
&+& \left. 12\,\mathcal{R}\mathrm{e} \left[ C_{V_R} C_T^* \right] \frac{m_{\tau}}{\sqrt{q^2}} \left( H_{T,0} H_{V,0} + H_{T,+} H_{V, -}  - H_{T, -} H_{V, +} \right) \right\rbrace  \, , \no 
 \eeqn
where $\lambda_{D^{(*)}} (q^2)
\equiv \lambda(m_B^2,m_{D^{(*)}}^2,q^2)
= \left[ \left( m_B - m_{D^{(*)}} \right)^2 - q^2 \right]\left[  \left( m_B+ m_{D^{(*)}} \right)^2-q^2 \right] $.

The helicity amplitudes, which encode the information from the hadronic form factors, can be found in the Appendix~\ref{appendix:helicity}.
The values of the quark and meson masses and other experimental inputs used in our analysis are listed in Table~\ref{table:expinput} of Appendix~\ref{app:addinput}.

Besides the semi-leptonic processes included in the fit, the pure leptonic decay $B_c \to \tau \bar{\nu}_\tau$ is crucial in determining the direction of potential NP effects, since it strongly constrains the axial ($C_{V_R} - (1+C_{V_L}$)) and,
especially, the pseudo-scalar ($C_{S_R}-C_{S_L}$) contributions \cite{Li:2016vvp,Alonso:2016oyd}: 
\begin{eqnarray}
\label{eq:Bctauu}
{\cal B} (B_c \to \tau \bar{\nu}_\tau)\, &=\, &
\tau_{B_c}\,\frac{m_{B_c}m_\tau^2 f_{B_c}^2G_F^2|V_{cb}|^2}{8\pi}\left(1-\frac{m_\tau^2}{m_{B_c}^2}\right)^2
\nonumber\\ && \hskip 1.5cm\times\,
\left| 1+ C_{V_L}-C_{V_R}+\frac{m_{B_c}^2}{m_\tau(m_b+m_c)}\, (C_{S_R}-C_{S_L})\right|^2  .
\end{eqnarray}

From these expressions, four classes of observables are obtained that are determined in experimental analyses:
\begin{itemize}

\item \textbf{The ratios $\boldsymbol{\mathcal{R}_{D^{(*)}}}$}

Experimental measurements of the ratios $\mathcal{R}_{D}$ and $\mathcal{R}_{D^{*}}$ have been published by BaBar~\cite{Lees:2012xj, Lees:2013uzd}, LHCb~\cite{Aaij:2015yra,Aaij:2017uff}, and Belle~\cite{Huschle:2015rga, Sato:2016svk, Hirose:2016wfn} (see also \cite{RDmoriondBelle}) using different techniques. These results have been averaged by the HFLAV collaboration, giving the values listed in Eq.~\eqref{eq:RDavg}
\cite{Amhis:2016xyh}. The results for each experiment and their average are also shown in Fig.~\ref{fig:RDRDstaravg}, with and without the result from Ref.~\cite{RDmoriondBelle,Abdesselam:2019XXX}.

As mentioned above, these ratios are advantageous both theoretically and experimentally, as they allow for the cancellation of uncertainties, specifically the CKM factors and leading form factor uncertainties on the theoretical side.

\begin{figure}
    \centering
    \includegraphics[width=7.3cm]{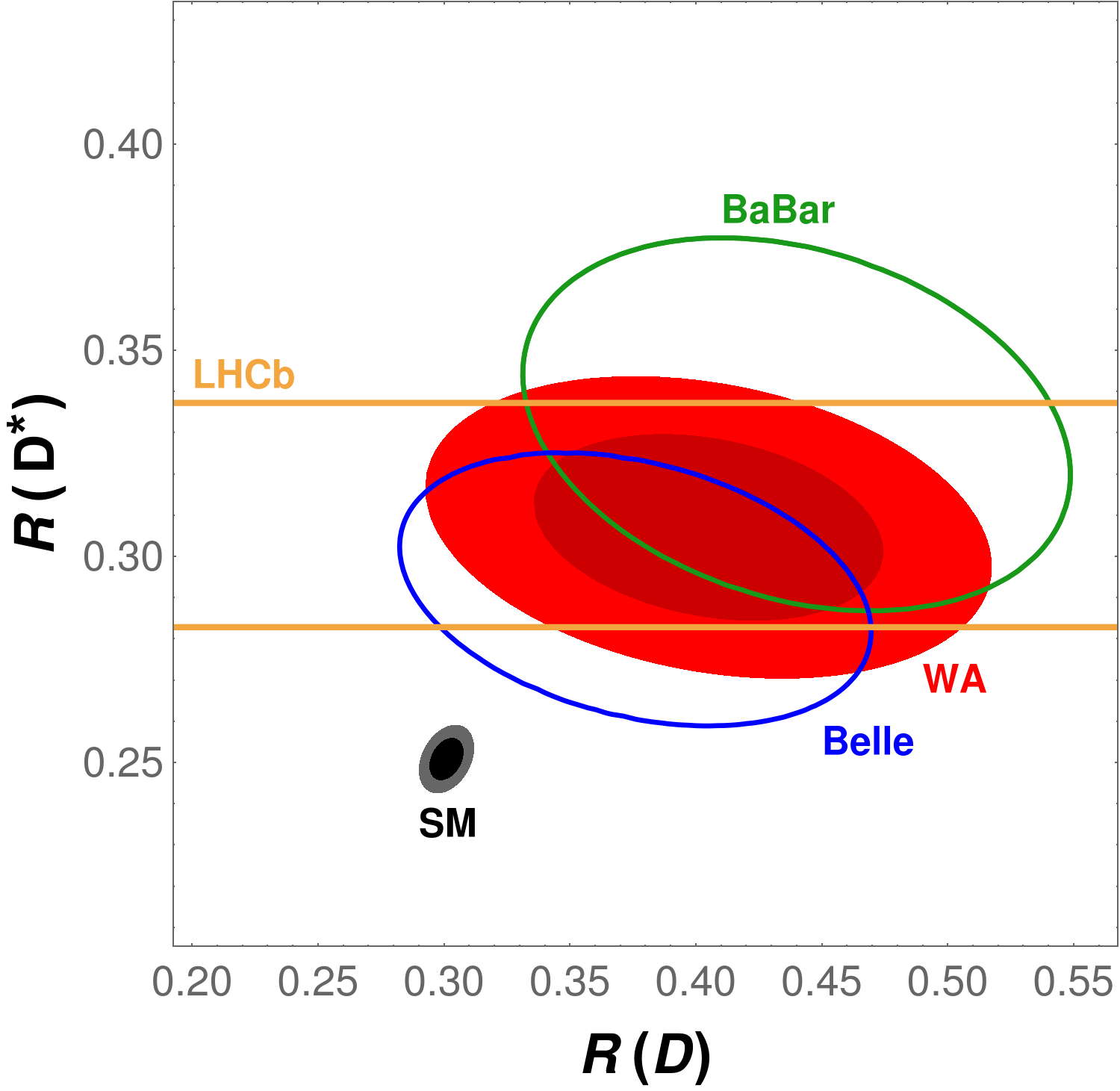}\qquad\qquad \includegraphics[width=7.3cm]{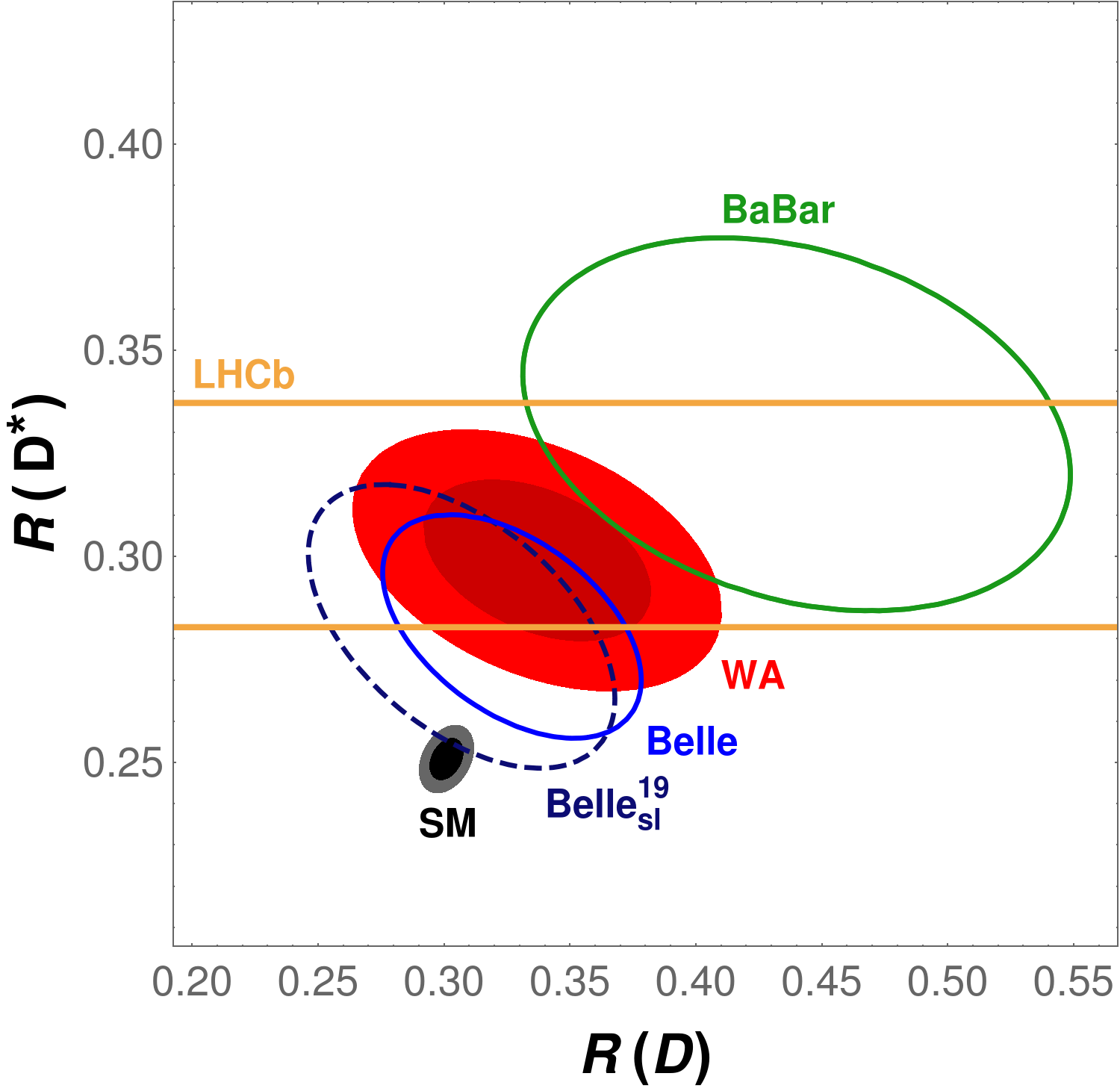}
    \caption{Measurements for $\mathcal{R}_{D}$ and $\mathcal{R}_{D^*}$, averaged for each experiment where applicable (contours corresponding to $68\%$~CL), our SM prediction and the global average (performed by HFLAV on the left, by us including the new preliminary Belle measurement~\cite{RDmoriondBelle,Abdesselam:2019XXX} on the right; filled ellipses correspond to 68 and $95\%$~CL).}
    \label{fig:RDRDstaravg}
\end{figure}

\item  \textbf{Differential distributions of the decay rates $\mathbf{\Gamma(B \to D}^{\textbf{(*)}}$ \boldmath $\tau \bar{\nu}_\tau) $}
 
Belle and BaBar have also provided data on the measured $q^2$ distributions for $B \to D^{(*)} \tau  \bar{\nu}_\tau$~\cite{Huschle:2015rga, Lees:2012xj}. We show the reported binned values in Appendix~\ref{app:addinput}, Table~\ref{table:angular}.
Since the global normalizations of these distributions are effectively already included via the values for $\mathcal{R}_{D^{(*)}}$ in these analyses, they are not independent degrees of freedom. This can be taken into account either by introducing a free normalization factor for the distributions as in Ref.~\cite{Celis:2016azn} or by normalizing the differential binned distributions in the following way:
\begin{equation}
\tilde{\Gamma}(B \to D^{(*)} \tau \bar{\nu}_\tau)_\text{bin} \equiv \frac{\Gamma (B \to D^{(*)} \tau \bar{\nu}_\tau)_\text{bin}}{\displaystyle\sum_\text{all bins} \Gamma(B \to D^{(*)} \tau \bar{\nu}_\tau)_{\text{bin}}}\, ,
\end{equation}
which keeps the information about the shape of the distribution, independently of the 
global normalization. The treatment of systematic uncertainties and correlations follows Ref.~\cite{Celis:2016azn}.

\item\textbf{The leptonic decay rate $\mathbf{B_c}$ \boldmath $\to \tau \bar{\nu}_\tau $}

While this decay is not expected to be measured in the foreseeable future, it can still be used as a constraint in the following way:
A 30-40\% upper bound can be derived from the $B_c$ lifetime~\cite{Beneke:1996xe, Celis:2016azn,Alonso:2016oyd}. A more stringent 10\% bound has been recently obtained from LEP data at the $Z$ peak \cite{Akeroyd:2017mhr}, and it may become even stronger by performing the analysis with the full L3 data \cite{Acciarri:1996bv}. However, this bound assumes the probability of a $b$ quark hadronising into a $B_c$ meson to be the same at LEP ($e^+e^-$), the Tevatron ($p\bar p$) and LHCb ($pp$), which exhibit very different transverse momenta. This is known to be a bad approximation in the case of $b$-baryons, see Ref.~\cite{Amhis:2016xyh}. The bound also makes use of the SM theoretical prediction for ${\cal B}(B_c \to J/\Psi \ell \bar{\nu}_\ell)$. See also Ref.~\cite{Blanke:2018yud} for a more detailed discussion.

In our fits, we will compare the two options of imposing the upper bounds
${\cal B} (B_c \to \tau \bar{\nu}_\tau) < 10\%\; (30\%)$. The bounds are used in a way that
only points in the parameter space that fulfill this constraint will be considered.\footnote{The uncertainty due to the precise numerical value of $V_{cb}$ (which might to a small extent also be affected by NP) is considered to be included in these bounds.}

\item\textbf{The longitudinal polarization fraction $\bm{F_L^{D^*}}$}

A measurement of the $D^*$ longitudinal polarization fraction, defined as
\begin{equation}
F_L^{D^*}= \frac{\Gamma_{\lambda_{D^*}=0} (B \to D^{(*)} \tau \bar{\nu}_\tau)}{\Gamma (B \to D^{(*)} \tau \bar{\nu}_\tau)}\, ,
\end{equation}
has been recently announced by the Belle collaboration~\cite{Abdesselam:2019wbt}. The explicit expression for $\Gamma_{\lambda_{D^*}=0} (B \to D^{(*)} \tau \bar{\nu}_\tau)$ is given in Appendix~\ref{app:FLDstar}. Being normalized to the total rate, this observable also enjoys the advantages of the other ratios. To study the implications of this measurement, we perform one fit with it and one without it.
\end{itemize}

\section{Fit and results}
\label{sec:FitResults}

In order to extract the information on the NP parameters $C_i$, we perform a standard $\chi^2$ fit. The $\chi^2$ function can be splitted in two parts,
\begin{equation}
\label{eq:chi2}
\chi^2 = \chi^2_{\rm exp} + \chi^2_{\text{FF}} \, ,
\end{equation}
where $\chi^2_{\rm exp}$  contains the experimental information discussed in the last subsection (again a sum of the three main contributions) and $\chi^2_{\rm FF}$ the information on the form factors discussed in Sec.~\ref{sec:FFs} in the form of pseudoobservables with the ``experimental'' information presented in Table~\ref{table:inputFF}.
Each individual $\chi^2$ is defined as:
\begin{equation}
\label{eq:chi2_i}
\chi^2(y_i) = F^T (y_i)  \, V^{-1} \, F (y_i)  \, , \qquad\quad F (y_i)  = f_{\text{th}} (y_i)  - f_{\text{exp}}\, ,  \qquad\quad  V_{ij} = \rho_{ij} \sigma_i \sigma_j\, ,
\end{equation}
with $y_i$ denoting the input parameters of the fit, i.e., $y_i = \{C_{V_L}, C_{S_L}, C_{S_R}, C_T$, $\rho^2$, $c$, $d$, $\chi_2(1)$, $\chi'_2(1)$, $\chi_3'(1)$, $\eta(1)$, $\eta'(1)$, $l_1(1)$, $l_2(1)\}$, $\rho_{ij}$ the correlation between the observables $i$ and $j$, and $\sigma_i$ the uncertainty of the observable $i$. In the above equation, $f_\text{th}$ represents the theoretical expression for a certain observable and $f_\text{exp}$ its experimental value. The contribution from the limit on the branching fraction of $B_c\to \tau \bar{\nu}_\tau$ is implemented as a Heavyside Theta function, its contribution being zero for parameter combinations where the limit is obeyed and infinity for those where it is not.
The uncertainty of a parameter $y_i$ is determined as the shift $\Delta y_i$ in that parameter, where the minimization of $\chi^2|_{y_i=y_i^{\rm min}+\Delta y_i}$ varying all remaining parameters in the vicinity of the minimum leads to an increase of $\Delta \chi^2=1$. 

\subsection{Standard Model}

We start by discussing the situation in the SM, corresponding to $C_i\equiv 0$. The global fit to the data discussed above does actually appear to be reasonable: we obtain $\chi^2_{\rm min}=65.5$ for, naively, 57~degrees of freedom (d.o.f.), corresponding to a naive confidence level (CL) of $\sim 20\%$. However, these numbers are misleading for the following reason: The systematic uncertainties added to the $d\Gamma/dq^2$ distributions have been chosen to be maximally conservative. Therefore, it can be expected that the corresponding $\chi^2$ contribution is reduced; this is indeed seen since the contribution from these distributions is $\chi^2_{\rm{min},d\Gamma}\sim 43$ for, again naively, 54~d.o.f.. Considering instead the contribution from $\mathcal{R}_{D^{(*)}}$, we do of course reproduce the well-known puzzle, \emph{i.e.}, we obtain $\chi^2=22.6$ for 2~d.o.f., corresponding to a $4.4\sigma$ tension. Note also that the limit from the $B_c$ lifetime is irrelevant in the SM fit.

These observations imply that also NP scenarios should not be judged simply by $\chi^2$ vs. d.o.f., but by the improvement they yield when compared to the SM.

\subsection{New Physics \label{ssec::NP}}

 Since the Wilson coefficients enter each observable bilinearly (the coefficient of the left-handed vector operator being $(1+C_{V_L})$), there is a degeneracy between a set of Wilson coefficients and a mirror minimum with
\begin{equation}
\label{eq:WilsonSecondMin}
C_{V_L}^{'} = -2 - C_{V_L} 
\qquad \text{and} \qquad
C_{i}^{'} = - C_{i}  \qquad  \text{for} \qquad i = S_R, \, S_L, \, T \, . 
\end{equation}
The two sets of Wilson coefficients give identical predictions for all observables and consequently have the same $\chi^2$ value.\footnote{This discrete degeneracy is what is left of the continuous rephasing invariance when considering complex contributions, i.e., the invariance under shifting all coefficients by the same complex phase.} In the following, we will always discuss the closest minimum to the SM scenario, i.e., with smaller $|C_{V_L}|$, and will omit the sign-flipped solution; this corresponds to considering only values $C_{V_L}\geq -1$.

\begin{figure}[tb]
\centering
\includegraphics[scale=0.49]{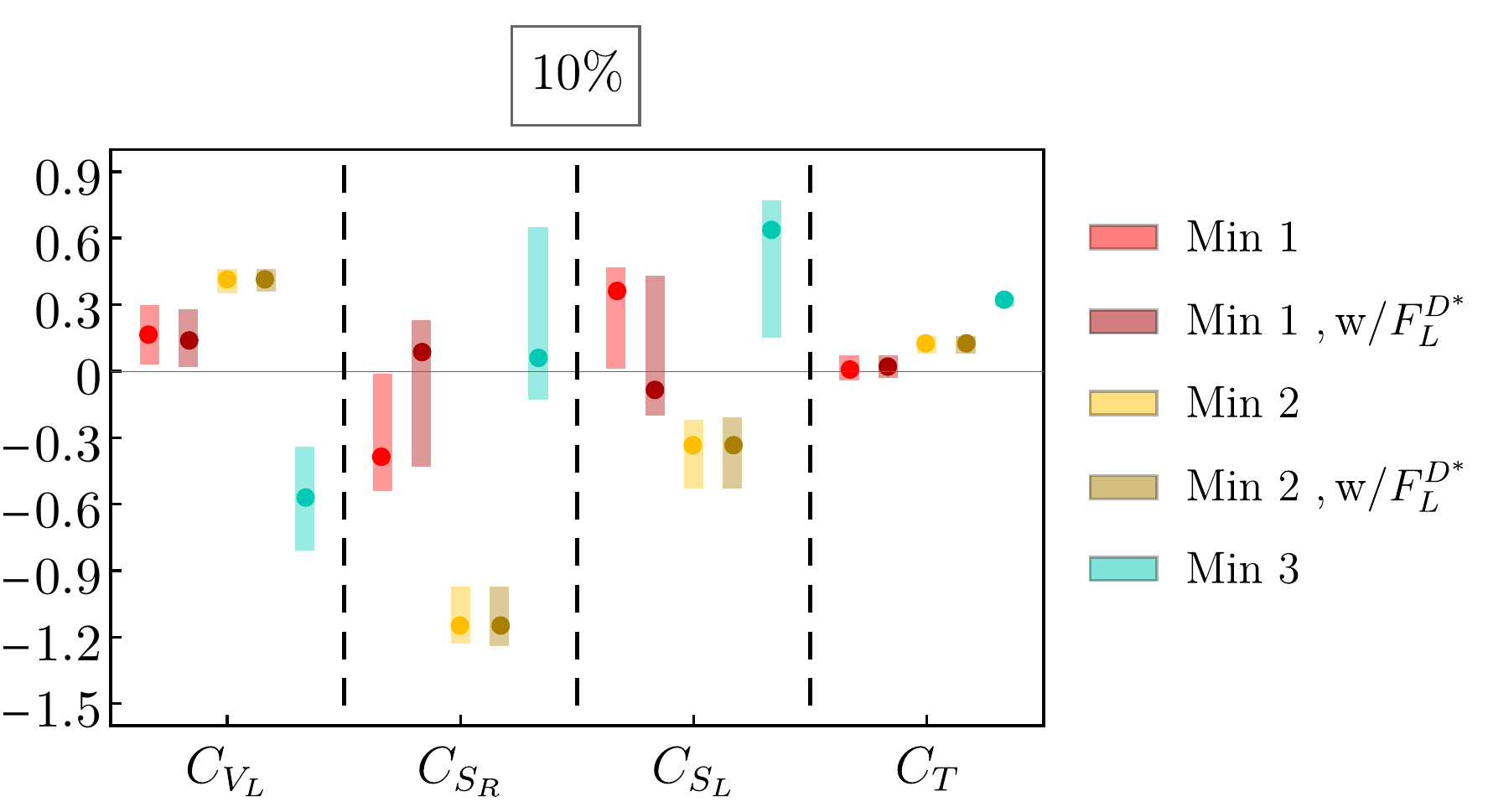}
\includegraphics[scale=0.49]{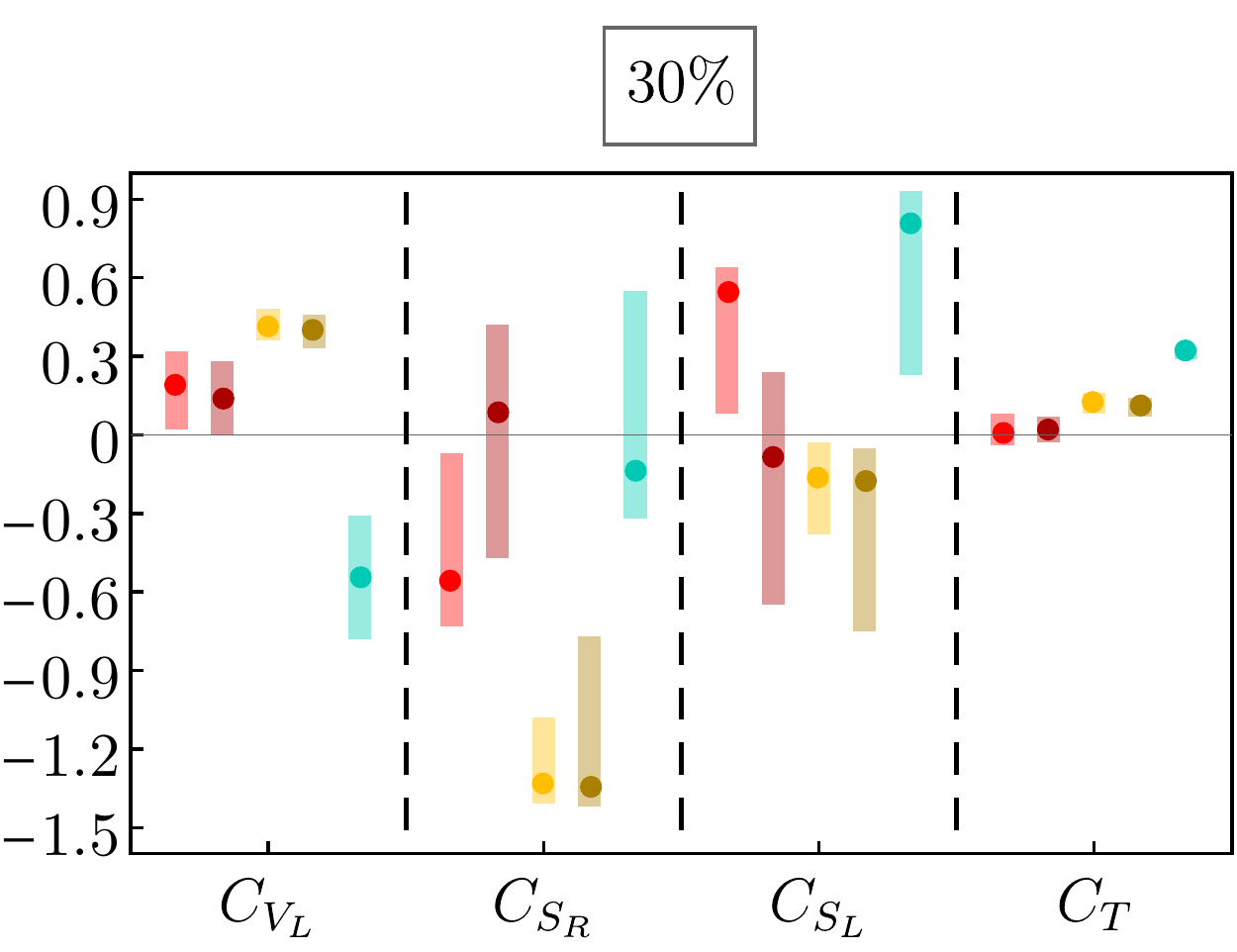}
\caption{Wilson coefficients for the minima obtained in the global fit with and without including the $F_L^{D^*}$ polarization. On the left (right) panel, ${\cal B}(B_c \to \tau \bar{\nu}_\tau) < 10\%$ ($ 30\%$). See Tables~\ref{table:minimaA} and~\ref{table:minimaB} for the explicit values.}
\label{fig:minima_plot}
\end{figure}

\begin{figure}
    \centering
    \includegraphics[scale=0.42]{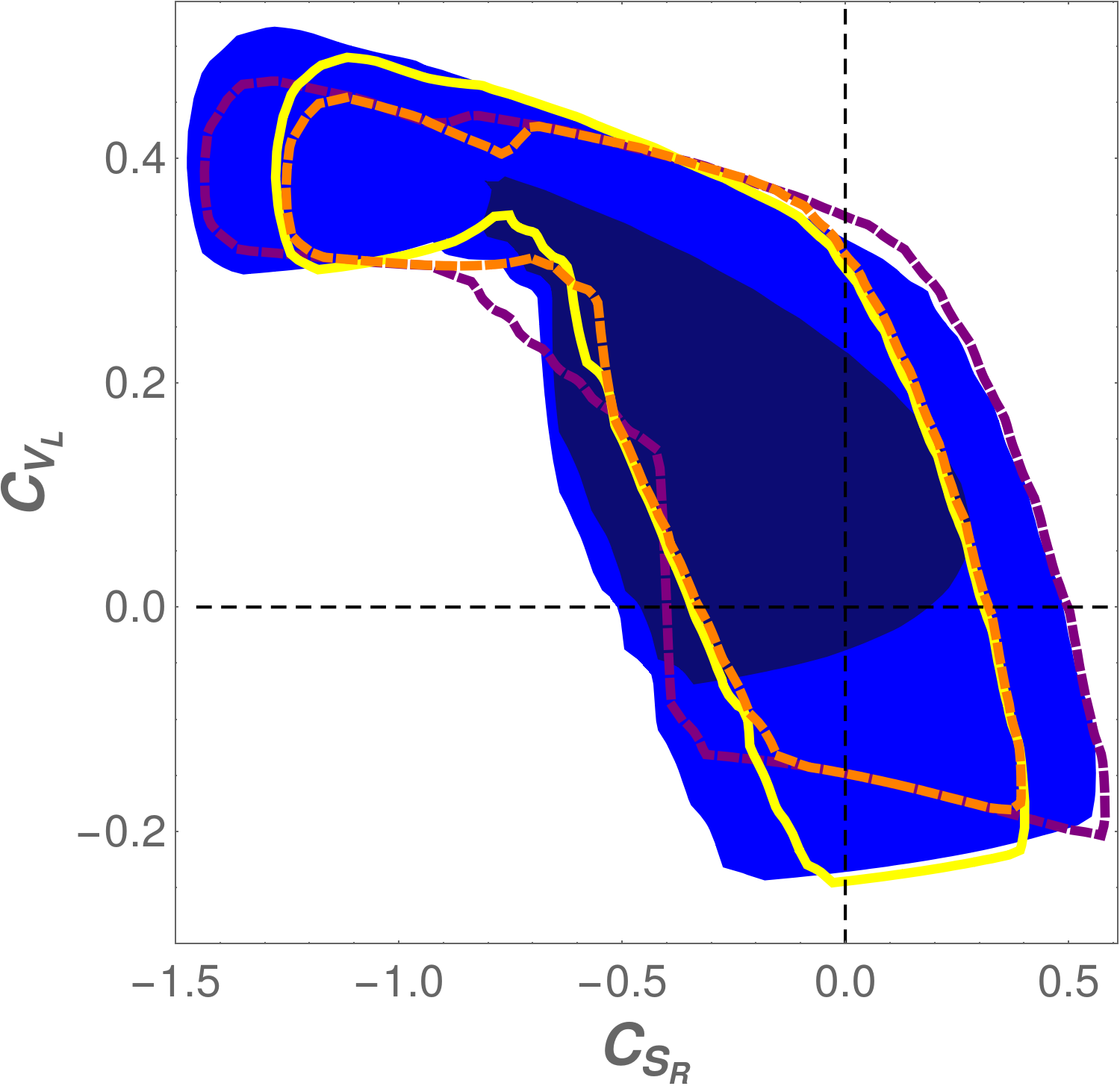}   \,
    \includegraphics[scale=0.42]{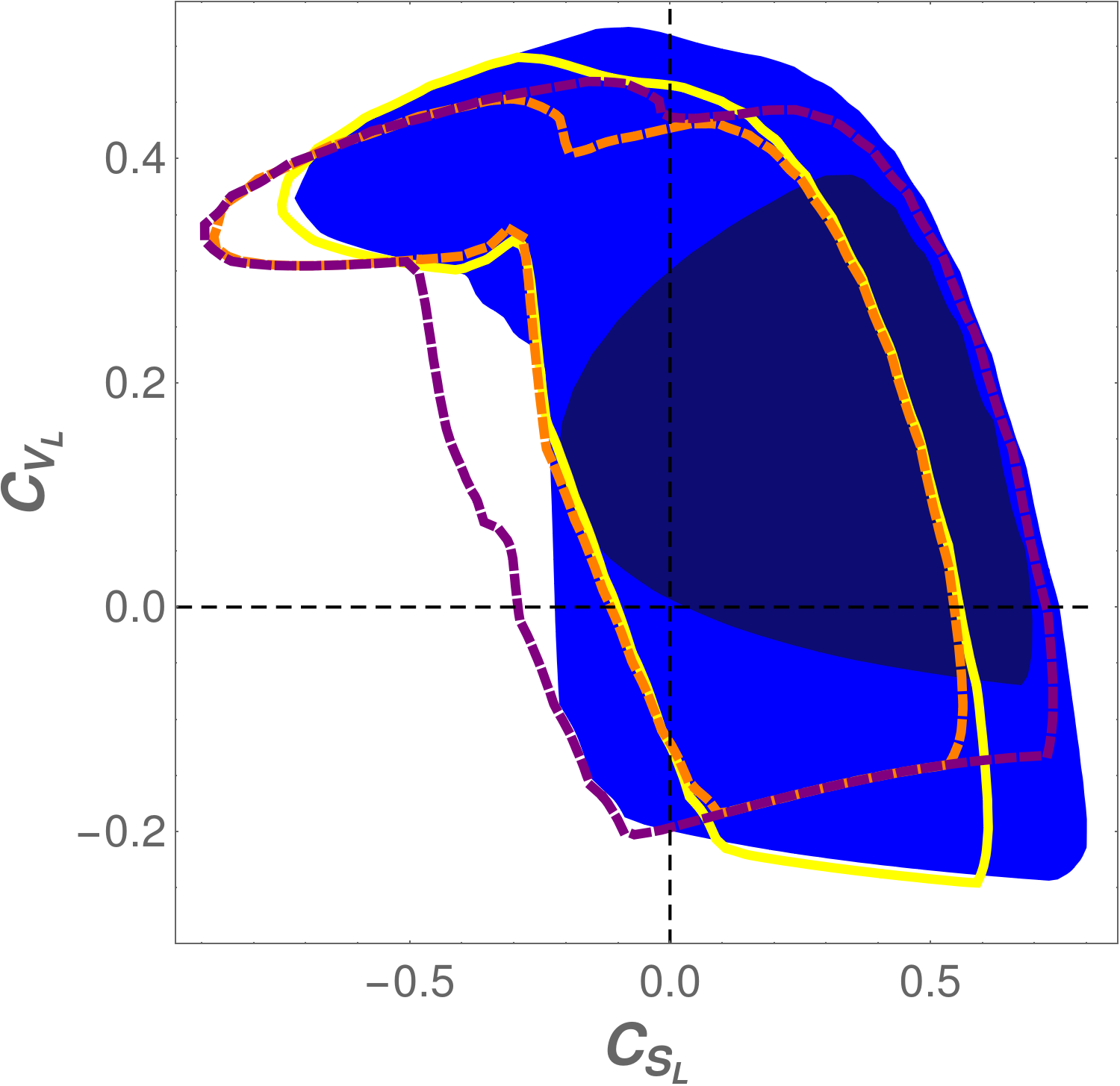}   \,
    \includegraphics[scale=0.42]{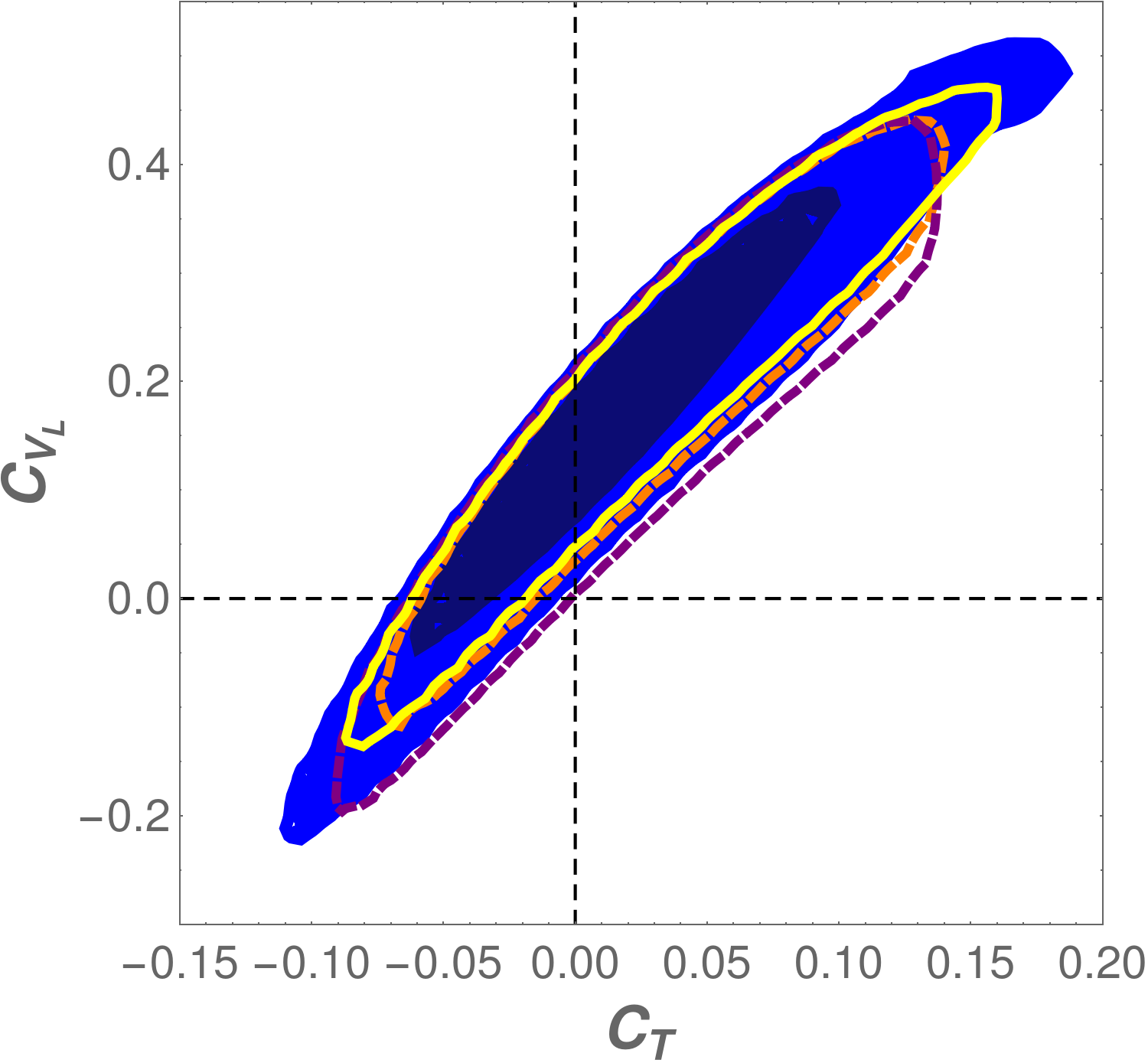}
    \includegraphics[scale=0.42]{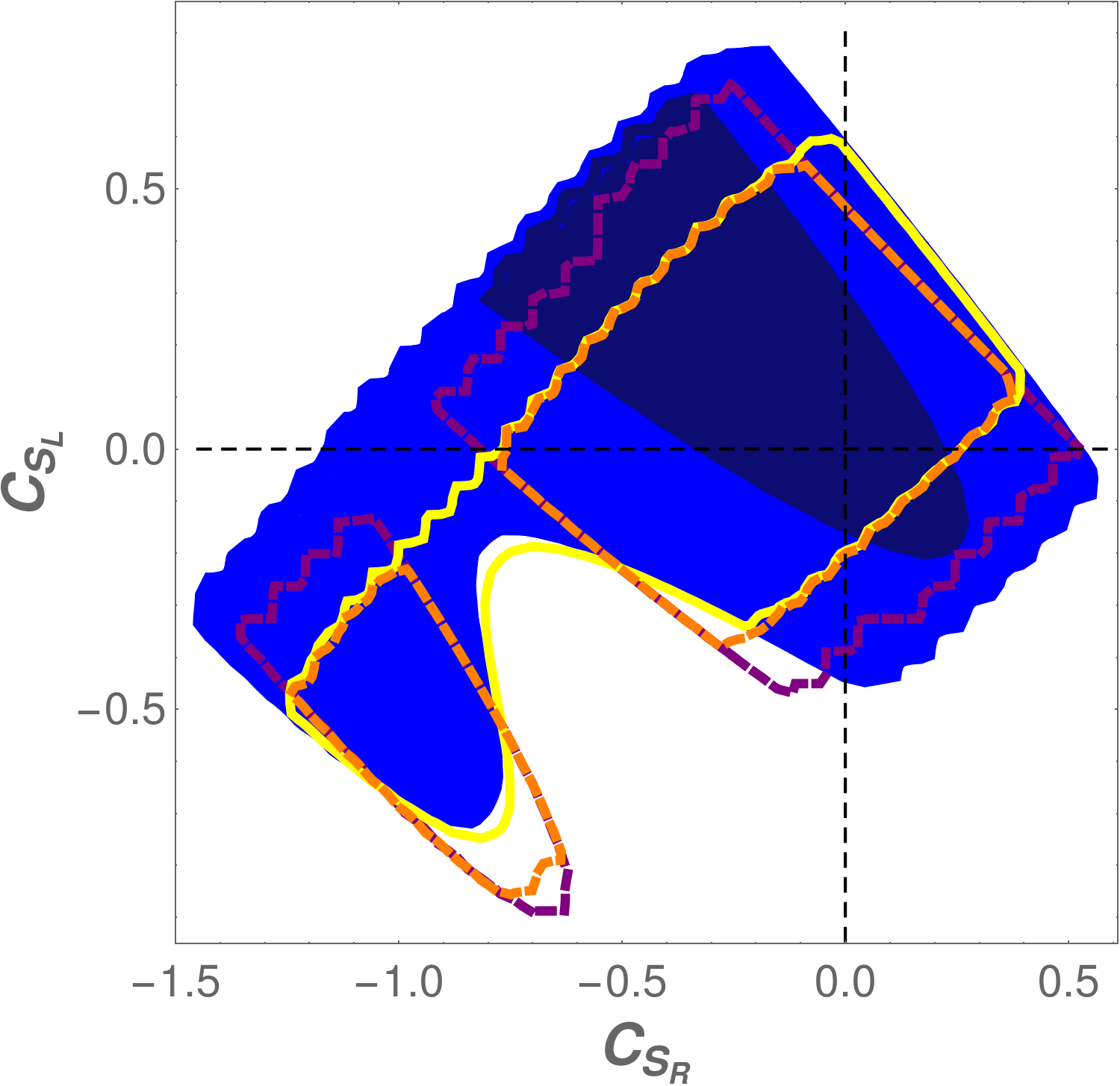}\,
    \includegraphics[scale=0.42]{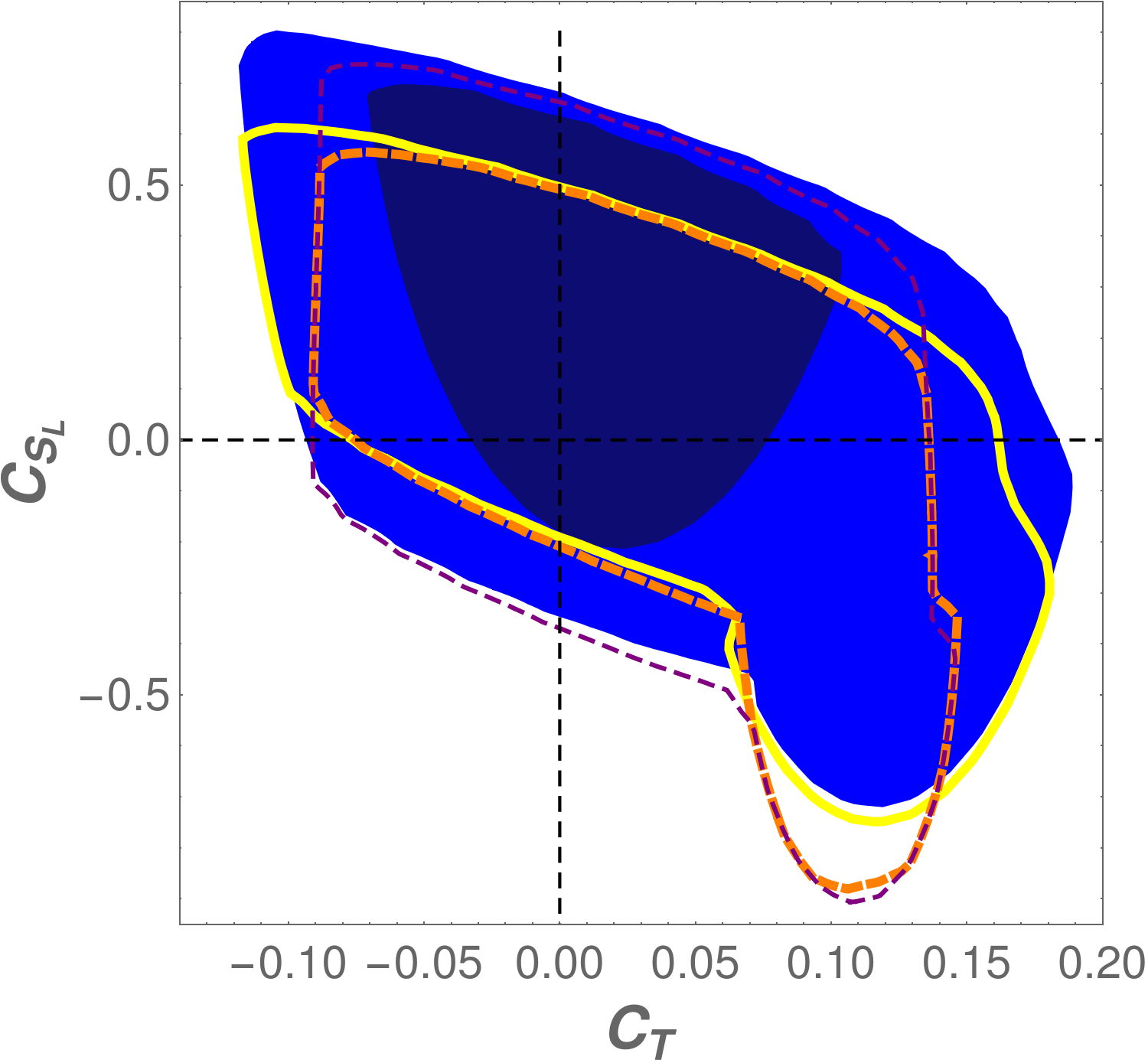}\,
    \includegraphics[scale=0.42]{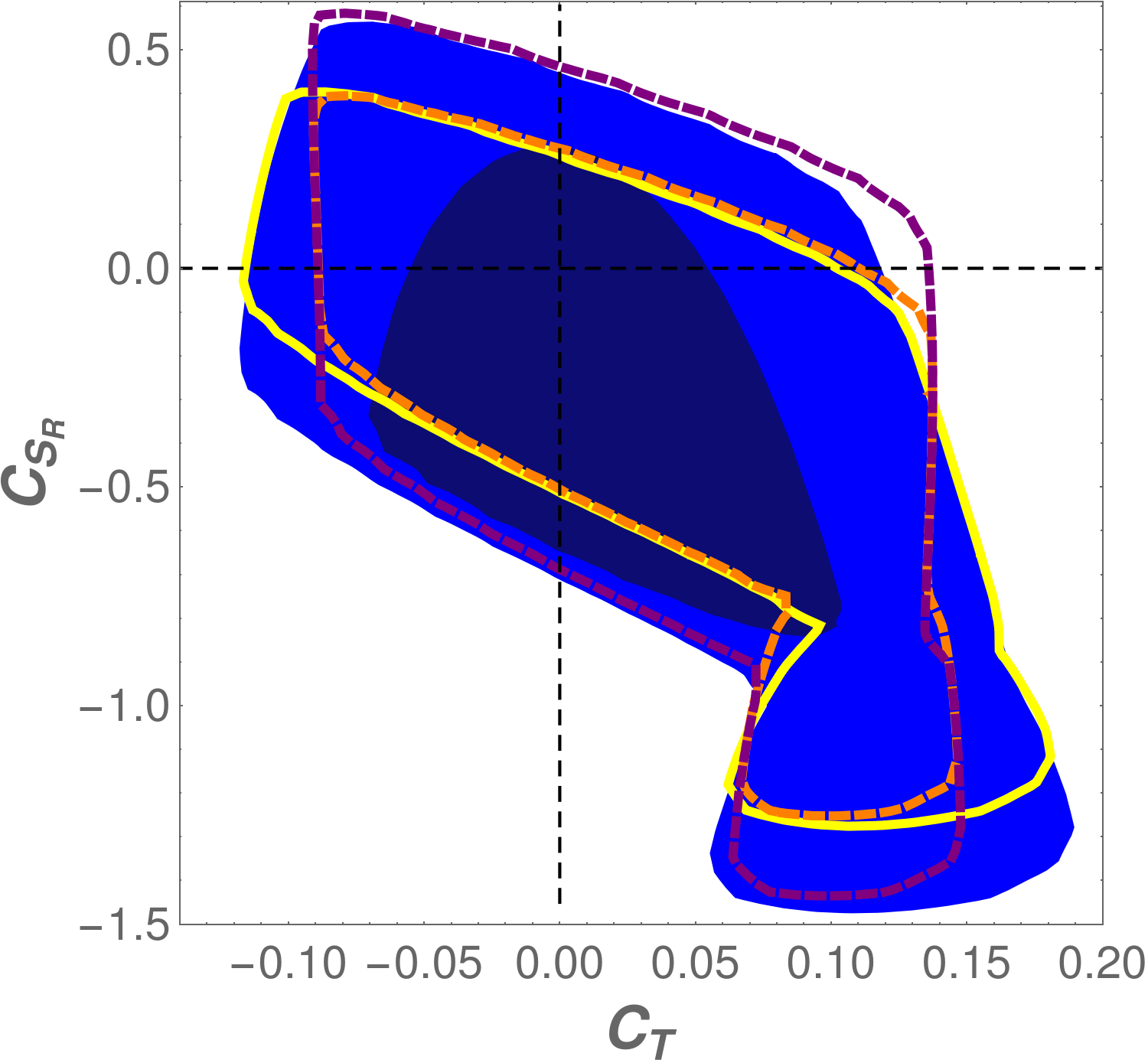}
    
    \caption{Allowed regions for all possible combinations of two Wilson coefficients for different scenarios: Blue areas (lighter 95\% and darker 68\% CL) show the minima without $F_L^{D^*}$ and with $\mathcal{B}(B_c \to \tau \bar{\nu}_\tau) \leq 30\%$. The yellow lines display how the 95\% CL bounds change when $\mathcal{B}(B_c \to \tau \bar{\nu}_\tau) \leq 10\%$. 
    The dashed lines show the effect of adding the observable  $F_L^{D^*}$ for both $\mathcal{B}(B_c \to \tau \bar{\nu}_\tau) \leq 30\%$ (purple) and for $\mathcal{B}(B_c \to \tau \bar{\nu}_\tau) \leq 10\%$ (orange).}
    \label{fig:Gloablmin_corr}
\end{figure}

The global fit to the data described in Section~\ref{sec:obs} without including the longitudinal polarization yields a unique global minimum (for $C_{V_L}>-1$) with $\chi^2_{\rm Min~1}=34.1$ for $53$ d.o.f.; in addition, we find two local minima, with $\chi^2_{\rm Min~2}=37.5$ and $\chi^2_{\rm Min~3}=58.6$, the latter of which is, however,  highly disfavoured by the differential distributions. We summarize the results for the NP parameters in Table~\ref{table:minimaA}.
Including the recently announced longitudinal polarization in the global fit, we find that the overall structure for the lower two minima remains the same; however, this observable reduces slightly the available parameter space for the NP parameters. The central values of the scalar NP parameters are smaller for the global minimum, while the $1\sigma$-ranges remain almost constant. The most striking effect is that the already less favoured local minimum disappears. The results for the NP parameters in this context can be found in Table~\ref{table:minimaB}. In both cases the form factor parameters reproduce their input distributions up to very small shifts. For illustration we show graphically in Fig.~\ref{fig:minima_plot} the NP parameters for the different minima obtained in the two scenarios.
%
There are important correlations between the NP parameters obtained from the fit. We illustrate them in the two-dimensional plots in Fig.~\ref{fig:Gloablmin_corr} for the different scenarios. The contours shown there are relative to the global minimum.

We note that the distributions for, especially, the scalar parameters are highly non-gaussian. Reasons are the way the upper limit on ${\cal B}(B_c\to\tau \bar{\nu}_\tau)$ is included and the fact that the first two minima overlap to some extent. The former is also the reason for the strong asymmetry in the uncertainties for $C_{S_{L,R}}$. Since only their sum and difference enter $B\to D$ and $B\to D^*$ decays, respectively, these parameters are furthermore highly correlated. The local minima are not very deep, resulting in complications in the determination of the uncertainties for the Wilson coefficients at these points.

\def\arraystretch{1.5}

\begin{table}[t]
\centering
\begin{tabular}{ c | c | c | c || c | c| c }
 & Min~1 & Min~2 & Min~3 & Min~1 & Min~2 & Min~3  \\
 \hline
 ${\cal B}(B_c \to \tau \nu)$ & \multicolumn{3}{c||}{10\%} & \multicolumn{3}{c}{30\%} \\
 \hline
$\chi^2_\text{min}/$d.o.f.  & $ 34.1/53 $ & $37.5/53$ & $58.6/53$ & $33.8/53$ & $36.6/53$ & $58.4/53$ \\
\hline
$C_{V_L}$ & $\phantom{-}0.17^{+0.13}_{-0.14}$
& $\phantom{-}0.41^{+0.05}_{-0.06}$& $-0.57^{+0.23}_{-0.24}$ & 
$\phantom{-}0.19^{+0.13}_{-0.17}$

& $\phantom{-}0.42^{+0.06}_{-0.06}$
& $-0.54^{+0.23}_{-0.24}$ \\
$C_{S_R}$ & $-0.39^{+0.38}_{-0.15}$
& $-1.15^{+0.18}_{-0.08}$
& $\phantom{-}0.06^{+0.59}_{-0.19}$
& $-0.56^{+0.49}_{-0.17}$
& $-1.33^{+0.25}_{-0.08}$

& $-0.14^{+0.69}_{-0.18}$ \\ 
$C_{S_L}$ & $\phantom{-}0.36^{+0.11}_{-0.35}$
&$-0.34^{+0.12}_{-0.19}$ & $\phantom{-}0.64^{+0.13}_{-0.49}$ & 
$\phantom{-}0.54^{+0.10}_{-0.46}$
& $-0.16^{+0.13}_{-0.22}$
& $\phantom{-}0.81^{+0.12}_{-{0.58}}$ \\
$C_T$ & $\phantom{-}0.01^{+0.06}_{-0.05}$&
$\phantom{-}0.12^{+0.04}_{-0.04}$

&$\phantom{-}0.32^{+0.02}_{-0.03}$ &
$\phantom{-}0.01^{+0.07}_{-0.05}$ 
& $\phantom{-}0.12^{+0.04}_{-0.04}$
& $\phantom{-}0.32^{+0.02}_{-0.03}$ \\
\end{tabular}
\caption{NP parameters for the minima obtained from the $\chi^2$ minimization and $1\sigma$ uncertainties. 
There are, in addition, three corresponding sign-flipped minima, as indicated in Eq.~\eqref{eq:WilsonSecondMin}. In the first three columns, the constraint ${\cal B}(B_c \to \tau \bar{\nu}_\tau)\leq 10\%$ has been applied, whereas in the last three columns, this requirement has been relaxed to ${\cal B}(B_c \to \tau \bar{\nu}_\tau)\leq 30\%$.}
\label{table:minimaA}
\end{table}

\def\arraystretch{1.5}
\begin{table}[h]
\centering
\begin{tabular}{ c | c | c || c | c}    
 & Min~1b & Min~2b & Min~1b & Min~2b \\
 \hline
 ${\cal B}(B_c \to \tau \nu)$ & \multicolumn{2}{c||}{10\%} & \multicolumn{2}{c}{30\%} \\
 \hline
$\chi^2_\text{min}/$d.o.f.  & $ 37.6/54 $ &$42.1/54$ &37.6 /54 & 42.0/54 \\
\hline
$C_{V_L}$ & $\phantom{-}0.14^{+0.14}_{-0.12}$ & $\phantom{-}0.41^{+0.05}_{-0.05}$ & $\phantom{-} 0.14^{+0.14}_{-0.14}$ & $\phantom{-}0.40^{+0.06}_{-0.07} $ \\
$C_{S_R}$ & $\phantom{-}0.09^{+0.14}_{-0.52}$ & $-1.15^{+0.18}_{-0.09}$ & $\phantom{-} 0.09^{+0.33}_{-0.56}$ & $-1.34^{+0.57}_{-0.08}$ \\ 
$C_{S_L}$ & $-0.09^{+0.52}_{-0.11}$  &$-0.34^{+0.13}_{-0.19}$ & $-0.09^{+0.68}_{-0.21}$& $-0.18^{+0.13}_{-0.57}$ \\
$C_T$ & $\phantom{-}0.02^{+0.05}_{-0.05}$ & $\phantom{-}0.12^{+0.04}_{-0.04}$  & $\phantom{-}0.02^{+0.05}_{-0.05}$ & $\phantom{-}0.11^{+0.03}_{-0.04}$  \\
\end{tabular}
\caption{NP parameters for the minima obtained from the $\chi^2$ minimization including $F_L^{D^*}$ and their $1\sigma$ uncertainties. There are, in addition, the corresponding sign-flipped minima, as indicated in Eq.~\eqref{eq:WilsonSecondMin}.}
\label{table:minimaB}
\end{table}
\def\arraystretch{1.5}

\begin{figure}
\centering
\includegraphics[scale=0.5]{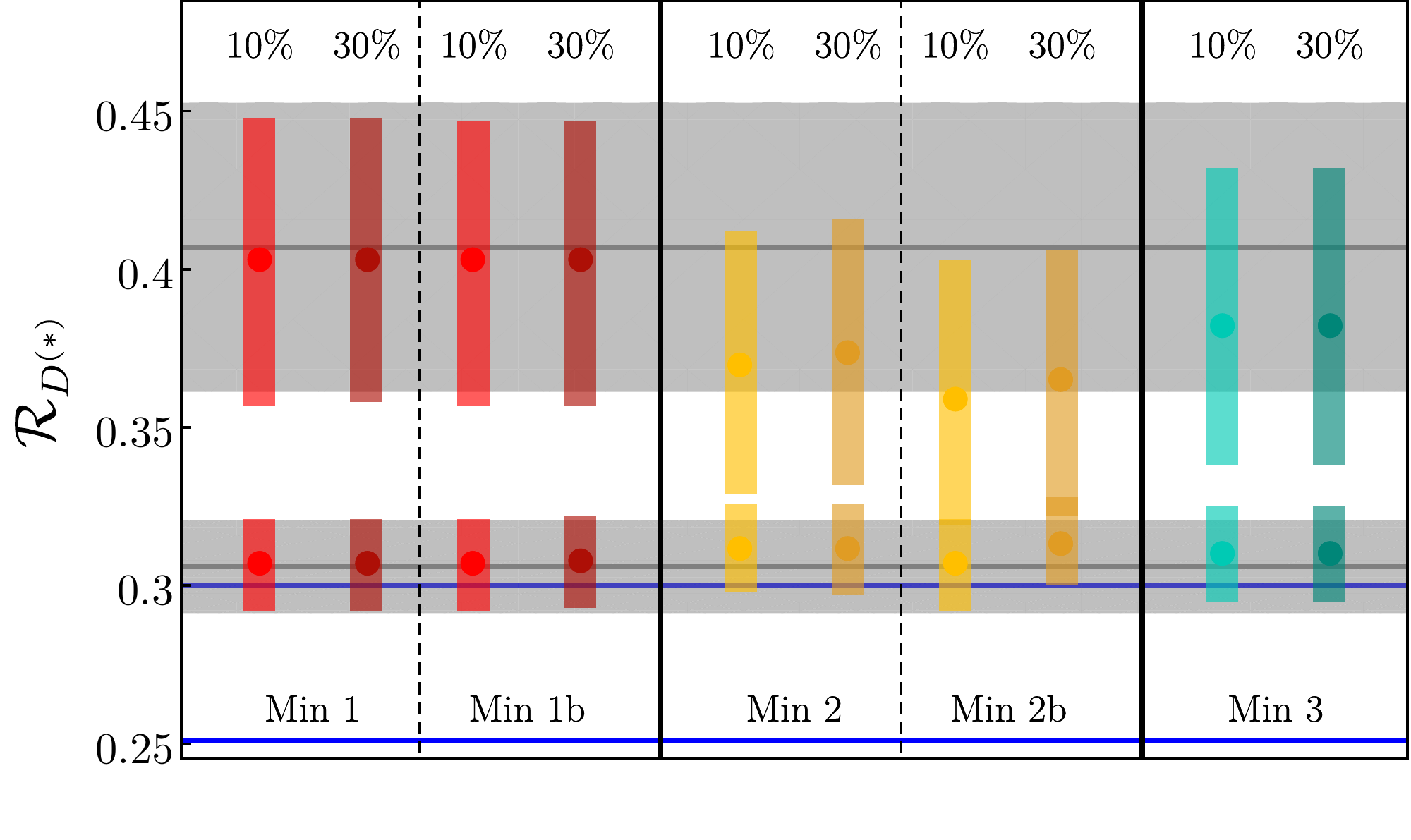}
\caption{Predictions for $\mathcal{R}_D$ (higher numerical values) and $\mathcal{R}_{D^*}$ (lower numerical values) for the minima obtained in the fit, both with and without including $F_L^{D^*}$, with $\mathcal{B}(B_c \to \tau \bar{\nu}_\tau )\leq 10\%$ and $\mathcal{B}(B_c \to \tau \bar{\nu}_\tau )\leq 30\%$. The experimental values are represented by the horizontal black lines, with their corresponding uncertainties (grey bands). The blue lines show the SM predictions, ${\cal R}_D = 0.300^{+0.005}_{-0.004}$ (upper blue line) and ${\cal R}_{D^*} = 0.251^{+0.004}_{-0.003}$ (lower blue line).}
\label{fig:predRs}
\end{figure}

The fit results for the ${\cal R}_{D}$ and ${\cal R}_{D^*}$ ratios at the different minima are presented in Fig.~\ref{fig:predRs}. As expected, the predictions obtained from the fit are compatible at the $1\sigma$ level with the experimental data, in the case of Min~1 and Min~1b essentially reproducing them.
From the fit results without including $F_L^{D^*}$, the following information can be extracted:

\begin{itemize}

\item The reduction of the global $\chi^2$ by 31.4 (31.7) for 4 NP parameters implies a strong preference of NP compared to the SM, taking the present data set at face value and $\mathcal{B}(B_c \to \tau \bar{\nu}_\tau )\leq 10\%$ (30\%).

\item There is no absolute preference of a single Wilson coefficient in the sense that for the global minimum each individual Wilson coefficient is compatible with zero within at most $1.1\sigma$.

\item On the other hand, considering scenarios with only a single Wilson coefficient present, there \emph{is} a clear preference for $C_{V_L}$: removing the other three Wilson coefficients increases $\chi^2$ only by $1.4$, corresponding to $0.14\sigma$. Hence, Min~1 is well compatible with a global modification of the SM, that is, $C_{V_L}$ being the only non-zero coefficient.

\item The other two minima are numerically further away from the SM; instead of a single  dominant contribution, there are several sizeable Wilson coefficients whose contributions partly cancel each other in some observables. These minima also imply different values for the fitted observables: Min~2 corresponds to a slightly worse fit for both, $\mathcal{R}_{D^{(*)}}$ and their $q^2$ distributions, while Min~3 fits $\mathcal{R}_{D^{(*)}}$ perfectly, but is essentially already excluded by the (rather coarse) measurements of the distributions available. 

\item All minima saturate the constraint ${\cal B}(B_c \to \tau \bar{\nu}_\tau) \leq 10 \%$ (30\%). 
Relaxing the upper bound allows for a larger splitting between the two scalar Wilson coefficients, and the contribution of the scalar operators gets enlarged. This constraint is consequently the main argument at low energies disfavouring a solution with only scalar coefficients. Any such solution would require a lower value for $\mathcal{R}_{D^{*}}$ by about $2\sigma$.

\item Having solutions with relevant contributions from all Wilson coefficients illustrates the importance of taking into account scalar and tensor operators in the fit.

\item The fit results for the form factor parameters  reproduce their input values displayed in Table~\ref{table:inputFF} up to tiny shifts. This implies that the uncertainties of the experimental data with tauonic final states are large compared to the hadronic uncertainties. Differently stated, while the ranges obtained for the NP parameters are obtained in fits varying all form factor parameters simultaneously with the NP ones, they are essentially determined by the experimental uncertainties at the moment.

\item Generalizing the fit to complex Wilson coefficients does not improve the minimal $\chi^2$ value, but opens up a continuum of solutions. Hence complex Wilson coefficients can explain the anomalies as well as real ones, but they do not offer any clear advantages regarding the fit quality, so they have not been considered here for simplicity. It should be mentioned, however, that in specific models the option of complex Wilson coefficients can open up qualitatively new solutions, as for example the model proposed in Ref.~\cite{Becirevic:2018afm}, where only the coefficients $C_{S_L,T}$ ($C_{S_L}\sim C_T$) are present, \emph{requiring} a non-vanishing imaginary part in order to accommodate the experimental data. This fact implies correlations with new observables like electric dipole moments, which can then be used to differentiate this model from solutions allowing for real coefficients \cite{Dekens:2018bci}.

\item As discussed above, for each minimum given in Table~\ref{table:minimaA} there is a degenerate solution, see Eq.~\eqref{eq:WilsonSecondMin}.

\end{itemize}

Including the recent measurement of the longitudinal polarization $F_L^{D^*}$ in the global fit, the above statements hold up to the following differences:
\begin{itemize}
\item Still there is no clear preference for a single Wilson coefficient. The central values for the scalar coefficients are smaller for the global minimum, such that the bound from the $B_c$ lifetime is not saturated even in the $10\%$ case. As a consequence, the minimum does not change when allowing for larger values of ${\cal B}(B_c\to \tau\bar\nu_\tau)$, only the allowed parameter ranges increase.
\item The second local minimum (previously referred to as Min~3) disappears.
\end{itemize}
It is not straightforward to compare our fit with the results from other analyses in the literature, because we are including the information from the $q^2$ distributions that has been ignored in previous fits with the exception of Ref.~\cite{Sakaki:2014sea,Freytsis:2015qca,Celis:2016azn, Bhattacharya:2016zcw}.
Besides that, some works include additional observables such as $\mathcal{R}_{J/\psi}$ or slightly different bounds on ${\cal B}(B_c \to \tau \bar{\nu}_\tau)$.
Nevertheless, comparing the findings of previous fits with our results is quite enlightening since it illustrates the relevance of the additional observables we are considering.

Generic fits to the  $\mathcal{R}_{D^{(*)}}$ world averages in Eq.~(\ref{eq:RDavg}), with the effective Hamiltonian of Eq. \eqref{eq:effH}~\cite{Celis:2012dk,Datta:2012qk,Duraisamy:2013kcw, Dutta:2013qaa,Sakaki:2013bfa,Duraisamy:2014sna, Freytsis:2015qca,Alonso:2015sja,Boubaa:2016mgn, Bardhan:2016uhr, Bhattacharya:2016zcw, Alonso:2016oyd, Choudhury:2016ulr, Celis:2016azn, Alok:2016qyh,Alok:2017qsi,Bernlochner:2017jka, Capdevila:2017iqn,Altmannshofer:2017poe,Buttazzo:2017ixm,Cai:2017wry,  Crivellin:2017zlb,Jung:2018lfu, Biswas:2018jun,Azatov:2018knx,Hu:2018veh,Angelescu:2018tyl,Blanke:2018yud, Bhattacharya:2018kig}, have shown the existence of many possible solutions, some of them involving only one or two Wilson coefficients. Including the ${\cal B}(B_c \to \tau \bar{\nu}_\tau)$ upper bound reduces the number of allowed possibilities, but several different scenarios remain still consistent with the data.
Dropping the binned $q^2$ distributions from our fit, we can easily reproduce all those solutions. However, most of them lead to differential distributions in clear conflict with the BaBar and Belle measurements. While a sizeable new-physics contribution to some Wilson coefficient can easily generate the needed enhancement of the $B\to D^{(*)}\tau \bar{\nu}_\tau$ rates, it tends to distort the shape of the differential distributions in a way than can no-longer accommodate the data, similarly to what happens for Min~3. Once the full experimental information on $\mathcal{R}_{D^{(*)}}$ (rates and binned distributions) is taken into account, the $\chi^2$ minimization only gives the three solutions shown in Table~\ref{table:minimaA}, and when including $F_L^{D^*}$ in the fit, the number of solutions is further reduced to two.

Finally, a few comments on the very recent measurement of ${\cal R}_{D}$ and ${\cal R}_{D^*}$ released in Moriond by Belle~\cite{RDmoriondBelle,Abdesselam:2019XXX} are in order. 
It should be kept in mind, however, that these results are still preliminary. Including the new average in the fit (see Fig.~\ref{fig:RDRDstaravg}), we find again qualitatively similar solutions as before, as can be seen by comparing the numerical results in Tables~\ref{table:minimaB} and~\ref{table:minima_Belle}. 
We show for simplicity only the solutions with ${\cal B}(B_c\to\tau\bar\nu_\tau)<10\%$; increasing this limit results again essentially in larger ranges for especially the scalar Wilson coefficients, although the new global minimum now \emph{does} saturate this limit, so also the central values do change. Again all individual coefficients are roughly compatible with zero at $1\sigma$. $C_{V_L}$ alone also still provides an excellent fit to all the data, now with a smaller central value of $\sim 0.08$. Interestingly, the fit with only $C_T$ is improved by the new results, which, however, does not correspond to a simple single-mediator scenario, as discussed below. However, related to that observation, also the fit in the scenario of Ref.~\cite{Becirevic:2018afm} improves by $\Delta \chi^2=-1.8$ (for ${\cal B}(B_c\to\tau\bar\nu_\tau)<30\%$).

\def\arraystretch{1.5}
\begin{table}[h]
\centering
\begin{tabular}{ c | c |  c }
 & Min 1b & Min 2b  \\
 \hline
$\chi^2_\text{min}/$d.o.f.  & $ 37.4/54 $ & $40.1/54$ \\
\hline
$C_{V_L}$ & $\phantom{-}0.09^{+0.13}_{-0.11}$
& $\phantom{-}0.35^{+0.04}_{-0.07}$ \\
$C_{S_R}$ & $\phantom{-}0.14^{+0.06}_{-0.67}
$ 
& $-1.27^{+0.66}_{-0.07}$
 \\ 
$C_{S_L}$ & $-0.20^{+0.58}_{-0.03} $ 
&$-0.30^{{+0.12}}_{-0.51}$ \\
$C_T$ & $\phantom{-} 0.007 ^{+ 0.046}_{- 0.044} $ &
$\phantom{-}0.091^{+0.029}_{-0.030}$ \\
\end{tabular}
\caption{Minima and $1\sigma$ uncertainties obtained from the global $\chi^2$ minimization, including the new preliminary result measured by Belle on the ${\cal R}_{D^{(*)}}$ ratios and the $F_L^{D^*}$ polarization, using ${\cal B}(B_c \to \tau \bar{\nu}_\tau )<10\%$. There are, in addition, the corresponding sign-flipped minima, as indicated in Eq.~\eqref{eq:WilsonSecondMin}.}
\label{table:minima_Belle}
\end{table}

\section{Interpretation of results}
\label{sec:interpretation}

\begin{figure}
\centering
\includegraphics[scale=0.25]{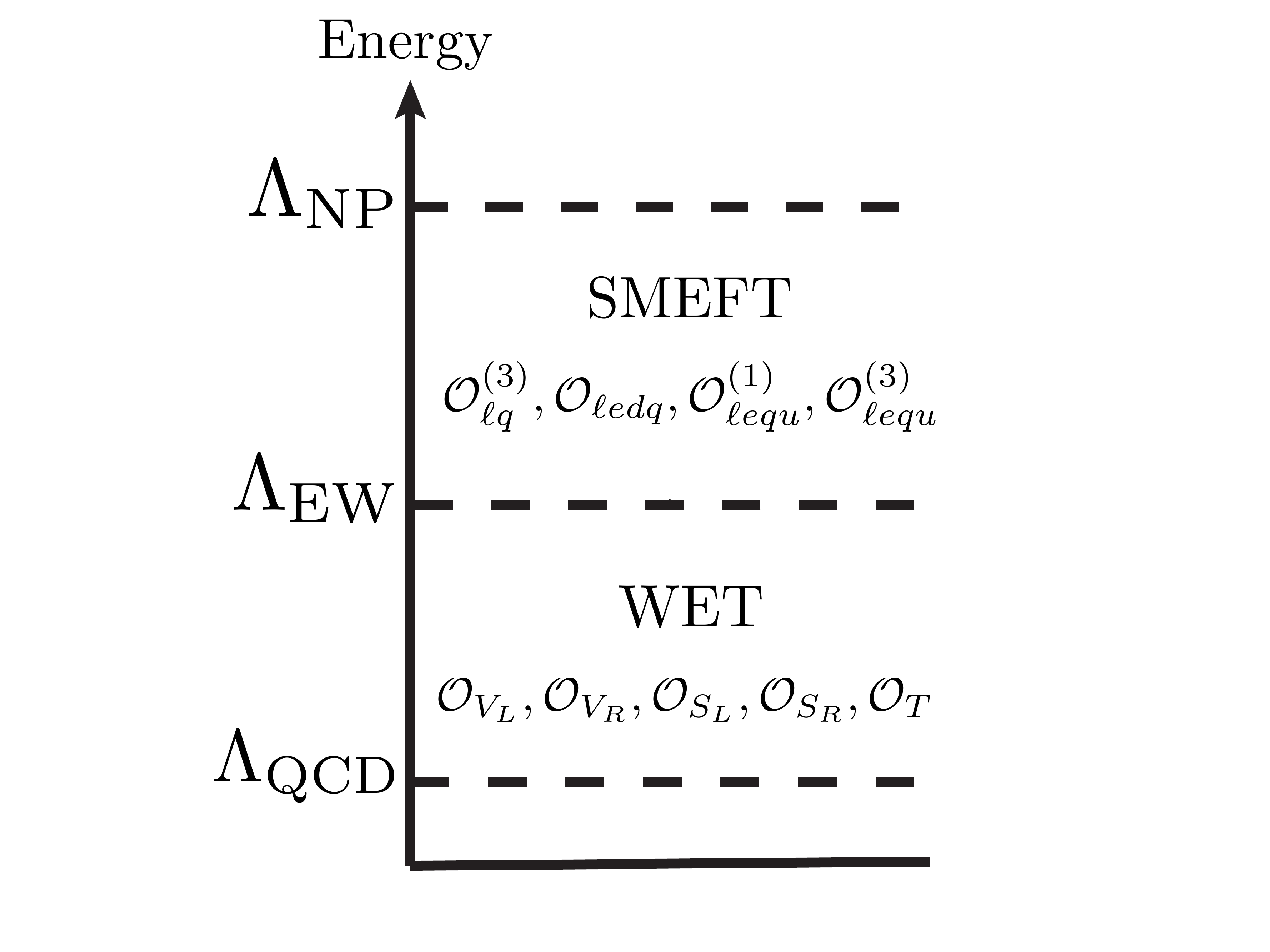}
\caption{Relevant scales for the study of the $B$ anomalies. The dashed lines indicate the thresholds between different EFTs.}
\label{scales}
\end{figure}

In Sec.~\ref{sec:FitResults} we have described the global fit to the available data on $b\to c\tau \bar{\nu}_\tau$ transitions in terms of the Wilson coefficients of an EFT framework defined at the $b$-quark mass scale. The EFT in this range is conventionally called Weak Effective Theory (WET) and is composed of the five lightest quarks and the three generations of leptons, and ruled by the $SU(3)_C \otimes U(1)_Q$ gauge symmetry. This is a valid approach assuming -- as strongly suggested by all available collider data -- that no new degree of freedom exists coupling to this channel with a mass around or lower than the $b$ quark. However, ultimately the goal is to gain insight into the high-energy structure of the theory. To that aim, renormalization-group techniques are used to relate the coefficients extracted in our analysis to those relevant at the scale of the potential new high-energy degree(s) of freedom. This process involves several scales and thresholds, see Fig.~\ref{scales}.

The relation to the coefficients at the electroweak scale is determined by QCD and are known~\cite{Buchalla:1995vs,Aebischer:2017gaw,Jenkins:2017jig,Jenkins:2017dyc}. 
Above the electroweak (EW) scale, the Lagrangian has not undergone spontaneous symmetry breaking and, therefore, the fermionic fields should be expressed in terms of weak eigenstates rather than mass eigenstates. Moreover, the top quark, the electroweak gauge bosons and the Higgs boson have to be considered as new degrees of freedom in the theory. The relevant framework at this scale is the full SM, with the addition of the effects of NP. For relatively low NP scales $\lesssim 1$~TeV, the relevant new degrees of freedom can be included explicitly. However, the suggested absence of new degrees of freedom below $\sim 1$~TeV allows us to parametrize any NP contribution in the framework of another effective theory. This can be the so-called SMEFT under the conditions specified in Section~\ref{sec:TheoricalFramework}, or a more general framework with a non-linear representation for the Higgs, see, \emph{e.g.}, Ref.~\cite{Feruglio:1992wf,Pich:2018ltt}. 

In SMEFT, the effective lagrangian can be expanded in inverse powers of the NP scale, $\Lambda_\text{NP}$, i.e.,
\begin{equation}
{\cal L}_{\text{NP}}\, =\, \sum_{d=6}\frac{1}{\Lambda_\text{NP}^{d-4}}\sum_i C_i^{(d)} {\cal O}_i^{(d)}\,,
\label{eq:NPHam}
\end{equation}
built from a series of higher-dimensional operators in terms of the SM fields and invariant under the SM gauge group $SU(3)_C \otimes SU(2)_L \otimes U(1)_Y$ \cite{Buchmuller:1985jz}. A convenient complete and non-redundant basis of dimension-six operators is the Warsaw basis \cite{Grzadkowski:2010es}. In order to relate both EFTs, the matching between the WET theory and the SMEFT has to be performed at the EW scale~\cite{Cirigliano:2009wk,Alonso:2014csa,Aebischer:2015fzz,Jenkins:2017jig,Jenkins:2017dyc,Aebischer:2018bkb}.
The matching onto the basis in the non-linear case \cite{Buchalla:2012qq,Buchalla:2013rka} is given in Ref.~\cite{Cata:2015lta}.

Finally, one has to consider the running  from $\Lambda_{\text{EW}}$ to $\Lambda_{\text{NP}}$~\cite{Jenkins:2013zja,Jenkins:2013wua,Alonso:2013hga,Gonzalez-Alonso:2017iyc,Hu:2018veh}. The corresponding equations can be solved numerically, but also analytically to very good approximation \cite{Buras:2018gto}.

As an illustration of the effect of the running, we show the relation between the WET Wilson coefficients at $\mu_b \approx 5$ GeV and the SMEFT Wilson coefficients at an hypothetical NP scale of $\Lambda = 1$ TeV, calculated in Ref.~\cite{Hu:2018veh,Gonzalez-Alonso:2017iyc}, which can be trivially inverted:
\beqn  
\no
C_{V_L} (\mu_b) &=& - 1.503\; \tilde{C}_{V_L} (\Lambda) \, , \\
\no
C_{S_L}(\mu_b)  &=& -1.257\; \tilde{C}_{S_L} (\Lambda) + 0.2076\; \tilde{C}_{T}  (\Lambda)\, , \\
C_{S_R} (\mu_b) &=& -1.254\; \tilde{C}_{S_R} (\Lambda)\, , \\
\no
C_T (\mu_b) &=& 0.002725\; \tilde{C}_{S_L} (\Lambda) - 0.6059\; \tilde{C}_{T} (\Lambda)\, . 
\label{eq:running}
 \eeqn
For a discussion of the notation used for the SMEFT Wilson coefficients in the Warsaw  basis see Appendix~\ref{appendix:Warsaw basis}.

With the coefficients at the potential NP scale at hand, one can try to go beyond the EFT framework and get an idea about which class of NP might be responsible for the observed pattern: at the scale $\Lambda$, the coefficients $C_i$ should result from integrating out the new heavy degrees of freedom.
In Table~\ref{table:NPQN}, the quantum numbers of all possible candidates able to participate in the $b \to c$ transitions are listed and their nature is identified (see also \cite{Freytsis:2015qca}). We note that, in some cases, a given NP mediator may contribute to more than one Wilson coefficient, thus resulting in correlations among them. In Appendix~\ref{app:UVLagrangian}, we list the effective Lagrangians obtained after integrating out each of the possible heavy degrees of freedom.
%
\begin{table}[tb]
    \centering
    \begin{tabular}{|c| c | c | c | c | c |}
    \hline
        Spin & Q.N. & Nature & Allowed couplings & SMEFT & WET \\
        \hline
0 & $S_1 \sim (\bar{3},1,1/3)$ & LQ & $\overline{q_L^c}\ell_L$, $\overline{d_R}u_R^c$, $\overline{u_R^c}e_R$ & $\tilde{C}_{V_L}, \tilde{C}_{S_L}$, $\tilde{C}_T$ & $C_{V_L}$, $C_{S_L}$, $C_{T}$ \\
0 & $ S_3 \sim (\bar{3},{3},1/3)$ & LQ & $\overline{q_L^c} \ell_L$ & $\tilde{C}_{V_L}$ & $C_{V_L}$\\
0 & $R_2 \sim (3,2,7/6)$ & LQ & $\overline{u_R}\ell_L$, $\overline{q_L}e_R$ & $\tilde{C}_{S_L}$,  $\tilde{C}_T$ & $C_{S_L}$, $C_T$\\
0 & $H_2 \sim (1,2,1/2)$ & SB & $\overline{q_L}d_R$, $\overline{\ell_L} e_R$, $\overline{u_R} q_L$ & $\tilde{C}_{S_R}$, $\tilde{C}_{S_L}$ & $C_{S_R}$, $C_{S_L}$, $C_T$\\
1 & $ V_2 \sim (\bar{3},2,5/6)$ & LQ &  $\overline{d_R^c}\gamma_\mu \ell_L$, $\overline{e_R^c}\gamma_\mu q_L$ & $\tilde{C}_{S_R}$ & $C_{S_R}$ \\
1 & $U_1 \sim (3,1,2/3)$ & LQ &  $\overline{q_L}\gamma_\mu \ell_L$, $\overline{d_R}\gamma_\mu e_R$ & $\tilde{C}_{V_L}$, $\tilde{C}_{S_R}$ & $C_{V_L}$, $C_{S_R}$\\
1 & $U_3 \sim (3,3,2/3)$ & LQ &  $\overline{q_L}\gamma_\mu \ell_L$ & $\tilde{C}_{V_L}$ & $C_{V_L}$\\
1 & $W^{\prime}_\mu \sim (1,3,0)$ & VB &  $\overline{\ell_L}\gamma_\mu \ell_L$, $\overline{q_L}\gamma_\mu q_L$ & $\tilde{C}_{V_L}$ & $C_{V_L}$\\
\hline
    \end{tabular}
    \caption{Spin, $SU(3)_C \otimes SU(2)_L \otimes U(1)_Y$ quantum numbers, nature (LQ = leptoquarks, SB = scalar boson and VB = vector boson)  and  allowed interactions of the possible candidates to mediate $b \to c$ transitions. In our notation, $\Psi_L^c \equiv (\Psi_L)^c$.}
    \label{table:NPQN}
\end{table}
%
We show in the last two columns of Table~\ref{table:NPQN} the set of Wilson coefficients to which the new degrees of freedom contribute, both in the SMEFT and in the WET. The RGE running changes the relative size of these coefficients, as seen above, and causes mixing
among the operators ${\cal O}_{S_L}$ and ${\cal O}_T$. 
When considering such specific classes of models, generally other constraints apply. Specifically, searches for the corresponding mediators can exclude a large part of the parameter space, or even the whole scenario (like the $W'$) \cite{Faroughy:2016osc,Feruglio:2018fxo,Greljo:2018tzh}. In the following we will not discuss these constraints, but simply give examples for how the required coefficients could be generated, irrespective of their actual viability.

We are now in a position to interpret the different solutions obtained in the fit shown in Table~\ref{table:minimaA} and Table~\ref{table:minimaB}. Let us focus first on the scenarios where $F_L^{D^*}$ is not included. The minimum with highest $\chi^2$, Min~3, presents relevant contributions from the operators ${\cal O}_{S_L}$ and ${\cal O}_T$. The origin of these Wilson coefficients could be explained, for instance, with the presence of the scalar leptoquarks $R_2 \sim (3,2,7/6)$ or $S_1 \sim (\bar{3},1,1/3)$, whose contributions to the Lagrangian at the NP scale are given in Appendix~\ref{app:UVLagrangian}. An additional mediator would be necessary to generate the sizeable contribution to $C_{V_L}$, however, in the former case. Min~2, which exhibits non-zero values for all Wilson coefficients, could be explained by combinations of several candidates, for instance $S_1$ and $H_2$. Also for Min~1 there are different possibilities, since the fit does not single out a specific coefficient. However, the simplest option remains the scenario where the only relevant contribution is proportional to the SM one, i.e., all Wilson coefficients but $C_{V_L}$ are compatible with zero at $1.1\sigma$. This possibility could be generated, for instance, by the effect of a $W^{'}$ boson, 
\begin{equation}
     {\cal L}_\text{eff} \supset  \displaystyle -\frac{\tilde g_{\ell \nu_\ell}^{\phantom{\dagger}} \tilde g^\dagger_{du}}{M_{W'}^2}\; (\bar \ell_L\gamma_\mu \nu_{\ell L})(\bar u_L\gamma^\mu d_L)\, ,
 \end{equation}
with $M_{W'}/(\tilde g_{\ell \nu_\ell}^{\phantom{\dagger}} \tilde g^\dagger_{du})^{1/2}\sim 2$~TeV. For a sequential $W'$ with SM couplings, one would need $M_{W'}\sim 0.2$~TeV, which is already ruled out by direct searches~\cite{ATLAS:2018lcz}.
More exotically, but more realistically given the aforementioned high-energy constraints, one could explain the modification on the ${\cal O}_{V_L}$ operator by introducing leptoquarks (LQs), such as the vector $U_3 \sim (3,3,2/3)$ or the scalar $S_1 \sim  (\bar{3},1,1/3)$ LQs. However, extra symmetries in the UV regime would have to be assumed in order to guarantee that other flavour transitions compatible with the SM are respected. 

\begin{figure}
\centering
\includegraphics[scale=0.5]{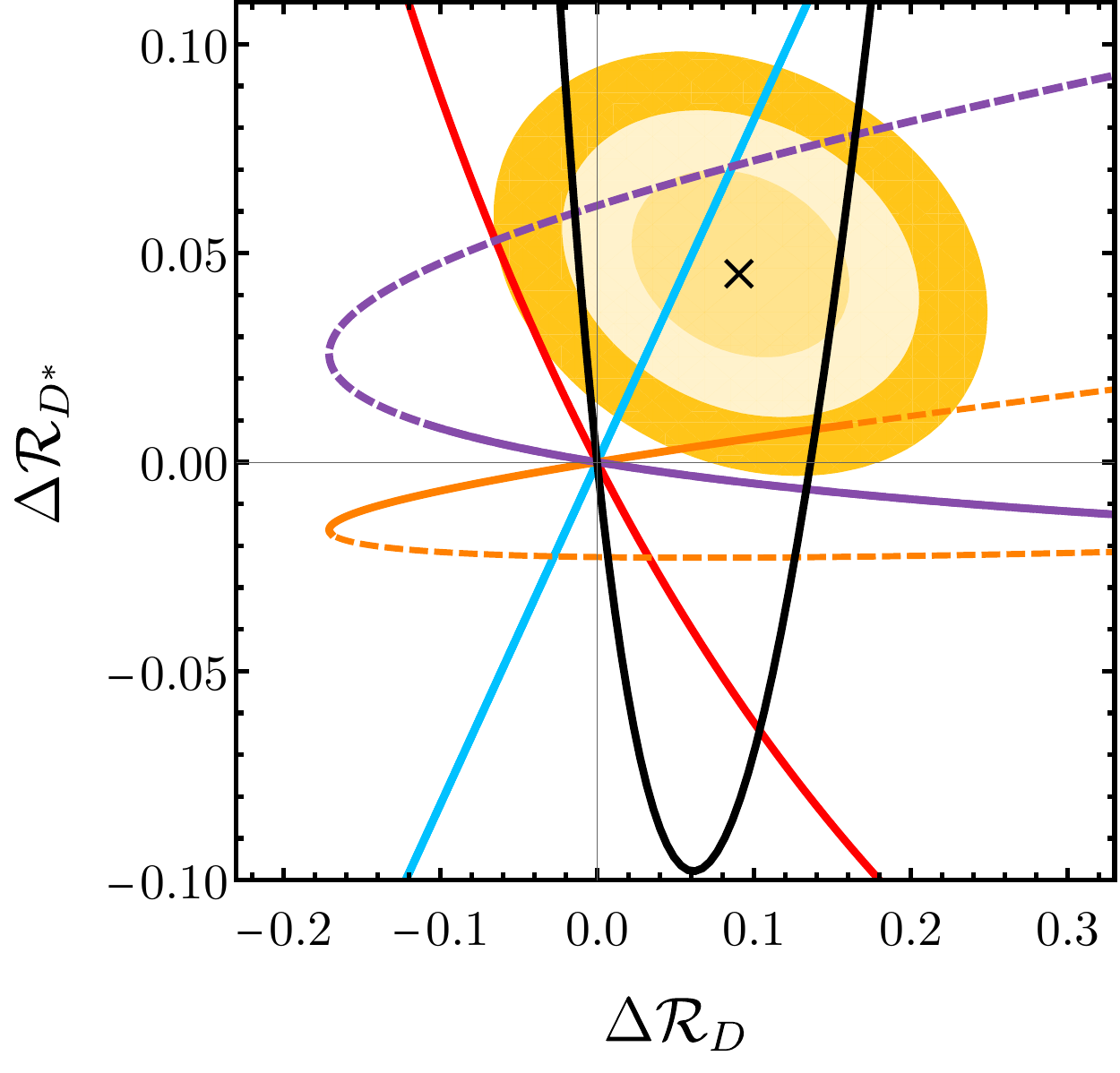}
\includegraphics[scale=0.5]{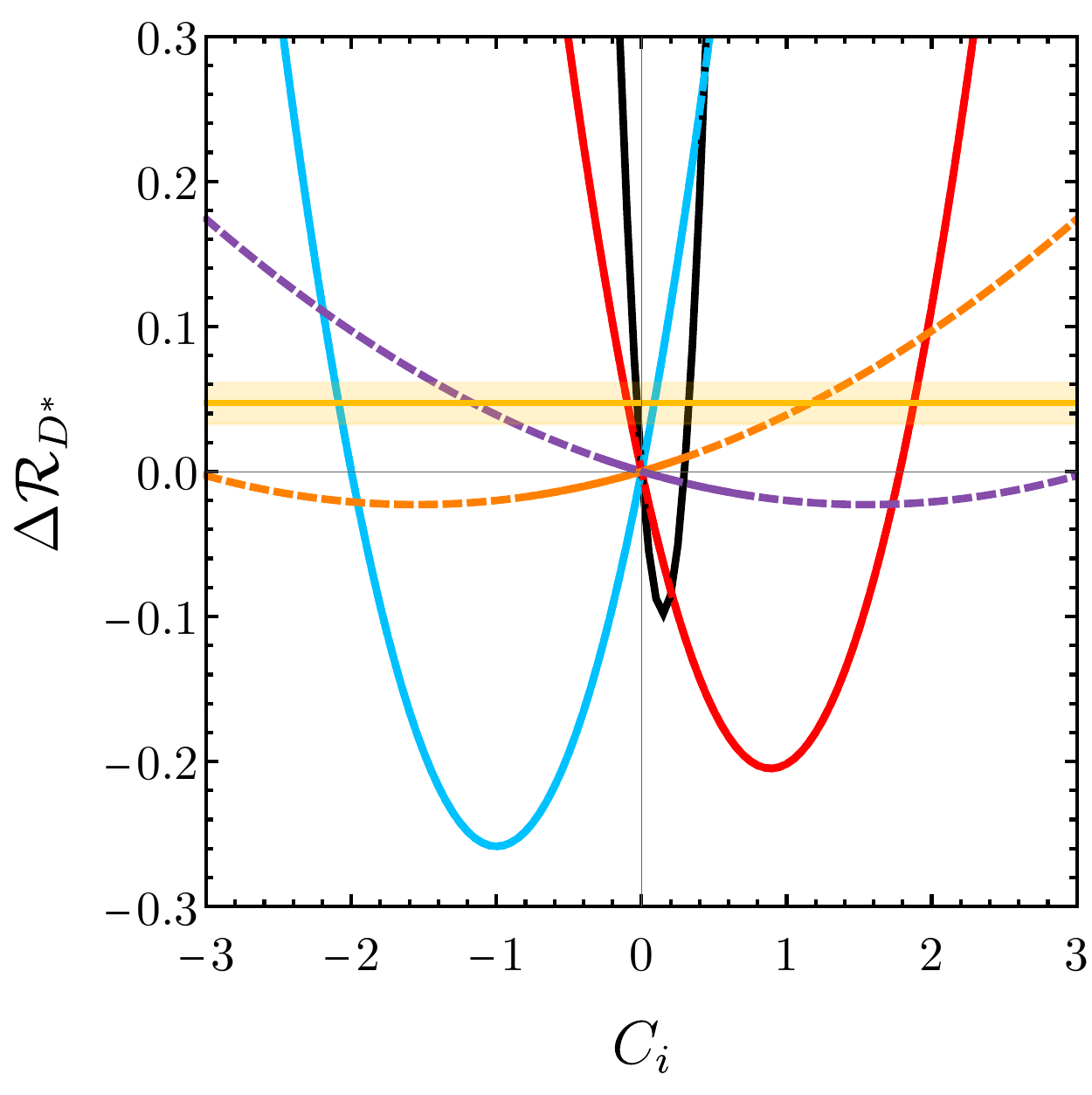}
\includegraphics[scale=0.9]{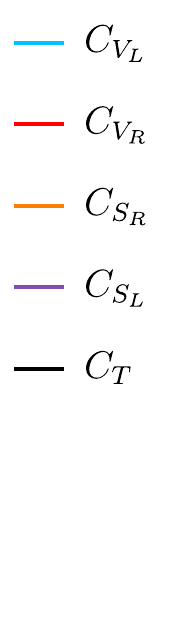}
\includegraphics[scale=0.5]{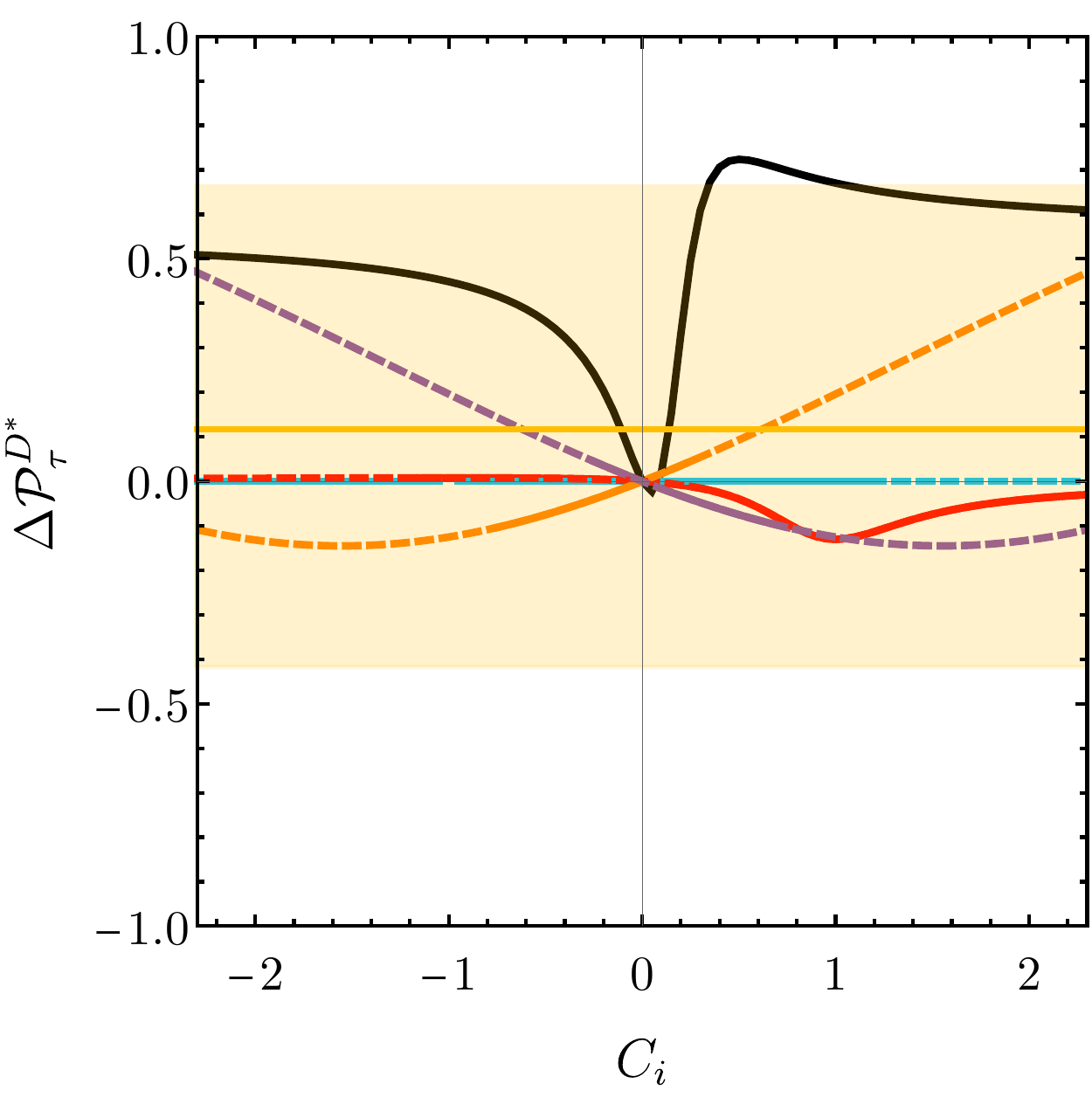}
\includegraphics[scale=0.5]{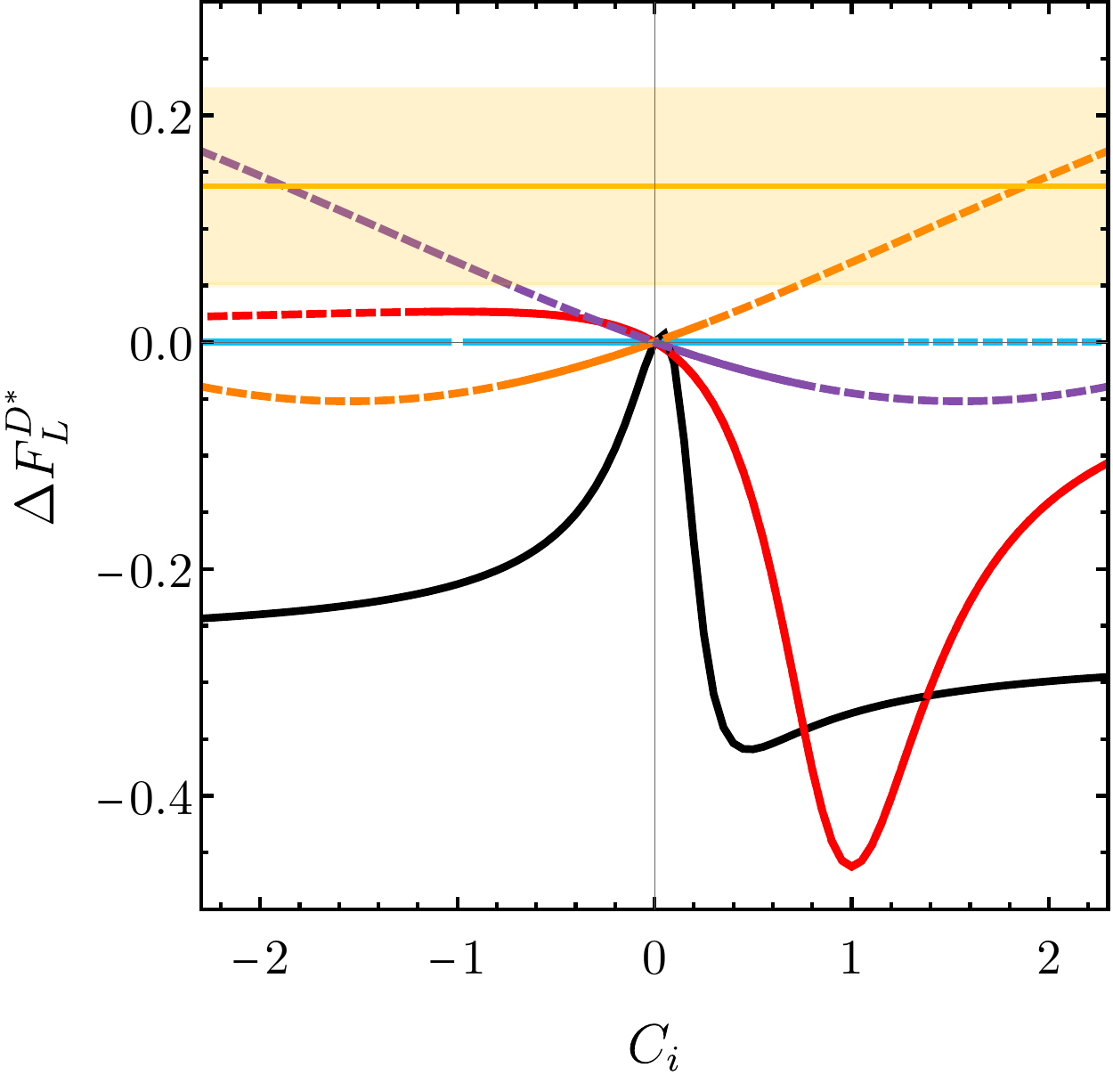}
\phantom{\includegraphics[width=0.09\linewidth]{LegendCi.pdf}}
\caption{Individual contributions of the Wilson coefficients of the WET Hamiltonian in different observables ($\Delta X \equiv X-X_{\text{SM}}$): correlation between $\Delta {\cal R}_D$ and $\Delta {\cal R}_{D^*}$, and $\Delta \mathcal{R}_{D^*}$, $\Delta \mathcal{P}_\tau^{D^{*}}$ and $\Delta {\cal F}_L^{D^{*}}$ as a function of the Wilson coefficients. Left-top panel: the experimental central value is denoted by a black cross and the $1\sigma, 2\sigma$ and $3\sigma$ uncertainties by yellow rings. Right-top and bottom panels:  experimental central values are displayed by a solid yellow line and their $1\sigma$ uncertainty by a yellow band.
Dashed lines indicate regions excluded by the constraint ${\cal B}(B_c \to \tau \bar{\nu}_\tau)< 10\%$.} 
\label{fig:indWilsons}
\end{figure}

In Fig.~\ref{fig:indWilsons} we show the dependence of selected observables on individual Wilson coefficients.
The left-top panel in Fig.~\ref{fig:indWilsons} shows that it is straightforward to achieve consistency with the experimental measurements for $\mathcal{R}_{D^{(*)}}$ by shifting only the Wilson coefficient $C_{V_L}$, i.e., modifying the SM coefficient. The polarization observables show a good potential to differentiate between different contributions. Particularly interesting is the longitudinal polarization fraction in $B\to D^*\tau\bar\nu_\tau$, shown in the bottom-right panel, for which the Belle collaboration recently announced a first measurement~\cite{Abdesselam:2019wbt}. As this sub-figure shows, it is difficult to accommodate it at $1\sigma$ for any of the individual Wilson coefficients~\cite{Iguro:2018vqb}. The only contributions allowing for a significantly larger value of this observable than in the SM are those from scalar operators; however, values accommodating $F_L^{D^*}$ are in conflict with the bound from ${\cal B}(B_c \to \tau \bar{\nu}_\tau)<10\%$ (dashed lines), and extending this bound to $30\%$ still does not allow to accommodate its central value. This figure therefore indicates why none of the fit scenarios yields values for $F_L^{D^*}$ in the $1\sigma$ range; we take this as a motivation to investigate the consistency of the different measurements in more detail.

\begin{figure}[tb]
    \centering
   \includegraphics[width=0.35\linewidth]{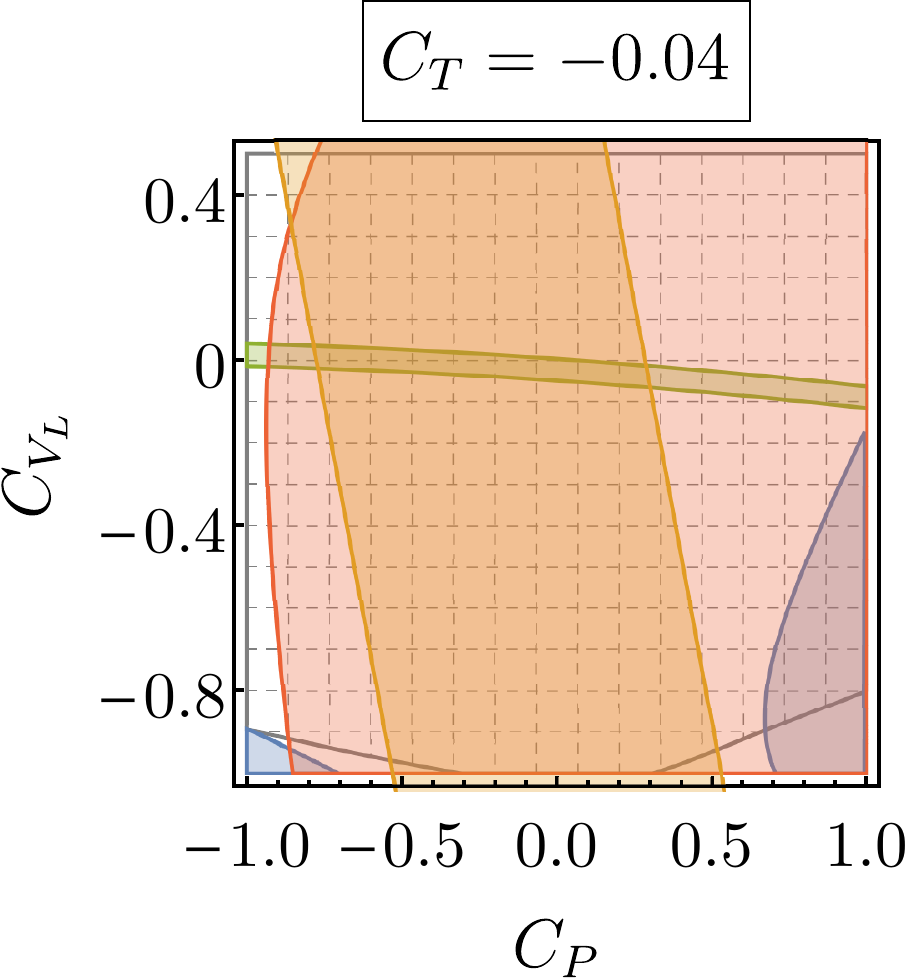}\hskip .5cm
    \includegraphics[width=0.35\linewidth]{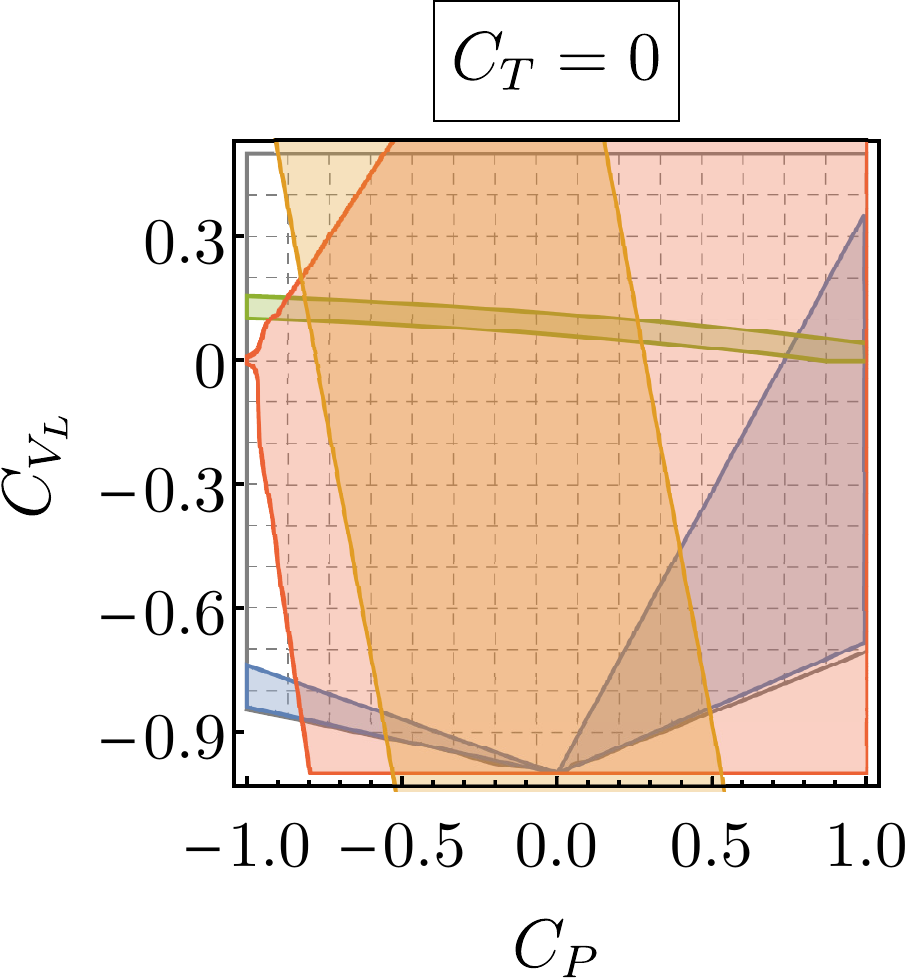}\vskip .5cm
    \includegraphics[width=0.35\linewidth]{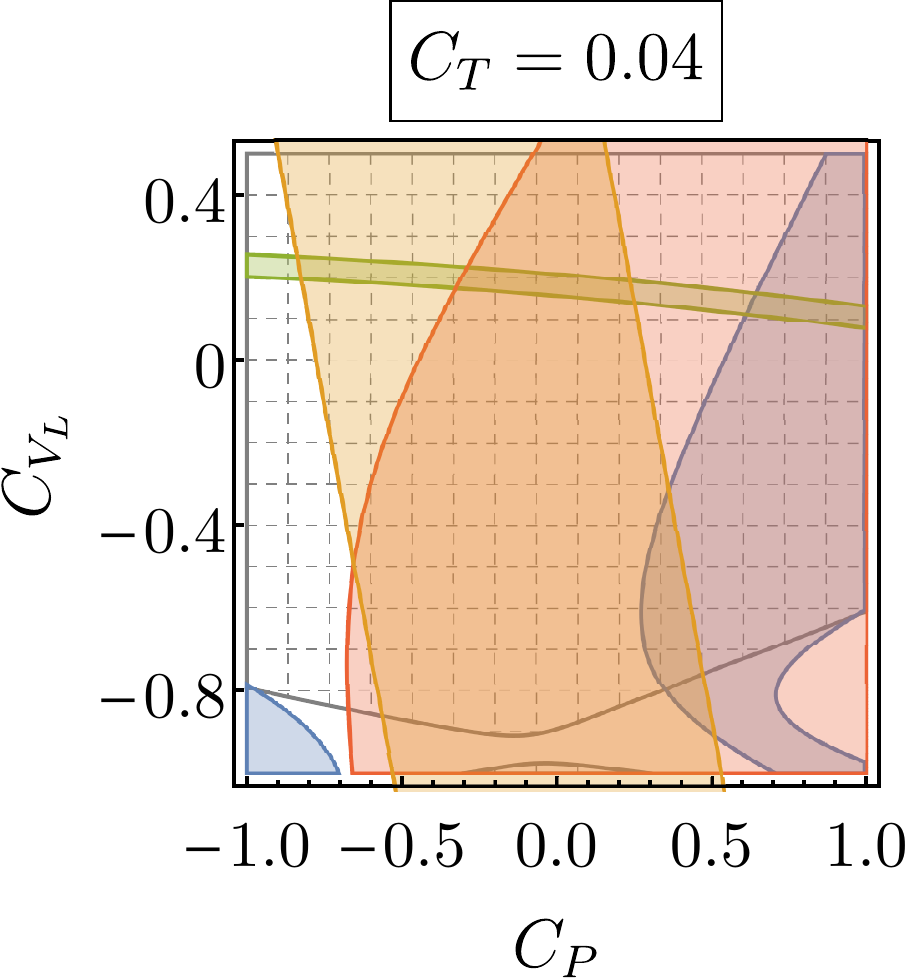}\hskip .5cm
    \includegraphics[width=0.35\linewidth]{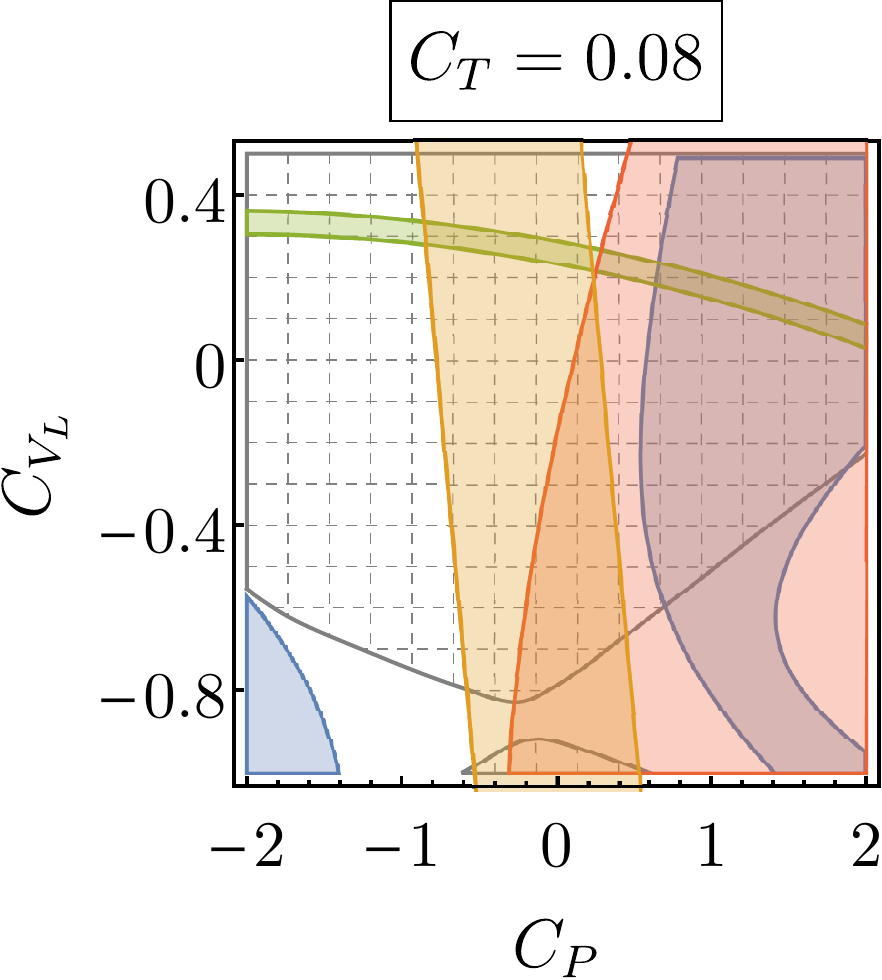}
    \caption{Allowed regions at $1\sigma$ from $F_L^{D^*}$ (blue), $\mathcal{R}_{D^*}$ (green), $\mathcal{P}_{\tau}^{D^*}$ (gray grid) and the $q^2$ distribution of $\Gamma(B \to D^* \tau \bar{\nu}_\tau)$ (red), together with  the region satisfying the bound $\mathcal{B} (B_c \to \tau \bar{\nu}_\tau)   < $  10\% (orange).} 
    \label{fig:CVReq0}
\end{figure}

In order to do so, we use the fact that only three combinations of the four Wilson coefficients enter $B\to D^*\tau\bar\nu_\tau$ observables as well as the leptonic $B_c$ decay:  $C_{V_L}$, $C_T$ and the pseudo-scalar coefficient $C_P \equiv C_{S_R}-C_{S_L}$. Every observable therefore results in a non-trivial constraint in the $C_P-C_{V_L}$ plane if $C_T$ is fixed to some value. We show the preferred parameter ranges obtained for the individual observables in Fig.~\ref{fig:CVReq0}, for a representative set of $C_T$ values. The combination of $\mathcal{R}_{D^*}$ and the bound on $\mathcal{B}(B_c\to \tau\bar\nu_\tau)$ determines a narrow strip in this parameter plane, dominated by the former for the bound on $C_{V_L}$ and the latter for the bound on $C_P$. The overlap of the other observables varies with the value for $C_T$; however, there is no value of $C_T$ for which all $1\sigma$ bands overlap. In fact, the $1\sigma$ range for $F_L^{D^*}$ cannot be reached by any NP parameter combination in this setup, when only imposing the $\mathcal{B}(B_c\to \tau\bar\nu_\tau)$ constraint of $10\%$ or even $30\%$ and at the same time requiring a positive shift in $\mathcal{R}_{D^*}$. Agreement can presently be achieved at the $2\sigma$ level; nevertheless, a confirmation of the present central values with higher precision could indicate the inconsistency between the data and any NP with flavour-universal $C_{V_R}$.

This potential incompatibility would suggest one of several possibilities:
\begin{enumerate}

\item[1)] One of our theoretical assumptions is incorrect and the SMEFT cannot be applied at the electroweak scale. This could happen if one or several of the following cases apply: (a) There is an insufficient gap between the electroweak and the NP scale, i.e., there are new degrees of freedom close enough to the EW scale to invalidate an EFT approach. (b) The electroweak symmetry breaking is non-linear, changing also the character of the observed Higgs-like particle. In that case $C_{V_R}$ could contribute to the fitted observables, because it would no-longer be necessarily flavour universal. (c) There are additional light degrees of freedom like right-handed neutrinos \cite{Asadi:2018wea,Greljo:2018ogz,Robinson:2018gza}, yielding additional operators.
		
Note that we also assumed the semi-leptonic decays with light leptons to be free from NP. However, the corresponding constraints are so strong that even relaxing this assumption would not significantly change our analysis \cite{Jung:2018lfu}.

\item[2)] An unidentified or underestimated systematic uncertainty in one or several of the experimental measurements.
\end{enumerate}
In any case, the upcoming experimental studies of not only the LHCb collaboration, but also the Belle~II experiment which started to take data will hopefully resolve this question soon.

\begin{figure}
\centering
\includegraphics[scale=0.42]{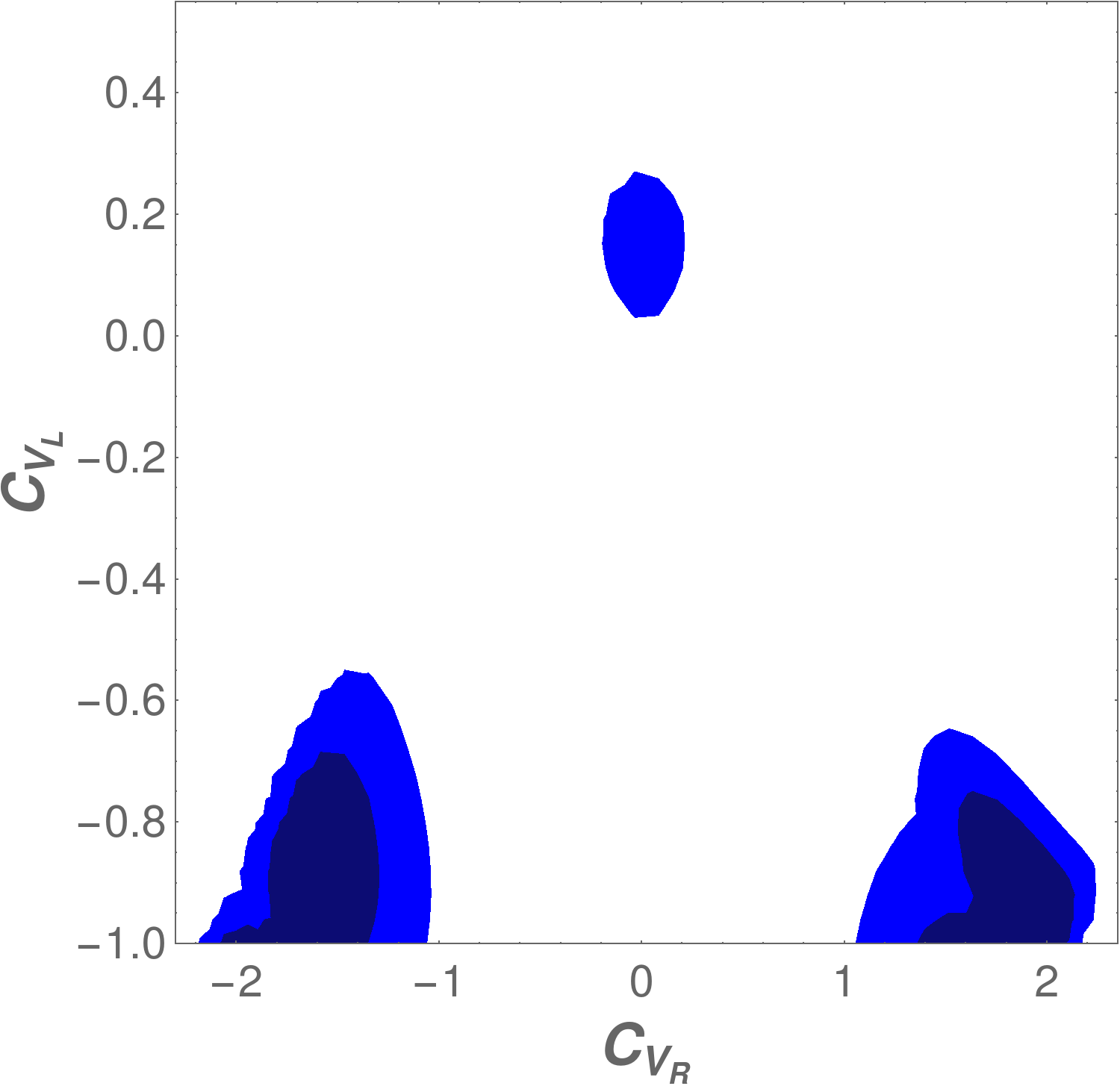}\,
\includegraphics[scale=0.42]{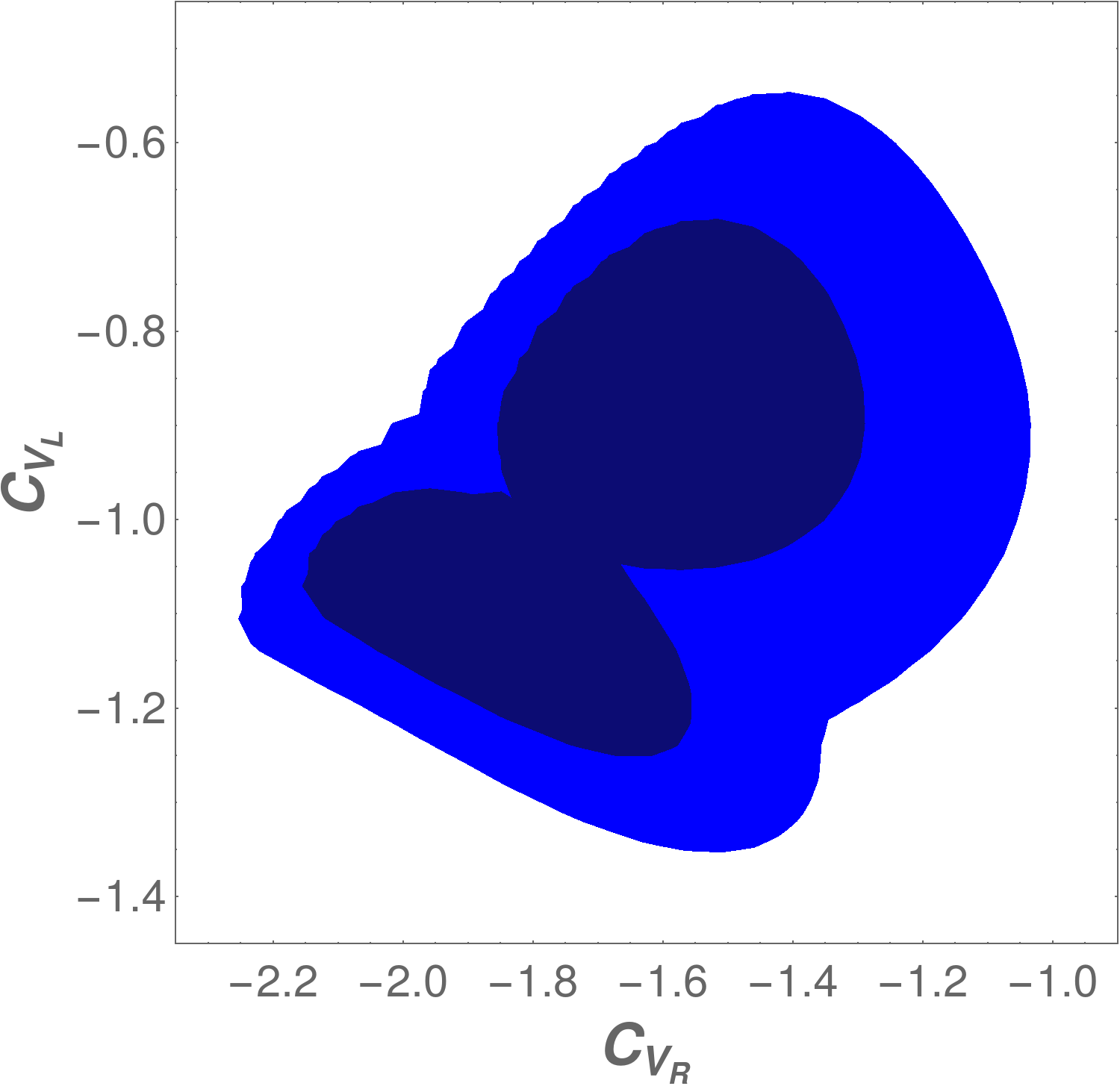}\,
\includegraphics[scale=0.42]{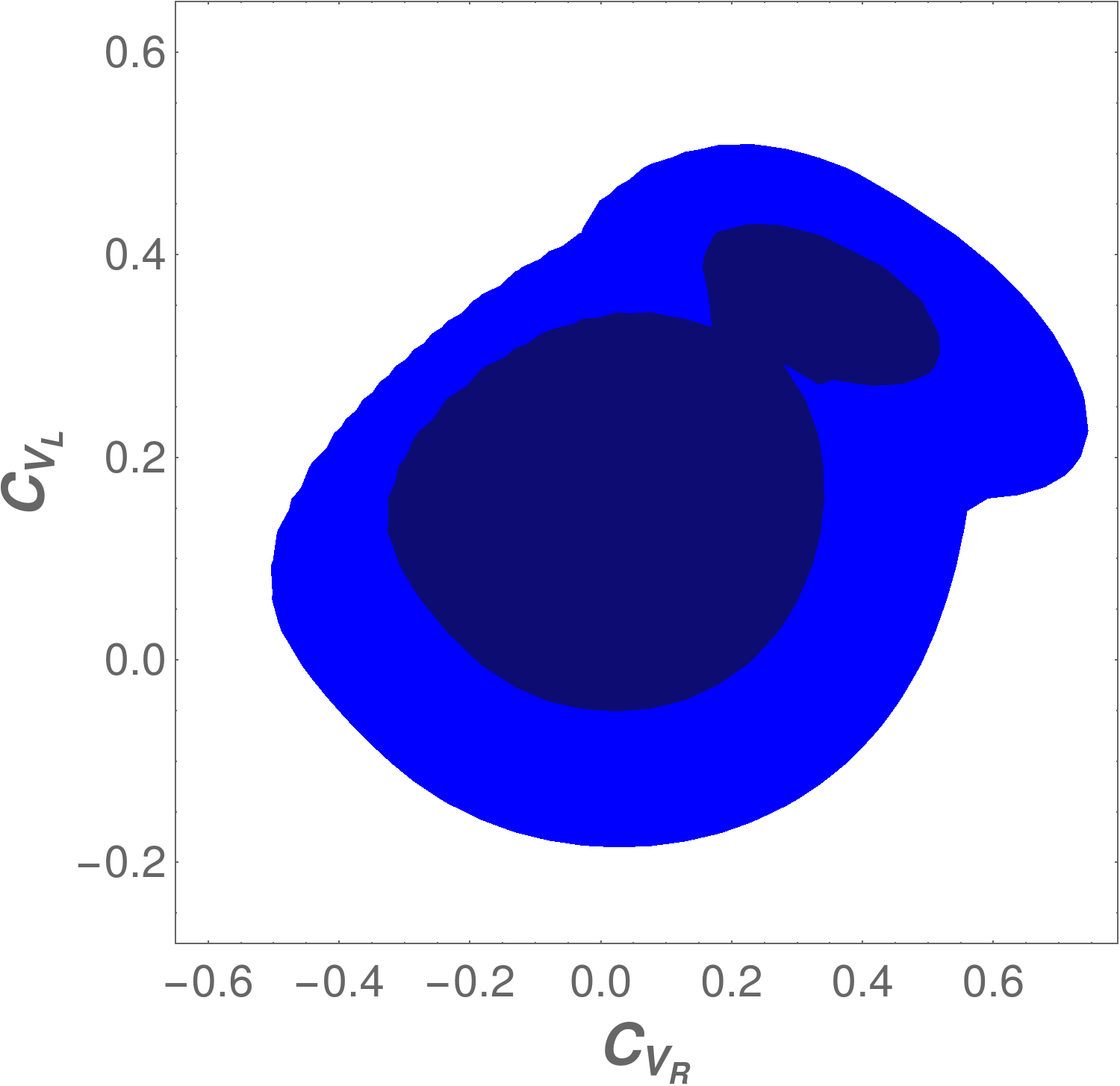}

\caption{Allowed regions in the $C_{V_R}-C_{V_L}$ plane, for the global fit including $F_L^{D^*}$\!, restricting  $\mathcal{B}(B_c \to \tau \nu) \leq 10\%$. Lighter and darker blue areas show regions with 95\% and 68\% CL, respectively. Left: All four minima shown in the chosen parameter convention with $C_{V_L}>-1$, relative to the global minimum. Center: the two minima with $C_{V_L}\sim -1$, without restricting $C_{V_L}>-1$, see text. Right: the two minima with $|C_{V_{R,L}}|<1$, relative to Min~6.}
  \label{fig:CVR}
\end{figure}

For completeness of our discussion, we have consequently performed the fit relaxing the condition of flavour universality on $C_{V_R}$. As a consequence of adding $C_{V_R}$ as an extra d.o.f. to fit, the number of solutions is enlarged. As shown in Fig.~\ref{fig:CVR}, one finds now four different solutions (plus their sign-flipped counterparts), given numerically in Table~\ref{table:minimaCVR}.%

\def\arraystretch{1.5}
\begin{table}[h]
\centering
\begin{tabular}{ c | c | c | c | c }    
& Min~4 & Min~5 & Min~6 & Min~7\\
\hline
$\chi^2_\text{min}/$d.o.f.  &  $32.5/53$  &  $33.3/53$ &  $37.6/53$ &  $38.9/53$  \\
\hline
$C_{V_L}$ &  $-0.91^{+0.10}_{-0.09}$& $-0.85^{+0.20}_{-0.10}$ & $\phantom{-}0.14^{+0.14}_{-0.12}$ & $ \phantom{-}0.35^{+0.08}_{-0.08}$ \\
$C_{V_R}$ &  $\phantom{-}1.89^{+0.19}_{-0.22}$ & $ -1.58^{+0.23}_{-0.22}$ & $\phantom{-}0.02^{+0.21}_{-0.24}$ & $\phantom{-}0.34^{+0.18}_{-0.18}$ \\
$C_{S_R}$ &  $ -0.44^{+0.12}_{-0.45}$ & $ -0.33^{+0.52}_{-0.16}$ & $\phantom{-}0.10^{+0.15}_{-0.59}$ & $-0.68^{+0.54}_{-0.14}$ \\ 
$C_{S_L}$ & $ -1.34^{+0.49}_{-0.12}$ & $\phantom{-} 0.56^{+0.23}_{-0.54}$ & $-0.12^{+0.65}_{-0.15}$ & $ -0.92^{+0.58}_{-0.11}$ \\
$C_T$ & $ -0.22^{+0.10}_{-0.11}$ & $\phantom{-}0.19^{+0.10}_{-0.10}$ & $\phantom{-} 0.01^{+0.09}_{-0.07}$ &  $-0.02^{+0.08}_{-0.07}$ \\
\end{tabular}
\caption{Minima with their $1\sigma$ uncertainties obtained from the global $\chi^2$ minimization, including $F_L^{D^*}$ and $\mathcal{B}(B_c\to\tau\bar\nu_\tau)<10\%$ in the fit while allowing for $C_{V_R}\neq 0$. There are, in addition, the corresponding sign-flipped minima, as indicated in Eq.~\eqref{eq:WilsonSecondMin}.
}
\label{table:minimaCVR}
\end{table}
\def\arraystretch{1.5}
The doubling of minima can be understood qualitatively in the following way: $B\to D$ is dominated by the combination of Wilson coefficients corresponding to the vector coupling $C_V = 1 + C_{V_L} + C_{V_R}$, while $B\to D^*$ is dominated by the axial-vector coupling $C_A=C_{V_R}-(1+C_{V_L})$. Their rates are correspondingly roughly given by $|C_{V,A}|^2$. For $C_{V_R}\equiv0$ we have $C_V=-C_A$, and the only remaining discrete symmetry is that discussed in Section~\ref{ssec::NP}, the second solution being eliminated by our choice $C_{V_L}>-1$. With a finite coefficient $C_{V_R}$, these two solutions become four ($\{C_A=\pm |C_A|,C_V=\pm |C_V|\}$), since now $|C_A|\neq |C_V|$; two of those are again eliminated by our choice for $C_{V_L}$, leaving two solutions per minimum with $C_{V_R}\equiv 0$. This degeneracy is broken by interference terms, notably ${\mbox{Re}}(C_A C_V^*)$ in $B\to D^*$, but also the interference with scalar and tensor operators. Nevertheless, this approximate degeneracy explains the doubling of solutions for finite $C_{V_R}$.

As can be seen from the comparison of Table~\ref{table:minimaCVR} with Table~\ref{table:minimaB}, the previous global minimum, Min~1b, remains a solution of this more general fit, now called Min~6. Min~7 is again relatively close to Min~6, however with a significant contribution from $C_{V_R}$ and hence qualitatively different from Min~2 in the previous fits. The new global minimum Min~4 and the close-lying Min~5 improve the agreement of the fit with the data significantly. However, in these scenarios the SM coefficient is almost completely cancelled and its effect replaced by several NP contributions. These are hence fine-tuned scenarios, and should be taken with a grain of salt.

\begin{figure}
\centering
\includegraphics[width=0.35\linewidth]{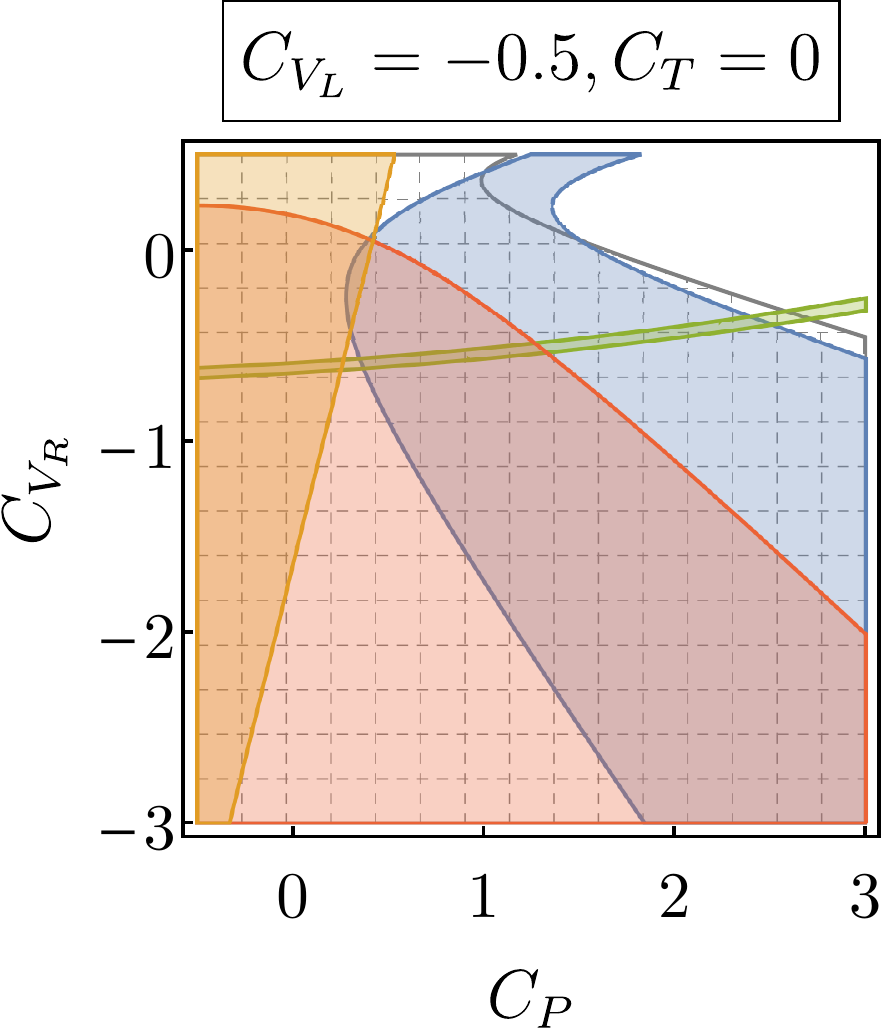}\hskip .5cm
\includegraphics[width=0.35\linewidth]{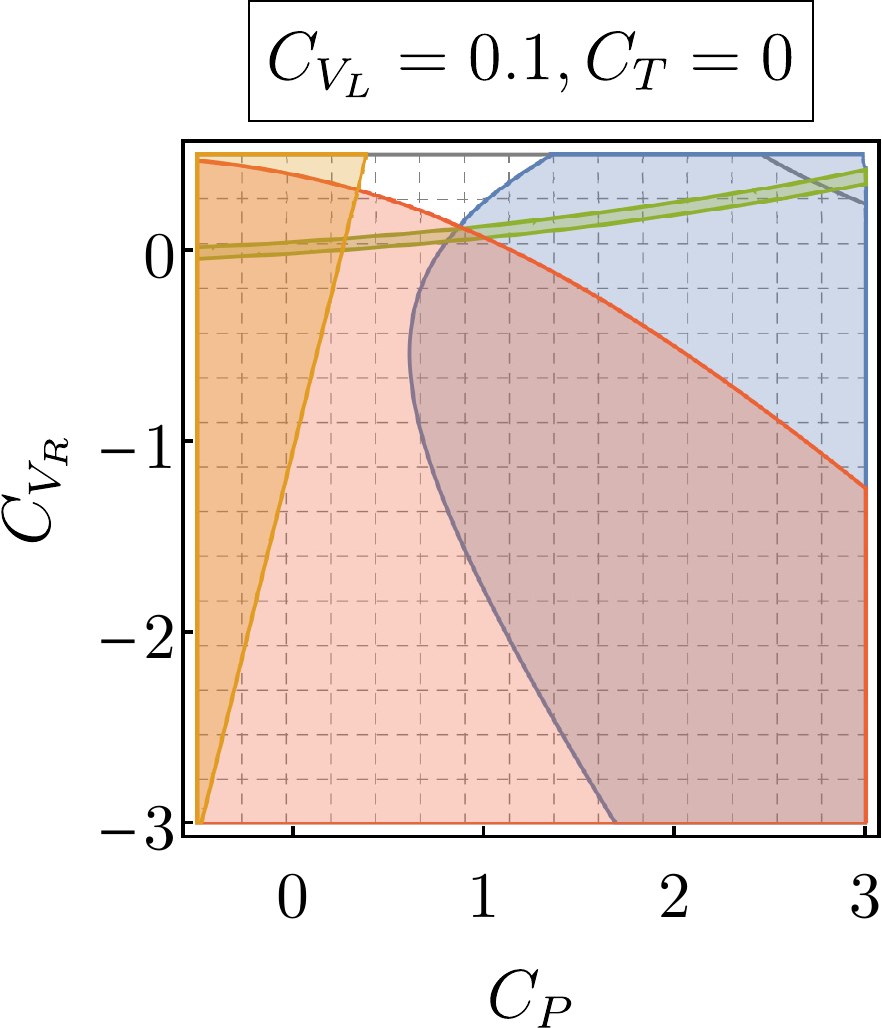}
\vskip .5cm
\includegraphics[width=0.35\linewidth]{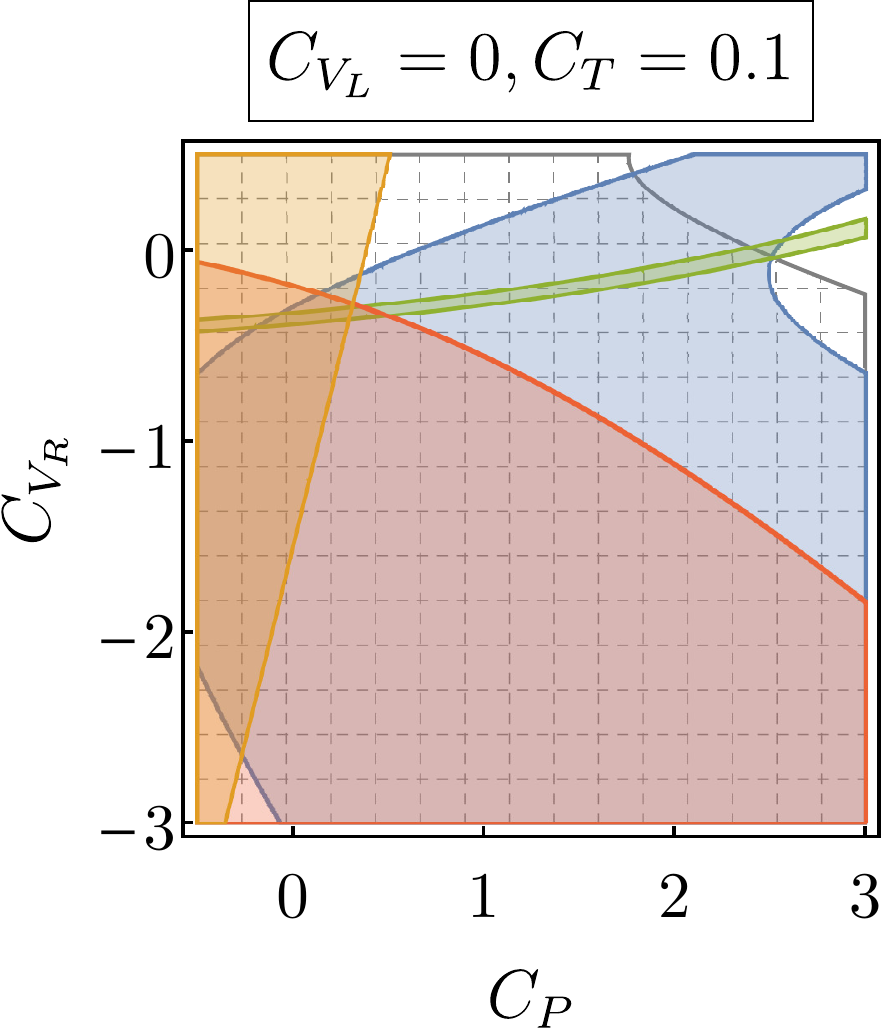}\hskip .5cm
\includegraphics[width=0.368\linewidth]{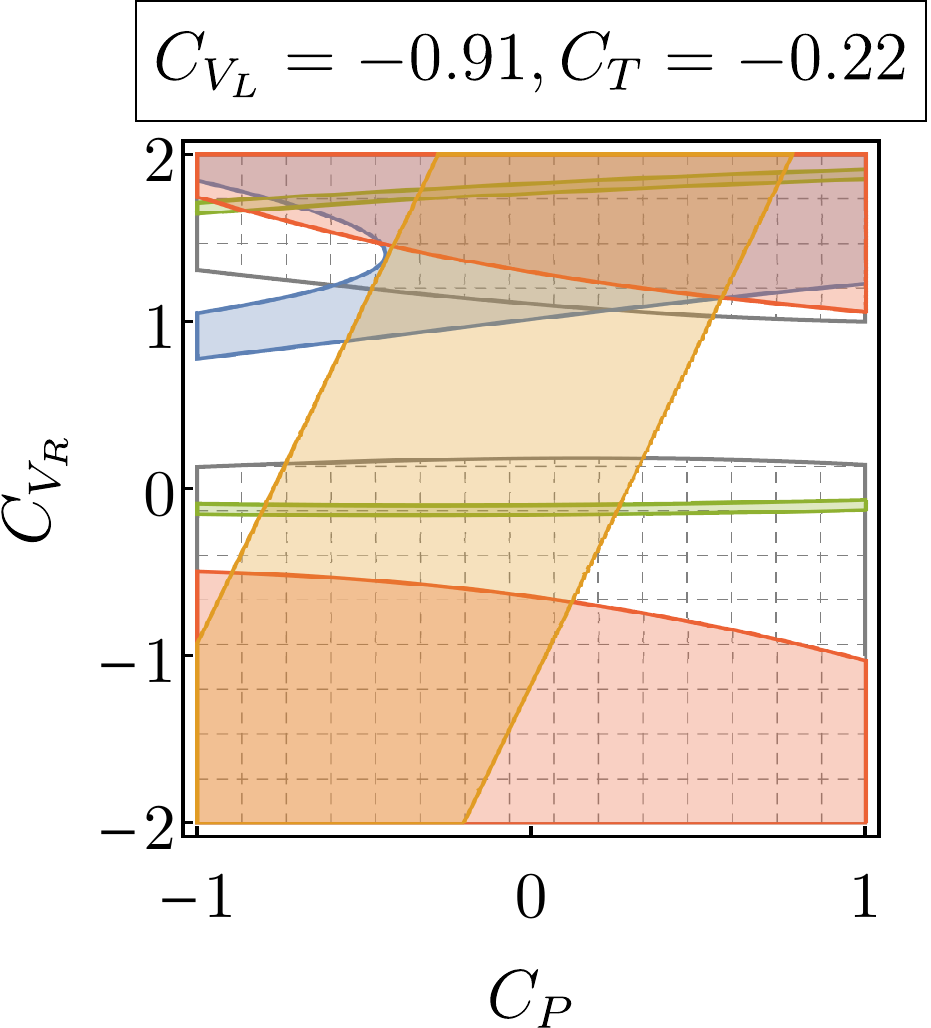}\hskip .5cm
\caption{Allowed regions at $1 \sigma$ from $F_L^{D^*}$ (blue), $\mathcal{R}_{D^*}$ (green), $\cal {P}_{\tau}^{D^*}$ (gray grid) and the $q^2$ distribution of $\Gamma(B \to D^* \tau \bar{\nu}_\tau)$ (red), together with  the region satisfying the bound ${\mathcal B}(B_c \to \tau \bar{\nu}_\tau) < $ 10\% (orange), with $C_{V_R} \neq 0$.}
  \label{fig:CVRneq0}
\end{figure}

We have also analyzed the individual observables in $B\to D^*$ and the bound on $\mathcal{B}(B_c\to \tau\bar\nu_\tau)$ for this case. 
This is illustrated in Fig.~\ref{fig:CVRneq0}, for different benchmark values of $C_{V_L}$ and $C_{T}$, in the plane $C_{V_R} - C_{P}$. The figure shows again the allowed regions at $1\sigma$ for the different observables.
In accordance with the above reduction for $\chi^2_{\rm min}$, we observe that in this case it is possible to have an overlap of all the bands. However, it is still not possible to reach the central value for the longitudinal polarization fraction, and as mentioned above, this scenario corresponds to a highly fine-tuned combination of parameters. 

%

\section{Predictions}
\label{sec:Predictions}

We use our global fits from Sec.~\ref{sec:FitResults} to predict selected observables that are either not measured yet, but expected to be measured soon, or presently measured with uncertainties that are larger than those from the fits. These additional measurements serve two purposes: firstly, they provide additional information that is theoretically related, but experimentally independent (to varying extent) from existing measurements, thereby helping to establish NP and excluding underestimated systematic uncertainties as the source for the anomaly. Secondly, they can provide experimental information on combinations of Wilson coefficients that are not or only weakly constrained so far, thereby allowing to distinguish different NP scenarios.

We will first present the predictions for observables of the key modes $B \to D^{(*)} \tau \bar{\nu}_\tau$, before focusing on other semi-leptonic decays, specifically $\Lambda_b \to \Lambda_c \, \tau \bar{\nu}_\tau$ and $B_c \to J/\psi \, \tau \bar{\nu}_\tau$.

\subsection{Predictions for $\boldsymbol{B \to D^{(*)}\tau \bar{\nu}_\tau}$ observables}
We start by analyzing the $q^2$ distributions of several angular observables. While these distributions can be very effective in distinguishing different NP scenarios, they are difficult to measure, due to the missing information on the neutrinos. 
The angular dependence of the differential decay width $B \to D^{(*)} \ell \nu$ can be parametrized by three independent angular coefficients, 
\begin{equation}
\frac{d^2 \Gamma^{D^{(*)}}}{dq^2  \, d\cos \theta_\ell}=a^{(*)}_\ell (q^2)
-b^{(*)}_\ell(q^2)\cos \theta_\ell + c^{(*)}_\ell(q^2)\cos^2\theta_\ell \, ,
\end{equation}
which are in principle experimentally accessible.
Here, $\theta_\ell$ is the angle between the $D^{(*)}$ and charged-lepton three-momenta in the $\ell$--$\nu$ center-of-mass frame.
An angular observable commonly defined in the literature is the forward-backward asymmetry, which is determined by the $b_\ell^{(*)}(q^2)$ coefficient according to the following expression:
\begin{equation}
{\cal A}_\text{FB}^{D^{(*)}}(q^2) \equiv  b^{(*)}_\ell(q^2) \left / \displaystyle \frac{d\Gamma^{D^{(*)}}}{dq^2} \right. = \left(\int_{-1}^0 d\cos \theta_\ell\; \frac{d^2\Gamma^{D^{(*)}}}{dq^2 d \cos \theta_\ell}-\int_{0}^1 d\cos \theta_\ell\; \frac{d^2 \Gamma^{D^{(*)}}}{dq^2 d\cos \theta_\ell}\right) \left / \frac{d\Gamma^{D^{(*)}}}{dq^2} \right. . 
\end{equation}
This observable yields complementary information, since it does not contribute for quantities integrated over the full range of $\cos\theta_\ell$.
One can also decompose the differential branching ratio according to the two possible polarizations of the charged ($\tau$) lepton, giving rise to another observable named $\tau$ polarization asymmetry:
\begin{equation}
{\cal P}_\tau^{D^{(*)}}(q^2) = \displaystyle \displaystyle \left( \frac{d\Gamma_{\lambda_{\tau}=1/2}^{D^{(*)}}}{dq^2}-\frac{d\Gamma^{D^{(*)}}_{\lambda_{\tau}=-1/2}}{dq^2}\right)  \bigg/ \displaystyle \frac{d\Gamma^{D^{(*)}}}{dq^2}\, ,
\end{equation}
where $\lambda_\tau$ is the helicity of the $\tau$ lepton, 
and $d\Gamma_{\lambda_{\tau}}^{D^{(*)}}/dq^2$ is the differential decay width of $B \to D^{(*)} \tau \bar{\nu}_\tau$ for a given helicity $\lambda_{\tau}$.

\begin{figure}
\centering
\includegraphics[width=0.45\linewidth]{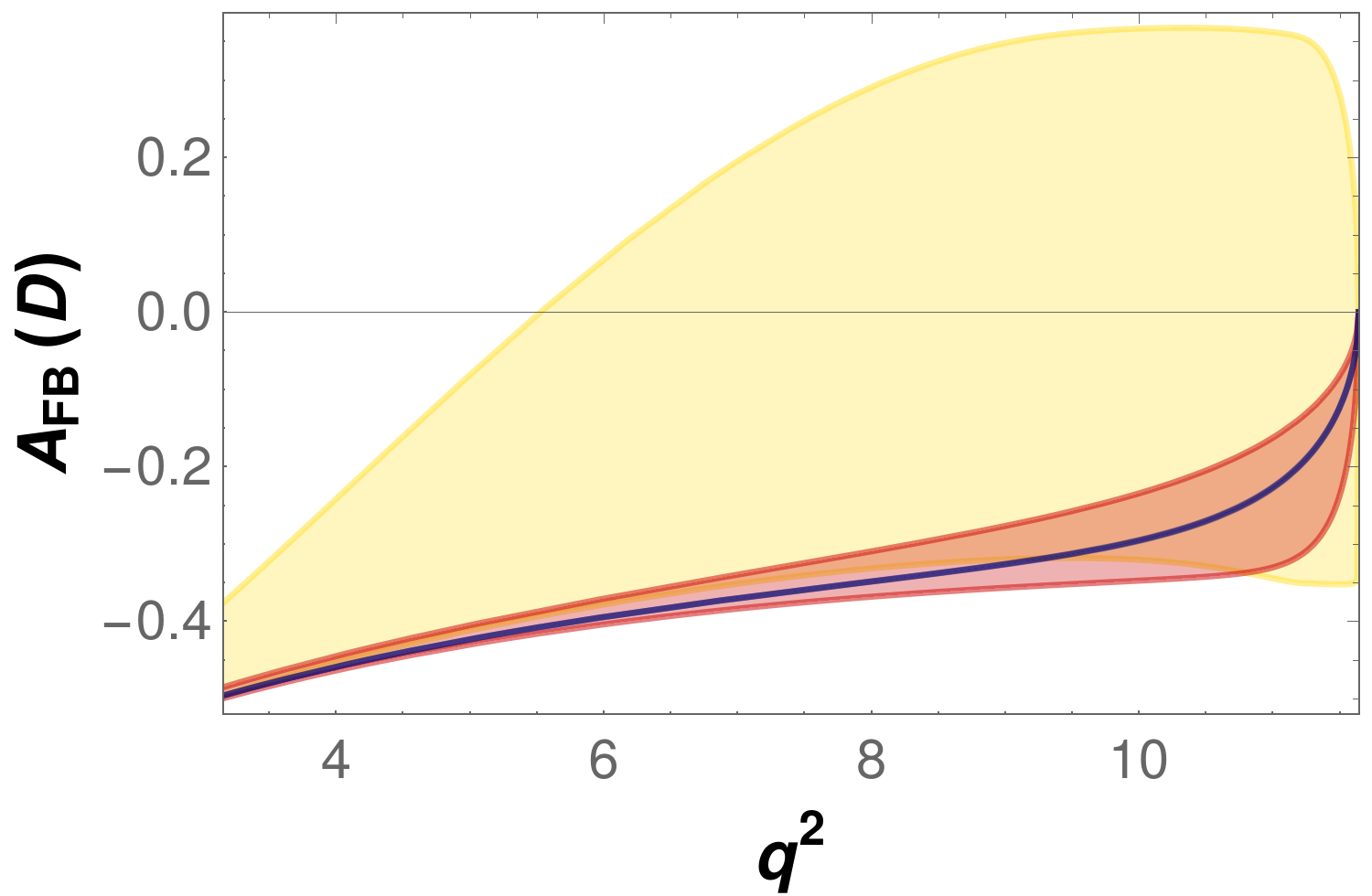}
\includegraphics[width=0.45\linewidth]{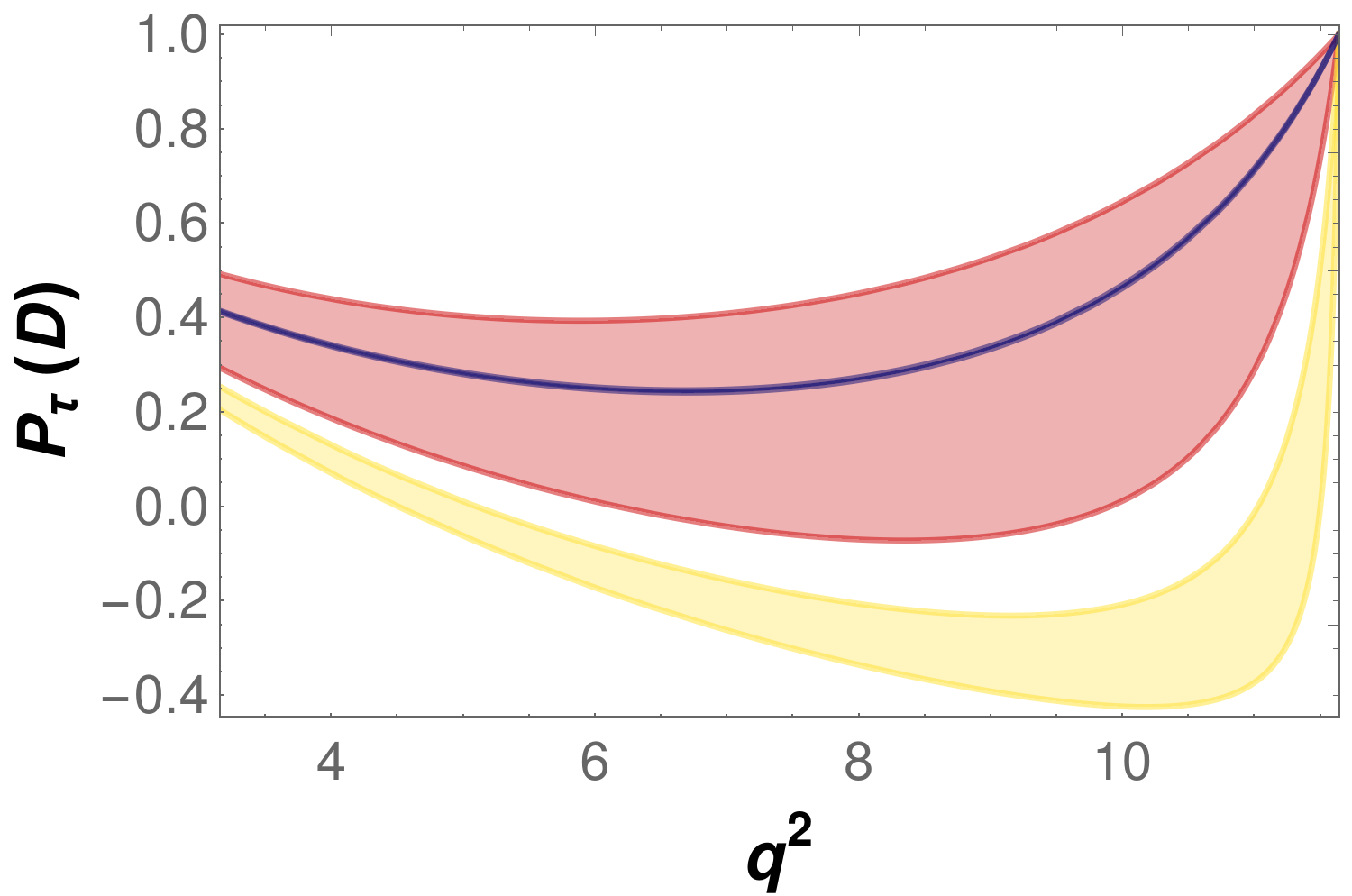}
\includegraphics[width=0.45\linewidth]{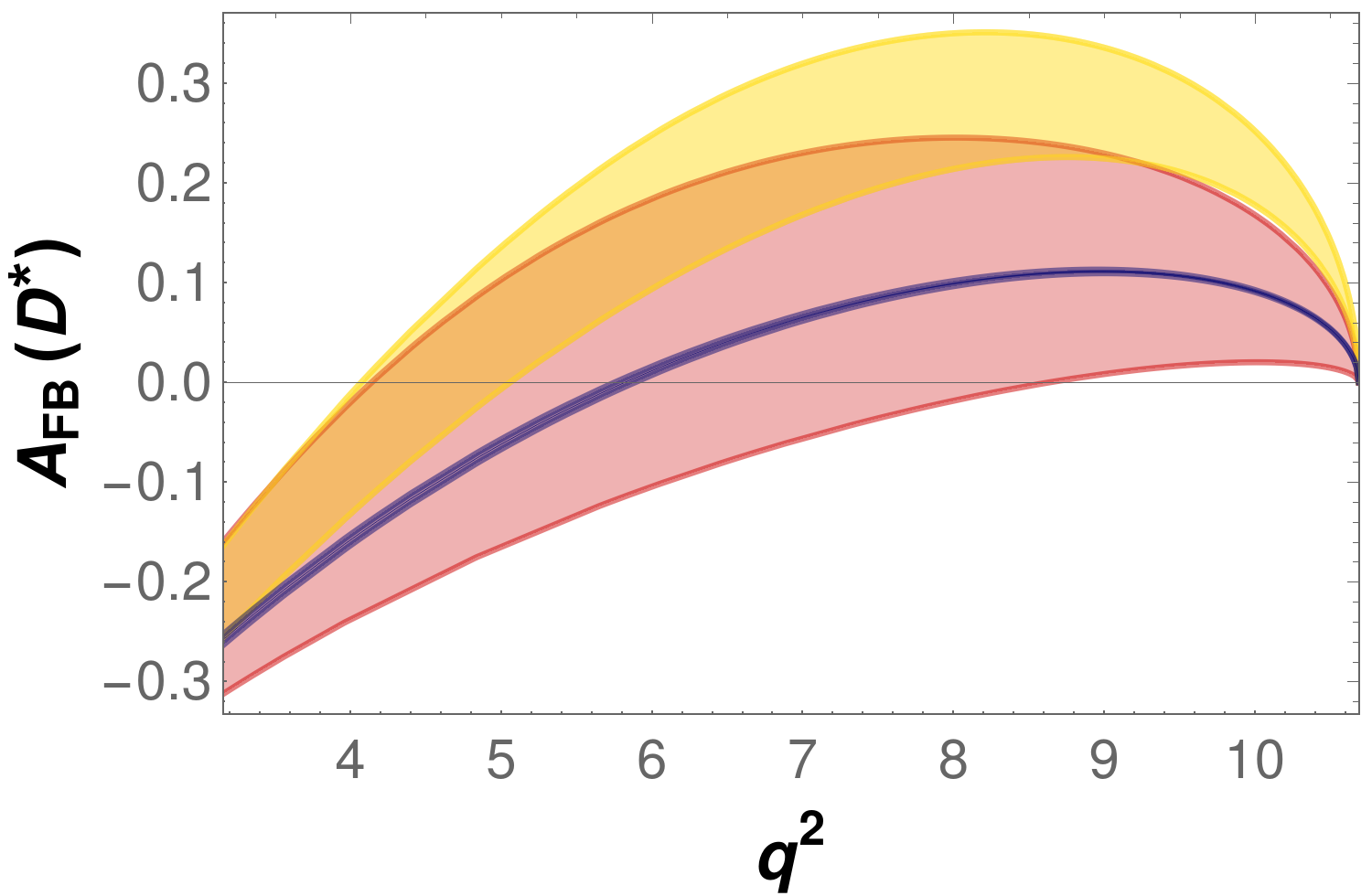}
\includegraphics[width=0.45\linewidth]{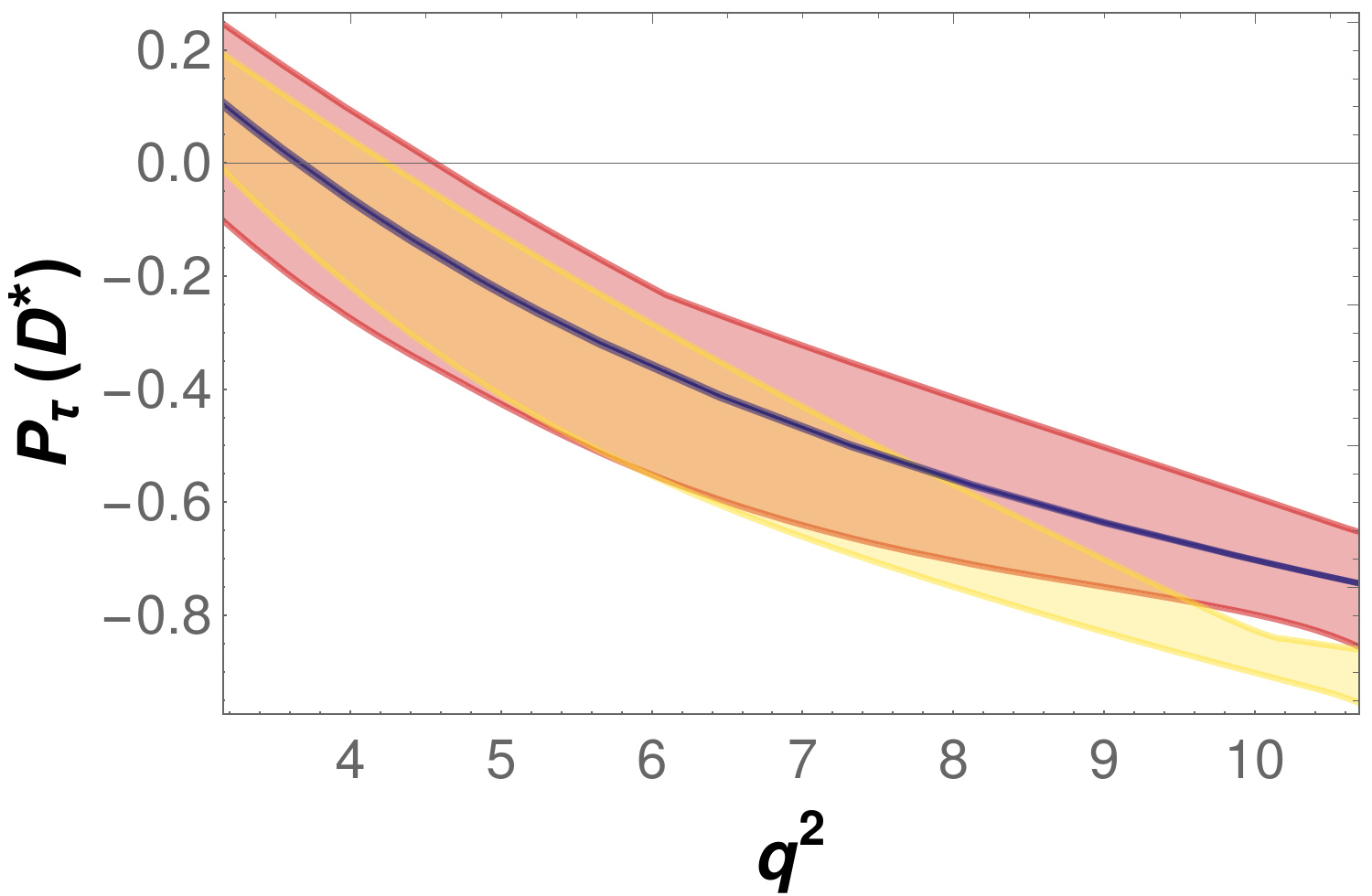}
\includegraphics[width=0.45\linewidth]{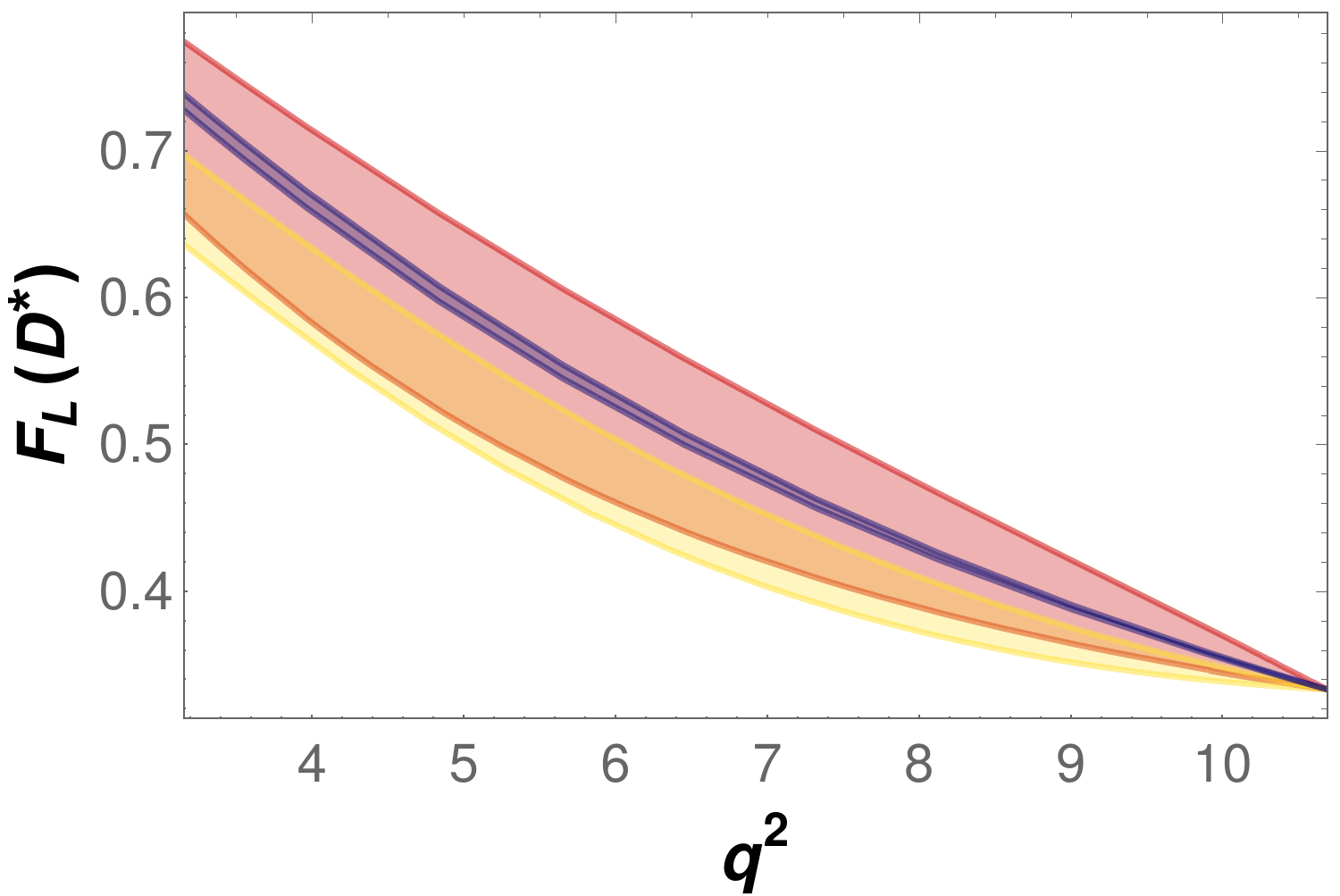}
\caption{Predictions and $1\sigma$ uncertainty on the $q^2$ dependence of the $B \to D^{(*)} \tau \bar{\nu}_\tau$ observables, for the solutions of the fit including the Moriond result and $F_L^{D^*}$. An upper bound of ${\cal B}(B_c \to \tau \bar{\nu}_\tau) \leq 30\%$ has been adopted.
The predictions of Min~1b, Min~2b and the SM are represented by a red, yellow and blue band, respectively.
}
\label{fig:Predictions_angular_q2}
\end{figure}

Analogously, one can extract from the angular distribution in the secondary $D^* \to D \pi$ decay the fraction of longitudinally polarised $D^*$ mesons by constructing the following observable:
\begin{equation}
    F_L^{D^*}(q^2) = \frac{d\Gamma_{\lambda_{D^*}=0}}{dq^2}  \bigg/  \frac{d\Gamma^{D^{*}}}{dq^2}\, .
\end{equation}
In Fig.~\ref{fig:Predictions_angular_q2}, we show the $q^2$ dependence of the $B \to D^{(*)} \tau \bar{\nu}$ observables defined above, for the two solutions obtained in the global fit including $F_L^{D^*}$, Min~1b and Min~2b, together with their SM prediction.

Using these observables, Min~2b could rather clearly be differentiated from both the SM and Min~1b. The same is not true for Min~1b and the SM, for the simple reason that this minimum is compatible with only shifting the SM coefficient at $1\sigma$. In that case the SM predictions are unchanged, which means that the width of the red bands is due to the possible presence of additional NP operators. Precise measurements of these distributions could hence show the existence of operators other than $\mathcal{O}_{V_L}$.

\begin{figure}
\centering
\includegraphics[scale=0.55]{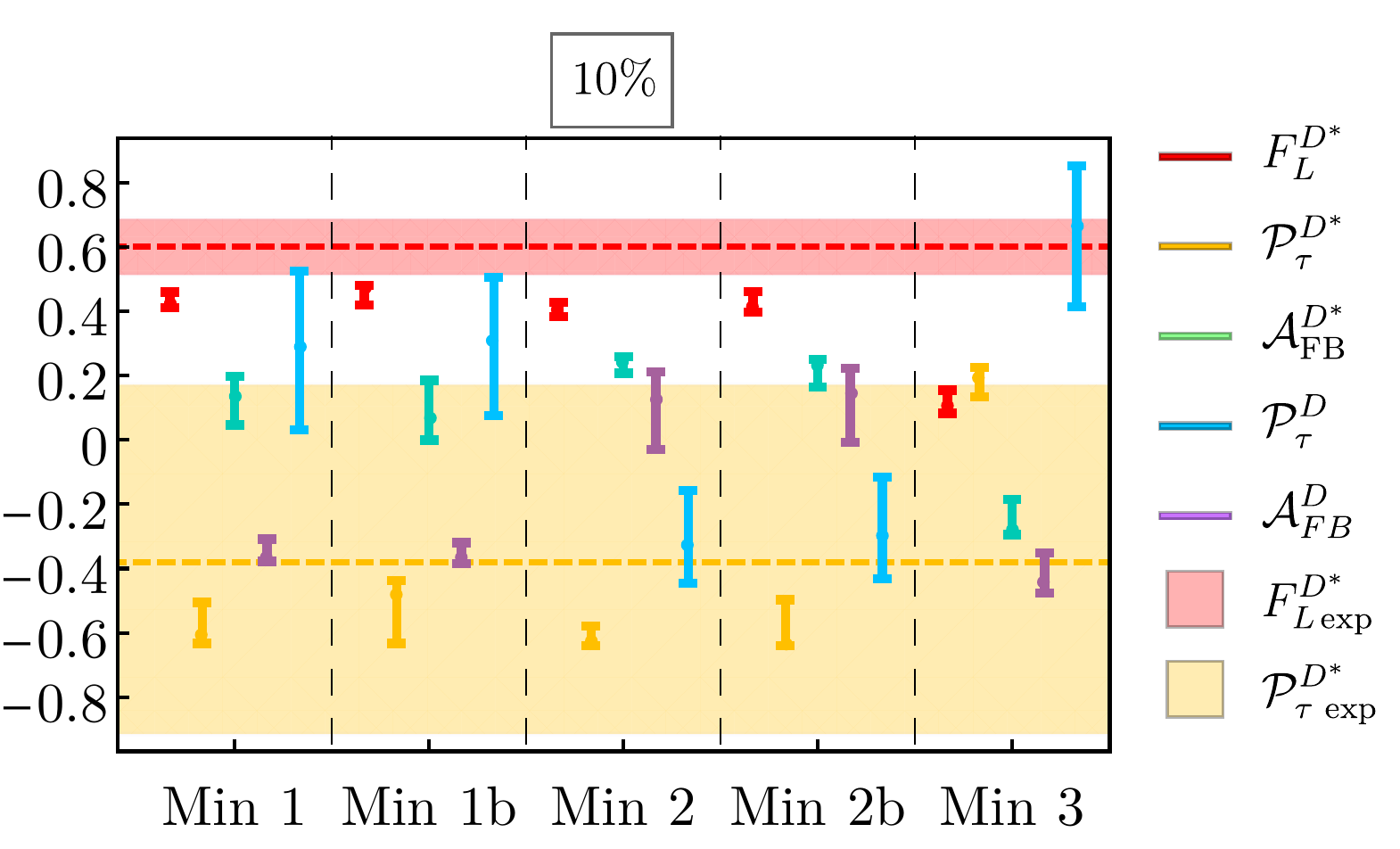}
\includegraphics[scale=0.55]{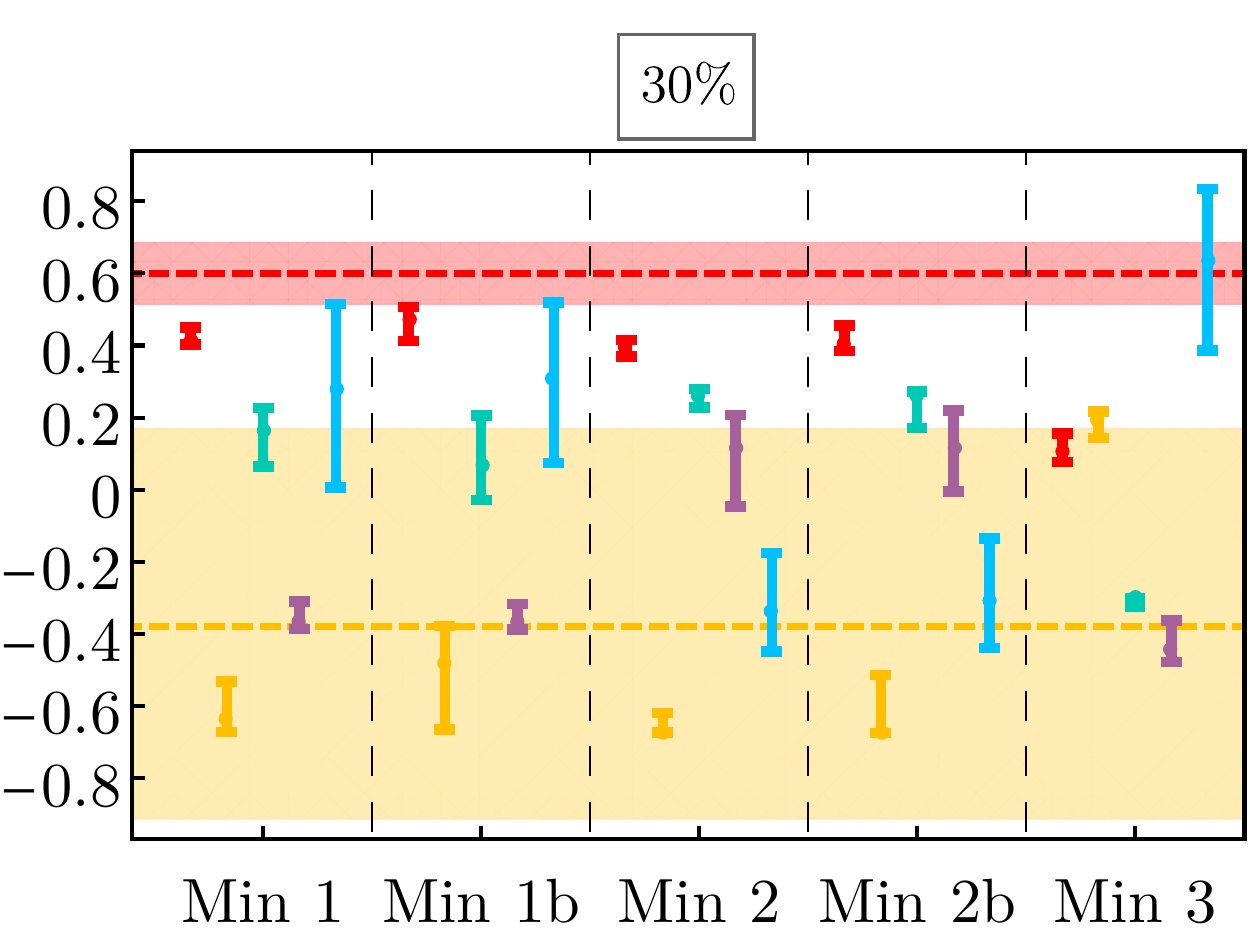}

\vspace{0.5cm}
\includegraphics[scale=0.55]{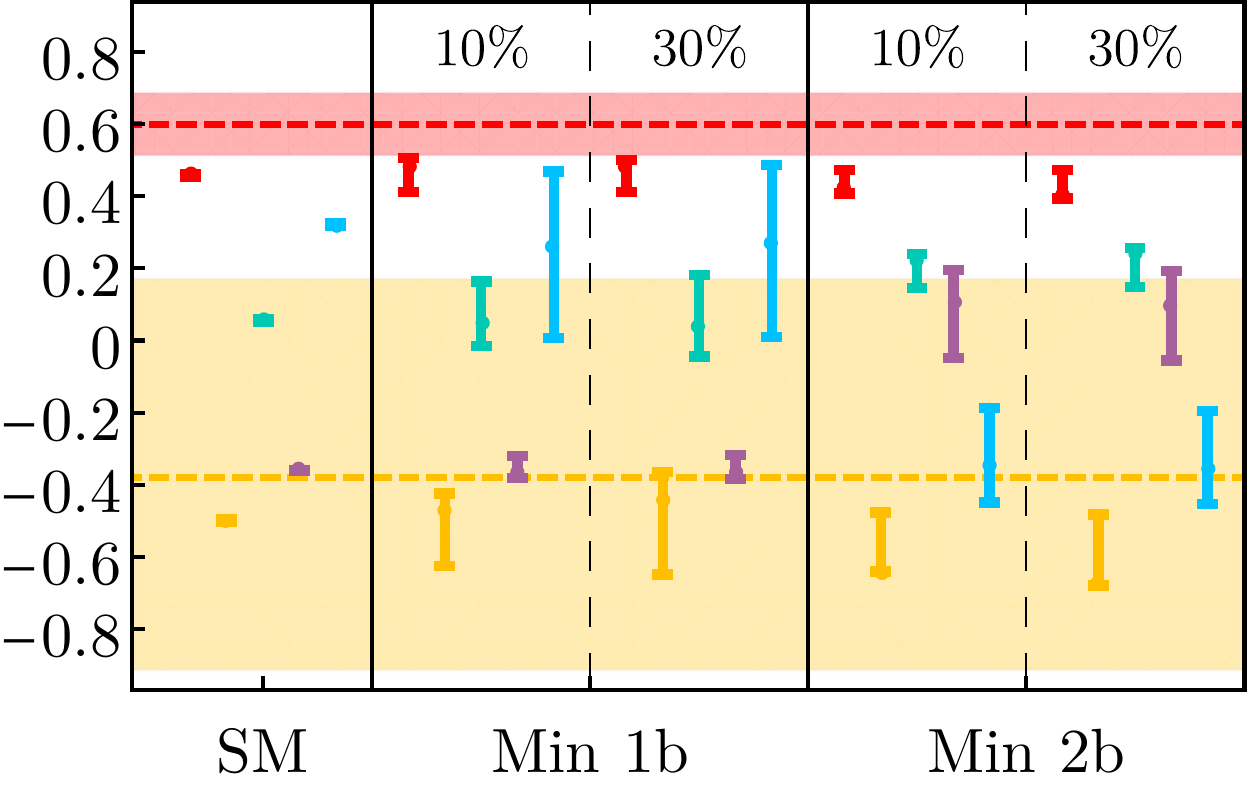}
\caption{The upper panels show the predictions of selected observables for the different minima without (Min~1, Min~2 and Min~3) and with (Min~1b, Min~2b) the inclusion of $F_L^{D^*}$ in the fit. The shaded areas show the experimental results at $1\sigma$ where applicable. On the left (right) panel, a bound of ${\cal B}(B_c \to \tau \bar{\nu}_\tau) \leq 10\%$ ($30\%$) has been applied. The lower panel shows the predictions of the same observables for the two minima obtained in the fit including $F_L^{D^*}$ and the preliminary Belle result, with a bound of ${\cal B}(B_c \to \tau \bar{\nu}_\tau) \leq 10\%$ and $30\%$, and for the SM (first column).}
\label{fig:observables_exp}
\end{figure}

Given the aforementioned difficulty with measuring $q^2$ distributions, typically the integrated observables are measured first, defined as
\begin{equation}
   \mathcal{O}\, =\, \frac{1}{\Gamma^{D^{(*)}}}\;
 \int_{m_\tau^2}^{q^2_\text{max}}  dq^2\; {\cal {O}}(q^2)
 \, , 
\end{equation}
where $\mathcal{O}(q^2)$ refers to the numerator in the ratios, i.e., numerator and denominator have to be integrated separately. The Belle collaboration has in fact released results for two integrated quantities,
the $\tau$ polarisation asymmetry ${\cal P}_\tau^{D^*}=-0.38 \pm 0.51 \text{ (stat) }^{+0.21}_{-0.16} \text{ (syst)}$ \cite{Hirose:2017dxl}, and the recently announced longitudinal polarisation of the $D^*$ meson, $F_L^{D^*}=0.60 \pm 0.08 \text{ (stat) }\pm 0.04 \text{ (syst)}$ \cite{Adamczyk:2019wyt,Abdesselam:2019wbt}.
In Fig.~\ref{fig:observables_exp}, we show the predictions for the integrated observables of $B \to D^{(*)} \tau \bar{\nu}_\tau$, together with their experimental values where available.
Clearly already the integrated observables provide a possibility to distinguish the different NP scenarios.
The fitted values for $F_L^{D^*}$ are closer to the experimental results for the fits including this observable, which is to be expected. However, they fail to reproduce the measurement within $1\sigma$, as discussed above, which renders a more precise measurement of this quantity an exciting prospect.

\FloatBarrier

\subsection{Predictions for other observables}
\label{sect:predictions}
\subsubsection{Predictions for  $\mathcal{R}_{\Lambda_c}$}
\label{sec:Rlambac}

Another observable that could shed light on the $\mathcal{R}_D^{(*)}$ puzzle is the $\Lambda_b \to \Lambda_c \tau \bar{\nu}_\tau$ decay, in particular the universality ratio 
\begin{equation}
    \mathcal{R}_{\Lambda_c}\,  =\, \frac{\mathcal{B}(\Lambda_b \to \Lambda_c \tau \bar{\nu}_{\tau})}{\mathcal{B}(\Lambda_b \to \Lambda_c \ell \bar{\nu}_{\ell})} \, .
\end{equation}
This decay mode has not been observed yet, but LHCb has the potential to perform this measurement in the near future.

On the theoretical side, the differential decay rate $\Lambda_b \to \Lambda_c \ell \bar{\nu}_\ell$ has been calculated in terms of the helicity amplitudes~\cite{Detmold:2015aaa, Datta:2017aue}:
 \beqn 
\label{eq:LambdabLambdac}
 \no
\frac{d \Gamma (\Lambda_b \to \Lambda_c \ell \nu )}{d q^2}\, &=&\, \frac{G_F^2 \abs{V_{cb}}^2}{348 \pi^3}\,\frac{q^2 \sqrt{Q_+ Q_-}}{m_{\Lambda_b}^3} \left( 1-\frac{m_{\ell}^2}{q^2} \right)^2 \left[ A_1^{VA} + \frac{m_{\ell}^2}{2 q^2}\, A_2^{VA} + \frac{3}{2}\, A_3^{SP}  \right. \\
&+&  \left. 2 \left( 1 + \frac{2 m_{\ell}^2}{q^2} \right) A_4^T + \frac{3 m_{\ell}}{\sqrt{q^2}}\, A_5^{VA-SP} + \frac{6 m_{\ell}}{\sqrt{q^2}}\, A_6^{VA-T} \right] \, ,
 \eeqn 
where $Q_\pm = (m_{\Lambda_b}\pm m_{\Lambda_c})^2 -q^2$.
The superindices $VA$ indicate vector and axial-vector contributions ($C_{V_R} \pm C_{V_L}$), $SP$ scalar and pseudoscalar ($C_{S_R} \pm C_{S_L}$), and $T$ tensor contributions ($C_T$). Being a baryonic decay, this mode is sensitive to different combinations of Wilson coefficients than $B\to D^{(*)}\tau\bar\nu_\tau$.
We use the parametrization of the QCD form factors from Ref.~\cite{Detmold:2015aaa, Datta:2017aue}, which take the simple form:
 \begin{equation}
f(q^2) = \frac{1}{1-q^2/(m_{\mathrm{pole}}^f)^2} \left[ a_0^f + a_1^f (z^f  (q^2))^2\right]  , \quad  \qquad z^f(q^2) = \frac{\sqrt{t_{+}^{f} - q^2} - \sqrt{t_{+}^f - t_0} }{\sqrt{t_{+}^{f} - q^2} + \sqrt{t_{+}^f - t_0}} \, .
\label{eq:Lambdab_cFF}
 \end{equation}
The numerical values of the corresponding form-factor parameters, extracted from lattice data~\cite{Detmold:2015aaa, Datta:2017aue}, are displayed in Table \ref{table:FFsRLambdac}.  Other relevant  experimental inputs are summarized in Table~\ref{table:expinput}.

\begin{table}
\centering
\begin{tabular}{| c | c  || c | c || c | c|}
\hline
$a_0^{f_+}$ & $\phantom{-}0.8146 \pm 0.0167$ & $a_0^{h_{+}}$ & $\phantom{-}0.9752 \pm 0.0303$ & $m_{\mathrm{pole}}^{f_{+, \perp}}$ & $6.332$ GeV \\ 
$a_1^{f_+}$ & $-4.8990  \pm 0.5425$ & $a_1^{h_{+}}$ & $-5.5000 \pm 1.2361$  & $m_{\mathrm{pole}}^{f_0}$  & $6.725 $ GeV\\
$a_0^{f_0}$ &  $\phantom{-}0.7439 \pm 0.0125$ &$a_0^{h_{\perp}}$  &$\phantom{-}0.7054 \pm 0.0137$ & $m_{\mathrm{pole}}^{g_{+, \perp}}$   & $6.768$ GeV \\
$a_1^{f_0}$ &  $ -4.6480  \pm 0.6084$ & $a_1^{h_{\perp}}$ & $ -4.3578 \pm 0.5114$  & $m_{\mathrm{pole}}^{g_0}$  & $ 6.276$ GeV\\
$a_0^{f_{\perp}}$ &  $\phantom{-} 1.0780  \pm 0.0256$ & $\phantom{-}a_0^{\tilde{h}_{\perp, +}}$ &$\phantom{-} 0.6728 \pm 0.0088$ & $m_{\mathrm{pole}}^{h_{+, \perp}}$ & $6.332$ GeV \\
$a_1^{f_{\perp}}$ &  $-6.4170  \pm 0.8480$ & $a_1^{\tilde{h}_{ +}}$ & $-4.4322 \pm 0.3882$ & $m_{\mathrm{pole}}^{\tilde{h}_{+, \perp}}$ & $6.768$ GeV \\
$\phantom{-}a_0^{g_{\perp, +}}$ &  $\phantom{-}0.6847  \pm 0.0086$ & $a_1^{\tilde{h}_{\perp}}$ &  $-4.4928 \pm 0.3584$ &  &  \\ 
$a_1^{g_+}$ &  $-4.4310  \pm 0.3572$ &  &  &  &  \\
$a_0^{g_{0}}$ &  $\phantom{-}0.7396  \pm 0.0143$ &  & & & \\
$a_1^{g_{0}}$ &  $-4.3660  \pm 0.3314$ & & & &  \\
$a_1^{g_{\perp}}$ &  $-4.4630  \pm 0.3613$ & & & & \\
\hline
\end{tabular}
\caption{Central values and uncertainties of the nominal form-factor parameters for $\Lambda_b \to \Lambda_c\ell\nu_\ell$ \cite{Detmold:2015aaa, Datta:2017aue}.}
\label{table:FFsRLambdac}
\end{table}

\begin{figure}
\centering
\includegraphics[scale=0.45]{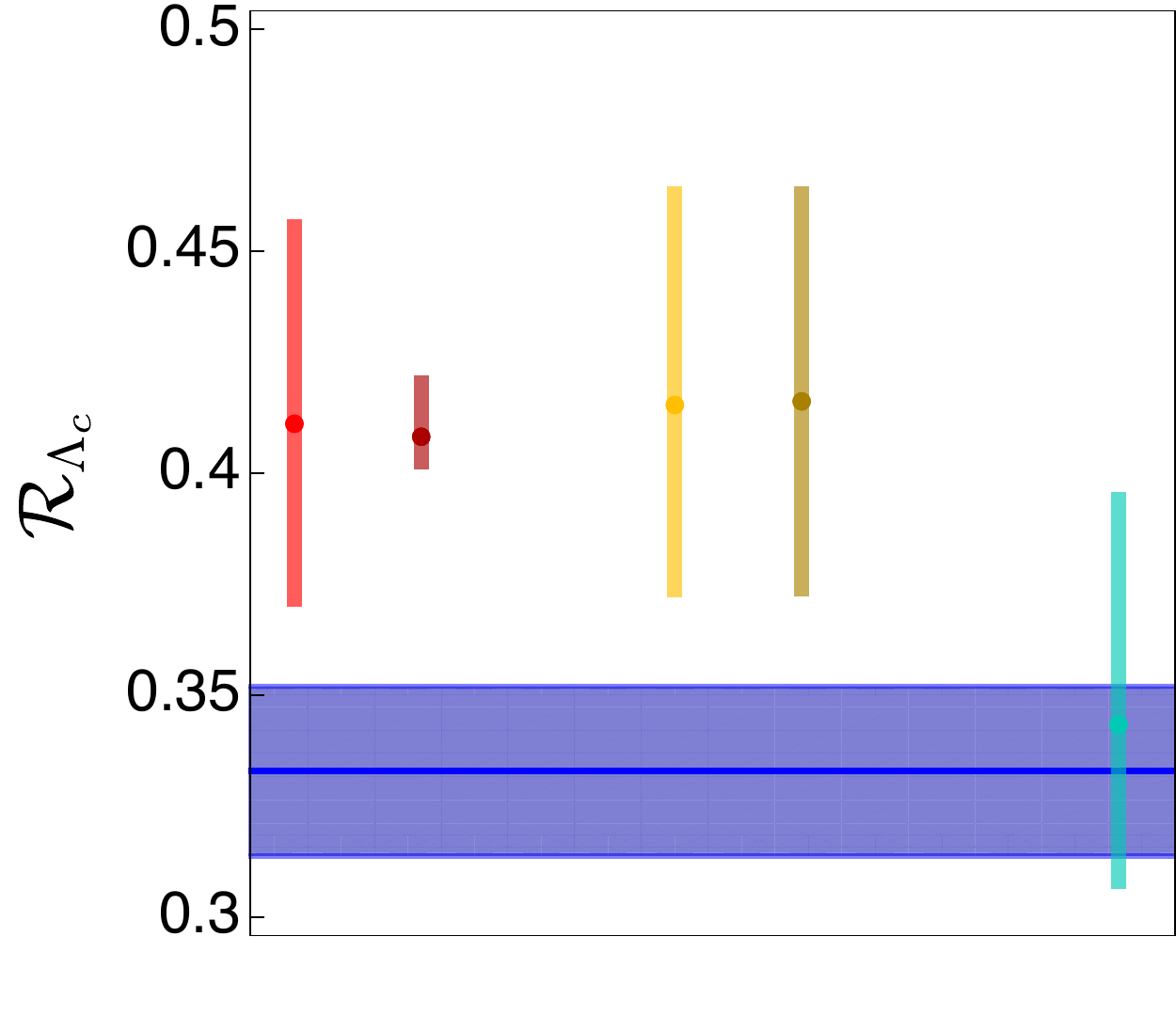}
\hskip .5cm
\includegraphics[scale=0.6]{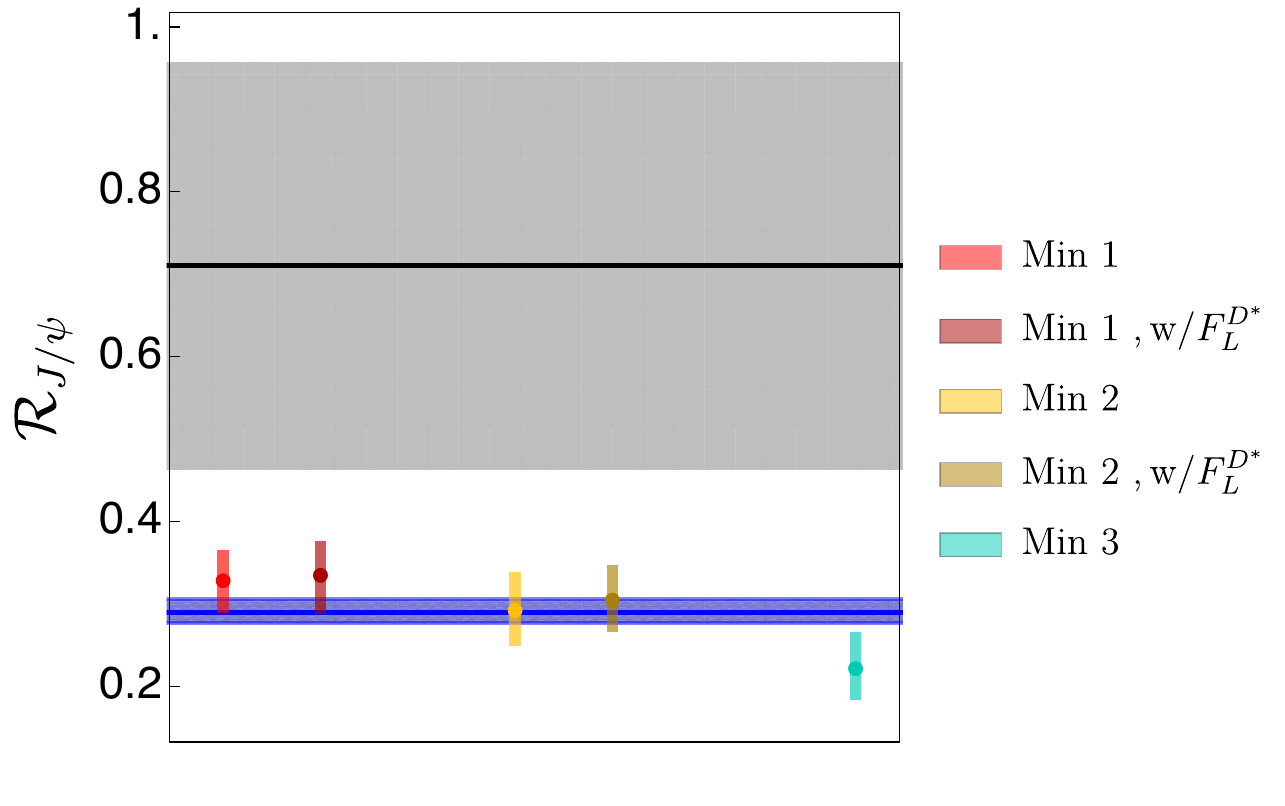}
\caption{Predictions for $\mathcal{R}_{\Lambda_c}$ (left) and $\mathcal{R}_{J/\psi}$ (right) for the minima of Table~\ref{table:minimaA} and Table~\ref{table:minimaB}, with an upper bound $\mathcal{B}( B_c \to \tau \nu) \leq 10\% $. The SM prediction is shown as a blue band. The experimental value of $\mathcal{R}_{J/\psi}$ is given by the gray band.} 
\label{fig:preds}
\end{figure}

Fig.~\ref{fig:preds} shows the predicted ratio $\mathcal{R}_{\Lambda_c}$ and its  uncertainty for the three minima of Table~\ref{table:minimaA} (Min~1, Min~2 and Min~3) and the two minima including $F_L^{D^*}$ of Table~\ref{table:minimaB} (Min~1b and Min~2b), with the upper limit $\mathcal{B}(B_c \to \tau \nu) \leq 10 \%$, and the SM prediction. The errors considered here just take into account the variation of the Wilson coefficients and the parametric error for the lattice input. Other systematic errors are not shown.
In all cases the predicted value of $\mathcal{R}_{\Lambda_c}$ is above the SM expectation. This agrees with the observation made in Ref.~\cite{Blanke:2018yud} that the measured enhancement of the ratios $\mathcal{R}_D^{(*)}$ implies an enhancement of $\mathcal{R}_{\Lambda_c}$ for any model of new physics described by the effective Hamiltonian (\ref{eq:effH}). The prediction closest to the SM is obtained with the unstable minimum Min3, which disappears when $F_L^{D^*}$ is included, because it involves a larger value of $C_T$.

\FloatBarrier

\subsubsection{Predictions for $\mathcal{R}_{J/\psi}$}
\label{sec:Jpsi}

The ratio
\begin{equation}
\mathcal{R}_{J/\psi} = \frac{\mathcal{B}(B_c \to J/\psi \tau \bar{\nu}_\tau)}{\mathcal{B}(B_c \to J/\psi \mu \bar{\nu}_\mu)} = 0.71 \pm 0.17 \pm 0.18\, ,
\end{equation}
has been recently measured by LHCb with the run-1 dataset ($3 \text{fb}^{-1}$) \cite{Aaij:2017tyk}. 
We have not included this observable in our fit because
the hadronic uncertainties are not at the same level as for the observables related to $B \to D^{(*)}$ transitions and the experimental error
is large. Instead, the predictions for this observable are computed and compared with the current data. The experimental uncertainties are expected to be significantly reduced with the larger statistics already accumulated at LHCb.

The differential decay rate for this transition can be expressed in a similar way than the $\bar{B} \to D^*$ distribution in Eq.~\eqref{eq:GammaDstar} \cite{Watanabe:2017mip}:
\beqn
 \no
\frac{d \Gamma (B_c \to J/\psi \ell \bar{\nu)}}{dq^2} &=& \frac{G_F^2 \abs{V_{cb}}^2}{192 \pi^3 m_{B_c}}\, q^2 \sqrt{\lambda_{J/\psi}(q^2)} \left(1- \frac{m_{\ell}^2}{q^2} \right) \times \\ \no
&& \left\lbrace (\abs{1+C_{V_L}}^2 + \abs{C_{V_R}}^2) \left[ \left( 1 + \frac{m_{\ell}^2}{2q^2} \right) \left(H_{V,+}^2 + H_{V,-}^2 + H_{V,0}^2 \right) + \frac{3}{2}\, \frac{m_{\ell}^2}{q^2} H_{V,t}^2 \right] \right. \\ \no
&-& 2 \,\mathcal{R}\mathrm{e} \left[(1+C_{V_L}) C_{V_R}^* \right] \left[ \left( 1 + \frac{m_{\ell}^2}{2q^2} \right) \left(H_{V,0}^2 + 2 H_{V,+} \cdot H_{V,-} \right)  + \frac{3}{2}\,
\frac{m_{\ell}^2}{q^2} H_{V,t}^2   \right] \\ \no
&+& \frac{3}{2} \abs{C_{S_L} - C_{S_R}}^2 H_S^2 + 8 \abs{C_T}^2 \left( 1+ \frac{2 m_{\ell}^2}{q^2} \right) \left( H_{T,+}^2 + H_{T,-}^2 + H_{T,0}^2 \right) \\
&+& 3\, \mathcal{R}\mathrm{e} \left[ \left( 1 + C_{V_L} - C_{V_R} \right) \left(C_{S_L}^* - C_{S_R}^* \right) \right]  \frac{m_{\ell}}{\sqrt{q^2}} H_S \cdot H_{V,t} \\  \no
&-& 12 \,\mathcal{R}\mathrm{e} \left[ \left( 1 + C_{V_L} \right) C_T^* \right]
\frac{m_{\ell}}{\sqrt{q^2}} \left(H_{T,0} \cdot H_{V,0} + H_{T,+} \cdot H_{V,+} - H_{T,-} \cdot H_{V,-} \right) \\ \no
&+&  \left. 12 \,\mathcal{R}\mathrm{e} \left[ C_{V_R} C_T^*\right] \frac{m_{\ell}}{\sqrt{q^2}} \left( H_{T,0} \cdot H_{V,0} + H_{T,+} \cdot H_{V,-} - H_{T,-} \cdot H_{V,+} \right)   \no \right\rbrace ,
\label{eq:Jpsi}
\eeqn
where $\lambda_{J/\psi}(q^2) = \left[( m_{B_c} - m_{J/\psi})^2 - q^2 \right]\left[( m_{B_c} +m_{J/\psi})^2 - q^2 \right]$ is the usual K\"{a}ll\'en function and $H_i$ are the hadronic helicity amplitudes, similar to the ones used for the decay rates of Sec.~\ref{sec:TheoricalFramework}, which can be found in Appendix~\ref{appendix:helicity}.

The predicted values of $R_{J/ \psi}$ for the minima of Tables~\ref{table:minimaA} and \ref{table:minimaB} as well as for the SM, are given in the right panel of Fig.~\ref{fig:preds}. Again the errors considered here just take into account the variation of the Wilson coefficients and the parametric error for the lattice input. For this observable, there are additional theoretical uncertainties associated with the parametrization of the form factors, which are difficult to quantify. Given the large errors, the predictions from all minima are in agreement with the experimental measurement. 
We note that the prediction from the global minimum is the one that approaches closest to the experimental measurement, albeit only slightly.

\FloatBarrier

\section{Conclusions}
\label{sec:Conclusions}

In this work we have analysed the new-physics parameter space able to explain the current anomalies in $b\to c\tau\nu$ data, taking the available experimental information at face value, i.e., disregarding the possibility that these anomalies could originate in underestimated systematic uncertainties or statistical fluctuations. We have performed a global fit to the available data in $b \to c \tau \bar{\nu}_\tau$ transitions, adopting an EFT approach with a minimal set of assumptions:  1) NP only enters in the third generation of fermions. 2) There is a sizeable energy gap between NP and the electroweak scale, the EFT operators are $SU(2)_L \otimes U(1)_Y$ invariant and the electroweak symmetry breaking is linearly realized. 3) All Wilson coefficients are real (CP is conserved). We have tested the impact of the latter assumption, but did not find an improved description of the data. In contrast to previous works, we considered the $q^2$ distributions measured by BaBar and Belle. Moreover, we study the effect of including the recently announced $F_L^{D^*}$ measurement by the Belle collaboration in the fit. A comparison with earlier analyses, either not including the $q^2$ distributions, the $F_L^{D^*}$ measurement, or considering smaller sets of operators, precisely illustrates the benefits of our fit: as described in Section~\ref{sec:FitResults}, most of the NP solutions found in previous fits are disfavoured once all the information considered in this work is added.


We performed the global fit in different scenarios. As a baseline, we considered the full dataset before the announcement of the $F_L^{D^*}$ measurement with the subset of operators implied by our assumptions, \emph{i.e.} with a flavour-universal coefficient $C_{V_R}$. We then performed extensive comparisons to datasets including the recent $F_L^{D^*}$ measurement, the preliminary Belle measurement of $\mathcal{R}_{D^{(*)}}$, and different bounds on ${\cal{B}}(B_c\to\tau\bar\nu_\tau)$, as well as a second parameter set, allowing for a non-universal $C_{V_R}$. 

In the baseline fit, three minima have been obtained, given in Table~\ref{table:minimaA}. The global minimum, referred to in the text as Min~1, has an excellent $\chi^2$; while none of the fitted Wilson coefficients are required to be non-zero for this minimum, the simplest interpretation of this solution is a global modification of the SM: setting all Wilson coefficients but $C_{V_L}$ to zero increases the $\chi^2$ only by $\Delta\chi^2=1.4$, implying an even better fit. 
The other two solutions are local minima which numerically exhibit stronger deviations from the SM, with larger contributions of the tensor and scalar operators. While the global minimum is compatible with a SM-like scenario, Min~2 and Min~3 require additional operators. For instance, they could involve scalar LQs with quantum numbers $R_2 \sim (3,2,7/6)$ or $ S_1 \sim (\bar{3},1,1/3)$.

The measurement of the $D^*$ longitudinal polarization fraction $F_L^{D^*}$ has quite a strong impact on our EFT analysis. It removes Min~3 as a solution for the fit, which was, however, already strongly disfavoured by the differential distributions.
Fig.~\ref{fig:CVReq0} illustrates the tension between the present measurement of $F_L^{D^*}$, the bound on ${\cal{B}}(B_c\to\tau\bar\nu_\tau)$, and the observation $\Delta\mathcal{R}_{D^*}>0\,$: the set of operators considered within our assumptions cannot accommodate all three observations at $1\sigma$ for any combination of Wilson coefficients. Indeed, including the $F_L^{D^*}$ measurement in the fit increases the minimal $\chi^2$ significantly also for the two lower-lying minima (Min~1b and Min~2b), see Table~\ref{table:minimaB}. 

We find that most of the minima saturate the upper bound $\mathcal{B}(B_c \to \tau \bar{\nu}_\tau) \leq 10\%$, and it is interesting to study the effect of changing this constraint on the fit. As shown in Tables~\ref{table:minimaA} and~\ref{table:minimaB}, adopting a more conservative upper bound of $\mathcal{B}(B_c\to \tau \bar{\nu}_\tau)~\leq~30\%$ we find the same number of minima; they are qualitatively similar to the previous ones, but with larger central values and ranges of the scalar Wilson coefficients, specifically their pseudoscalar combination. While even this larger upper bound is saturated in most of our fits, the overall decrease in $\chi^2$ is small.

The fact that $F_L^{D^*}$ cannot be accommodated within $1\sigma$ for $C_{V_R}=0$ could have important consequences, should the present value be confirmed with higher precision. This led us to investigate the scenario with non-zero $C_{V_R}$ as a possible resolution of this tension on the theory side.
We find that its inclusion
helps to reduce the tension among the experimental $B\to D^*$ data, and it is now possible to satisfy all constraints at $1\sigma$, as illustrated in Fig.~\ref{fig:CVRneq0}. The global fit including $C_{V_R}$ leads to four different minima, as Fig.~\ref{fig:CVR} shows.
Two of these minima have a significantly lower $\chi^2$ than the previous fits, however, they correspond to fine-tuned solutions where the SM coefficient becomes very small and its effect is substituted by several sizeable NP contributions, especially $C_{V_R}$. This scenario seems therefore not to be a satisfactory resolution of the tension.

We have also presented predictions for selected $b \to c \tau \bar{\nu}_\tau$ observables, such as $\mathcal{R}_{\Lambda_c}$, $\mathcal{R}_{J/\psi}$ or the forward-backward asymmetries and $\tau$ polarization in $B\to D^{(*)}\tau \bar{\nu}_\tau$, which have not been included in the fits because either they have not been measured yet or their current experimental values have too large uncertainties. We have studied these observables for the different solutions emerging from our fits, finding that they provide complementary information to the existing data. This is displayed in Figs.~\ref{fig:Predictions_angular_q2}, \ref{fig:observables_exp} and~\ref{fig:preds}. The future measurement of these observables could both establish NP in these modes and allow for a discrimination among the currently favoured scenarios.

We conclude that the anomaly in $b\to c\tau\bar\nu_\tau$ transitions remains and can be addressed by NP contributions. Apart from $\mathcal{R}_{D^{(*)}}$, also the differential $q^2$ distributions, $F_L^{D^*}$ and ${\cal{B}}(B_c\to\tau\bar\nu_\tau)$ are important to constrain NP, leaving only two viable minima in the global fit. Our general EFT approach does not allow to identify uniquely the potential mediator, since the global minimum can be generated by several combinations of parameters. 
The generality of our analysis on the other hand allows to use the obtained parameter ranges in more general SMEFT analyses. An improved measurement of $F_L^{D^*}$ close to its present central value holds the exciting potential to invalidate this general approach, which would have major implications, like a Higgs sector different from the SM one, the existence of NP particles relatively close to the electroweak scale, or new light degrees of freedom. As we have shown, additional measurements will be able to clarify these questions.

\section*{Note added in proof}

After the submission of our paper for publication, the HFLAV collaboration released a new world average of the  ${\cal R}_D$ and ${\cal R}_{D^*}$  ratios \cite{Amhis:2016xyh}: \footnote{The additional shift with respect to our preliminary average given in Eq.~(7) is due to a different treatment for $\mathcal{B}(B\to D^{(*)}\ell\nu)$: in the new HFLAV average a measurement by the BaBar collaboration \cite{Aubert:2008yv} is omitted, because it does not allow for a separation of the different isospin modes.}
\beqn 
\mathcal{R}_{D}^{\text{avg,new}} &= 0.340\pm 0.027 \pm 0.013 \qquad\quad  \text{  and  } \qquad\quad
\mathcal{R}_{D^*}^{\text{avg,new}} &= 0.295\pm 0.011 \pm 0.008\,,
\eeqn
with a correlation of -0.38. These averages give a $3.7\sigma$ discrepancy with respect to our SM prediction instead of the $3.1\sigma$ calculated by HFLAV. The slightly larger significance with respect to the value quoted by HFLAV is due to our different SM prediction and has three aspects: slightly smaller central value and uncertainty for $R_{D^*}$, as well as the inclusion of the correlation between the SM predictions for $R_D$ and $R_{D^*}$. Regarding the central value, note also the $\sim1\sigma$ lower central value  of the SM prediction for $R_{D^*}$ in \cite{Gambino:2019sif} compared to \cite{Bigi:2017jbd} after taking into account new data for $B\to D^*\ell\nu$.

In Table~\ref{table:new} we update the results of our baseline fit with the new HFLAV averages (assuming a lepton universal $C_{V_R}$ and including the longitudinal $D^*$ polarisation, $F_L^{D^*}$).

\def\arraystretch{1.5}
\begin{table}[h]
\centering
\begin{tabular}{ c | c |  c }
 & Min 1b & Min 2b  \\
 \hline
$\chi^2_\text{min}/$d.o.f.  & $ 37.4/54 $ & $40.4/54$ \\
\hline
$C_{V_L}$ & $\phantom{-}0.09^{+0.13}_{-0.12}$
& $\phantom{-}0.34^{+0.05}_{-0.07}$ \\
$C_{S_R}$ & $\phantom{-}0.086^{+0.12}_{-0.61}
$ 
& $-1.10^{+0.48}_{-0.07}$
 \\ 
$C_{S_L}$ & $-0.14^{+0.52}_{-0.07} $ 
&$-0.30^{{+0.11}}_{-0.50}$ \\
$C_T$ & $\phantom{-} 0.008 ^{+ 0.046}_{- 0.044} $ &
$\phantom{-}0.093^{+0.029}_{-0.030}$ \\
\end{tabular}
\caption{Minima and $1\sigma$ uncertainties obtained from the global $\chi^2$ minimization, including the new HFLAV world average on the ratios ${\cal R}_D$ and ${\cal R}_{D^*}$~\cite{Amhis:2016xyh} and the $F_L^{D^*}$ polarization, using ${\cal B}(B_c \to \tau \bar{\nu}_\tau )<10\%$. There are, in addition, the corresponding sign-flipped minima, as indicated in Eq.~\eqref{eq:WilsonSecondMin}.}
\label{table:new}
\end{table}

As the reader may notice and can be expected, the results shown in Table~\ref{table:new} are very similar to those shown in Table~\ref{table:minima_Belle}, with central values for the Wilson coefficients, in general, slightly closer to the SM. 

\section*{Acknowledgements}

We would like to thank the authors of Ref.~\cite{Blanke:2018yud} for their useful comments on the $\Lambda_b\to\Lambda_c\ell\nu$ results.
This work has been supported in part by the Spanish Government and ERDF funds from
the EU Commission [Grant FPA2017-84445-P], the Generalitat Valenciana [Grant Prometeo/2017/053], the Spanish Centro de Excelencia Severo Ochoa Programme [Grant SEV-2014-0398] and the DFG cluster of excellence “Origin and Structure of the Universe”.
The work of Clara Murgui has been supported by a La Caixa--Severo Ochoa scholarship.
The work of Ana Pe\~nuelas is funded by Ministerio de Ciencia, Innovación y Universidades, Spain [Grant FPU15/05103].

\appendix

\section{Additional experimental and theoretical inputs}
\label{app:addinput}

The binned distributions for $B\to D^{(*)}\tau\nu$ from BaBar and Belle are given in Table~\ref{table:angular} and additional experimental inputs used in our analysis are summarized in Table~\ref{table:expinput}.

\begin{table}
\centering
\begin{tabular}{| c | c  | c || c | c | c |}
\hline
\multicolumn{3}{|c||}{Belle} & \multicolumn{3}{c|}{BaBar} \\
\hline
$q^2\; (\mathrm{GeV}^2) $ & $B \to D \tau \nu$ & $B \to D^* \tau \nu$ & $q^2\; (\mathrm{GeV}^2) $ & $B \to D \tau \nu$ & $B \to D^* \tau \nu$ \\
\hline
4.0 - 4.53 &  $24.0 \pm 16.3 $ &    $5.4 \pm 9.3 $  &4.0-4.5 &  $23.8 \pm 12.1 $ &  $0.6 \pm 7.1 $ \\ 
4.53-5.07 &   $27.8 \pm 15.2$ &  $3.4 \pm 8.1 $ & 4.5-5.0 &   $15.8 \pm 11.8$ &  $23.6 \pm 9.5 $ \\ 
5.07-5.6 &  $22.0\pm 14.0$ &  $-3.8 \pm 6.8 $ & 5.0-5.5 &  $27.0\pm 10.5$ &  $22.4 \pm 7.7 $ \\ 
5.6 - 6.13  &  $28.4 \pm 14.4$ &  $12.1 \pm 8.4 $ & 5.5-6.0  &  $45.1 \pm 13.1$ &  $20.8\pm 7.8 $ \\
6.13-6.67 & $16.2 \pm 14.8$ &  $8.0 \pm 9.4 $ & 6.0-6.5 & $46.9 \pm 13.3$ &  $20.0 \pm 7.5$ \\
6.67-7.2 & $44.5 \pm 15.5$ & $24.7 \pm 8.2 $ & 6.5-7.0 & $39.7 \pm 13.6$ & $38.8 \pm 8.6 $ \\ 
7.2-7.73 & $14.2 \pm 16.3$ &  $2.7 \pm 7.8 $ & 7.0-7.5 & $31.7 \pm 12.4$ &  $44.4 \pm 9.2 $ \\
7.73-8.27 & $-3.1 \pm 15.3$ & $28.7 \pm 9.2 $ & 7.5-8.0 & $47.4 \pm 14.9$ & $49.3 \pm 10.3 $ \\ 
8.27-8.8 & $16.1 \pm 15.2$ &  $30.8 \pm 8.5 $ & 8.0-8.5 & $33.7 \pm 14.0$ &  $40.0 \pm 9.4 $ \\ 
8.8-9.33 & $37.2 \pm 15.5$ &  $24.9 \pm 7.6 $ & 8.5-9.0 & $17.7 \pm 13.2$ &  $37.3 \pm 9.5 $ \\ 
9.33-9.86 & $19.3 \pm 15.2$ &  $15.0 \pm 6.8 $ & 9.0-9.5 & $-0.7 \pm 13.1$ &  $38.4 \pm 9.8 $ \\ 
9.86-10.4 & $37.0 \pm 15.5$ &  $14.8 \pm 5.1 $ & 9.5-10.0 & $6.9 \pm 14.3$ &  $31.7 \pm 11.0 $ \\ 
10.4-10.93 & $-1.0 \pm 14.2$ &  $16.3 \pm 5.1 $ & 10.0-10.5 & $35.4 \pm 16.0$ &  $31.9 \pm 10.5 $ \\ 
10.93-11.47 & $20.0 \pm 13.1$ &  - & 10.5-11.0 & $2.8 \pm 12.1$ &  $16.7 \pm 10.4 $ \\ 
11.47-12.0 & $3.4 \pm 10.9$ &  -  & 11.0-11.5 & $1.7 \pm 11.3$ &  -  \\ 
 & & & 11.5-12.0 & $6.5 \pm 8.9$ &  -   \\
\hline
\end{tabular}
\caption{Measured $q^2$ distributions for $B \to D^{(*)} \tau \nu$ events by Belle \cite{Huschle:2015rga} (left) and
BaBar \cite{Lees:2013uzd} (right).}
\label{table:angular}
\end{table}

\begin{table}[h]
\centering
\begin{tabular}{ c | c  | c }
Parameter & Value & Comments \\
\hline
$m_{D^{+}}$  &  $(1869.65  \pm 0.05)\cdot 10^{-3} $ GeV &  \cite{Tanabashi:2018oca}  \\ 
 $m_{D^0}$ &   $(1864.83  \pm  0.05)\cdot 10^{-3} $ GeV &  \cite{Tanabashi:2018oca}  \\ 
 $m_{D^*}$ &  $ (2006.85   \pm 0.05)\cdot 10^{-3} $ GeV &  \cite{Tanabashi:2018oca}  \\  
$m_{D^{*+}}$  &   $(2010.26  \pm  0.05)\cdot 10^{-3} $ GeV &  \cite{Tanabashi:2018oca}  \\ 
$m_{B^+}$ & $ (5279.32 \pm 0.14 )\cdot 10^{-3}$ GeV &  \cite{Tanabashi:2018oca}  \\ 
$m_{B^0}$ & $(5279.63   \pm 0.15)\cdot 10^{-3} $ GeV &  \cite{Tanabashi:2018oca}  \\
$m_{B_c}$ & $ (6274.9 \pm 0.8)\cdot 10^{-3}  $ GeV &  \cite{Tanabashi:2018oca}  \\ 
$\tau_{B_c}$ & $ (0.507 \pm 0.009)\cdot  10^{-12}  $ s & \cite{Tanabashi:2018oca}  \\ 
$f_{B_c}$ & $ (0.434 \pm 0.15) $ GeV & \cite{Colquhoun:2015oha}  \\ 
\hline
$m_{\Lambda_b}$  &  $(5619.690  \pm 0.17) 10^{-3} $ GeV &  \cite{Tanabashi:2018oca}  \\ 
$m_{\Lambda_c}$ &   $(2286.465  \pm 0.14) 10^{-3} $ GeV &  \cite{Tanabashi:2018oca}  \\ 
\hline
$\mathcal{R}_{J/\psi}$ & $0.71 \pm 0.17 \pm 0.18$ & \cite{Aaij:2017tyk}\\
$m_{J/\psi}$  &  $(3096.900  \pm 0.06) 10^{-3} $ GeV &  \cite{Tanabashi:2018oca} \\ 
\end{tabular}
\caption{Experimental inputs used in the analysis.}
\label{table:expinput}
\end{table}


The correlation matrix of the input HQET parameters given in Table \ref{table:inputFF}, used to determine the hadronic form factors, is given in Table \ref{table:FFinputcorrelation}.

\begin{table}
\centering
\begin{tabular}{ c | c  c   c  c  c c c c c c }
 & $\rho^2$ & $c$ & $d$ & $\chi_2(1)$ &  $\chi_2(1)'$ & $\chi_3(1)'$ & $\eta(1)$ &$\eta(1)'$ & $l_1(1)$ & $l_2(1) $ \\
 \hline
$\rho^2$  & 1 &  &  &  &  &  & & &  &   \\ 
$c$ &  0.82 &  1 &  &  &  &  & &   &  &   \\ 
$d$ & -0.57 &  -0.91 & 1 &  &  & &  &  &  &   \\ 
$\chi_2(1)$  &  -0.29  & -0.22 & 0.13  &  1 &  &  &  &  &  &  \\ 
$\chi_2(1)'$ &  0.01 & 0.13 &  -0.13  & 0.00 & 1 &  &  &  &  &  \\
$\chi_3(1)'$ & 0.89  &   0.75 & -0.51 & 0.00 &  -0.01 & 1 &  &  &  &  \\
$\eta(1)$ &  0.09 & 0.13 & -0.14  & -0.01  & 0.01  & 0.01 & 1 &   &   &   \\
$\eta(1)'$ &  -0.08 &  0.04 & -0.08 & 0.03& 0.00 &  -0.07 & 0.28  & 1 &  &    \\
$l_1(1)$ &  - 0.03 & 0.01 & -0.05 & 0.00 &   0.00 & 0.01  & 0.34 & -0.15 & 1 &  \\
$l_2(1)$ &  -0.01 &  0.00 & 0.00 & 0.00 & -0.01 &-0.01 & 0.00 & 0.00 &  0.01  & 1  \\
\end{tabular}
\caption{Correlation matrix of the inputs in Table \ref{table:inputFF},
used to determine the form factors in the HQET parametrization.}
\label{table:FFinputcorrelation}
\end{table}

\section{Helicity amplitudes}
\label{appendix:helicity}

The helicity amplitudes of  $\bar{B} \to M \tau \bar{\nu}_\tau \  (M=D, D^*)$ transitions, $H_{i, \lambda}^{\lambda_M}$, are defined through the hadronic matrix elements \cite{Sakaki:2013bfa}
\beqn
\no
H_{V_{L,R}, \lambda}^{\lambda_M} &=& \epsilon_{\mu}^*(\lambda) \bra{ M(\lambda_M)}  \bar{c} \gamma^{\mu} (1 \mp \gamma_5) b \ket{\bar{B}} , \\ 
H_{S_{L,R}, \lambda}^{\lambda_M} &=&  \bra{ M(\lambda_M)}  \bar{c} \gamma^{\mu} (1 \mp \gamma_5) b \ket{\bar{B}}  , \\ \no
H_{T, \lambda \lambda'}^{\lambda_M} &=& - H_{T, \lambda' \lambda}^{\lambda_M} =  \epsilon_{\mu}^* (\lambda) \epsilon_{\nu}^*(\lambda') \bra{ M(\lambda_M)}  \bar{c}\sigma^{\mu \nu} (1-\gamma_5) b \ket{\bar{B}}  , 
\label{eq:hadampl}
\eeqn
where $\lambda_M$ ($=s$ for $D$ and $0,\pm 1$ for $D^*$)
and $\lambda$ ($=0,\pm 1,t$)
are the helicities of the $D^{(*)}$ meson and the intermediate boson, respectively, in the $B$ rest frame. The amplitudes for $\bar{B} \to D $ transitions are:  
\beqn
\no
H_{V, 0}^s (q^2) &\equiv& H_{V_L, 0}^s (q^2) =  H_{V_R, 0}^s (q^2)  =  \sqrt{\frac{\lambda_D(q^2)}{q^2}}\, F_1(q^2) \, , \\ \no
H_{V, t}^s(q^2) &\equiv& H_{V_L, t}^s(q^2) = H_{V_R, t}^s(q^2)  =   \frac{m_B^2-m_D^2}{\sqrt{q^2}}\, F_0(q^2)\, , \\ 
H_S^s(q^2) &\equiv& H_{S_L}^s (q^2) =  H_{S_R}^s (q^2) \simeq \frac{m_B^2 - m_D^2}{m_b - m_c}\, F_0(q^2) \, , \\ \no
H_T^s(q^2) &\equiv&  H_{T, +-}^s(q^2) = H_{T, 0t}^s(q^2) = - \frac{\sqrt{\lambda_D(q^2)}}{m_B+m_D}\, F_T(q^2) \, , 
\label{eq:hadampD}
\eeqn
and for $\bar{B} \to D^*$:
\beqn
\no
H_{V, \pm} (q^2) &\equiv& H_{V_L, \pm}^{\pm} (q^2) =  - H_{V_R, \mp}^{\mp} (q^2) =  (m_B+m_{D^*})\, A_1(q^2) \mp \frac{\sqrt{\lambda_{D^*} (q^2)}}{m_B + m_{D^*}}\, V(q^2) \, , 
\\ \no
H_{V,0} (q^2) &\equiv &  H_{V_L,0}^0 (q^2) = -H_{V_R,0}^0 (q^2)
\\ \no
&=& \frac{m_B+m_{D^*}}{2 m_{D^*} \sqrt{q^2}} \left[- (m_B^2-m_{D^*}^2 - q^2)\, A_1(q^2) + \frac{\lambda_{D^*}(q^2)}{(m_B+m_{D^*})^2 } \, A_2(q^2)\right] , 
\\ \no
H_{V, t} (q^2) &\equiv& H_{V_L, t}^0 (q^2) = -H_{V_R, t}^0 (q^2) = - \sqrt{\frac{\lambda_{D^*}(q^2)}{q^2}}\, A_0(q^2) \, , 
\\ 
H_S(q^2) &\equiv& H_{S_R}^0(q^2) = -  H_{S_L}^0(q^2) \simeq - \frac{\sqrt{\lambda_{D^*}(q^2)}}{m_b+m_c}\, A_0(q^2)\, , 
\\ \no
H_{T, \pm}(q^2) &\equiv& \pm H_{T, \pm t}^{\pm} (q^2) = \frac{1}{\sqrt{q^2}} \left[\pm (m_B^2 - m_{D^*}^2)\, T_2(q^2) + \sqrt{\lambda_{D^*}(q^2)}\, T_1(q^2) \right] , \\ \no
H_{T, 0} (q^2) &\equiv& H_{T, +-}^0(q^2) = H_{T, 0t}^0 (q^2) = \frac{1}{2 m_{D^*}} \left[ - (m_B^2+3\, m_{D^*}^2- q^2)\, T_2(q^2) + \frac{\lambda_{D^*}(q^2)}{m_B^2 - m_{D^*}^2}\, T_3(q^2) \right] . \no 
\label{eq:hadampDstar}
\eeqn

The form factors $F_0(q^2)$, $F_1(q^2)$ and $F_T(q^2)$ appearing in the $D$ matrix elements 
are defined by
\begin{eqnarray}
F_1(q^2) &=& \frac{1}{2\sqrt{m_B m_D}}\,\big[ (m_B + m_D)\, h_+ (q^2) - (m_B - m_D)\, h_- (q^2)\big] \, ,\no \\
F_0(q^2) &=& \frac{1}{2\sqrt{m_B m_D}}\left[\frac{(m_B+m_D)^2-q^2}{m_B + m_D}\, h_+ (q^2)- \frac{(m_B-m_D)^2-q^2}{m_B - m_D}\, h_- (q^2)\right] ,\\
F_T(q^2)&=& \frac{m_B + m_D}{2\sqrt{m_B m_D}}\, h_T(q^2) \, , \no
\end{eqnarray}
while the $D^*$ helicity amplitudes involve the functions
\begin{eqnarray}
V(q^2) &=& \frac{m_B+m_{D^*}}{2\sqrt{m_B m_{D^*}}}\, h_V(q^2), 
\no\\
A_1(q^2) &=& \frac{(m_B + m_{D^*})^2-q^2}{2\sqrt{m_B m_{D^*}}(m_B + m_{D^*})}\, h_{A_1}(q^2), 
\no\\
A_2(q^2)&=& \frac{m_B +m_{D^*}}{2\sqrt{m_B m_{D^*}}} \left[ h_{A_3}(q^2) + \frac{m_{D^*}}{m_B}h_{A_2}(q^2)\right],  
\\
A_0(q^2)&=& \frac{1}{2\sqrt{m_B m_{D^*}}} \left[ \frac{(m_B+m_{D^*})^2-q^2}{2m_{D^*}}\, h_{A_1}(q^2)-\frac{m_B^2-m_{D^*}^2+q^2}{2m_B}\, h_{A_2}(q^2)
\right. \no\\ &&\hskip 1.9cm\left. \mbox{}
-\frac{m_B^2-m_{D^*}^2-q^2}{2m_{D^*}}\, h_{A_3}(q^2)\right] ,
\no
\end{eqnarray}
and
\begin{eqnarray}
T_1(q^2)&=& \frac{1}{2\sqrt{m_B m_{D^*}}}\left[ (m_B + m_{D^*})\, h_{T_1}(q^2)-(m_B - m_{D^*})\, h_{T_2}(q^2) \right] , 
\no \\
T_2(q^2) &=& \frac{1}{2\sqrt{m_B m_{D^*}}}\left[ \frac{(m_B + m_{D^*})^2-q^2}{m_B + m_{D^*}}\, h_{T_1}(q^2)-\frac{(m_B-m_{D^*})^2-q^2}{m_B - m_{D^*}}\, h_{T_2}(q^2)\right] , 
\\
T_3(q^2)&=&\frac{1}{2\sqrt{m_B m_{D^*}}} \left[ (m_B-m_{D^*})\, h_{T_1}(q^2)-(m_B+m_{D^*})\, h_{T_2}(q^2)-2\,\frac{m_B^2-m_{D^*}^2}{m_B}\, h_{T_3}(q^2)\right] .
\no
\end{eqnarray}
The reduced functions $\hat h_i(q^2) = h_i(q^2)/\xi(q^2)$ take the form \cite{Bernlochner:2017jka}
\beqn
\no
\hat{h}_{+} &=& 1 + \hat{\alpha}_s \left[ C_{V_1} + \frac{\omega + 1}{2} \left( C_{V_2} +  C_{V_3} \right) \right] + \left( \varepsilon_c + \varepsilon_b \right)  \hat{L}_1 \, , \\ \no
\hat{h}_{-} &=&  \hat{\alpha}_s \, \frac{\omega + 1}{2}  \left( C_{V_2} - C_{V_3} \right) + \left( \varepsilon_c - \varepsilon_b \right) \hat{L}_4 \, , \\ 
\hat{h}_{S} &=& 1 +  \hat{\alpha}_s C_S + \left( \varepsilon_c + \varepsilon_b \right) \left( \hat{L}_1  -\hat{L}_4 \,\frac{\omega -1}{\omega + 1 } \right)  , \\ \no
\hat{h}_{T} &=& 1 +  \hat{\alpha}_s  \left(C_{T_1} - C_{T_2} + C_{T_3} \right) + \left( \varepsilon_c + \varepsilon_b \right) \left( \hat{L}_1 - \hat{L}_4 \right) , 
\eeqn
for $B \to D$, and
\beqn
\no
\hat{h}_{V} &=&  1 + \hat{\alpha}_s C_{V_1} + \varepsilon_c \left( \hat{L}_2 - \hat{L}_5 \right) + \varepsilon_b \left(  \hat{L}_1 - \hat{L}_4 \right)  , \\ \no
\hat{h}_{A_1} &=&  1 + \hat{\alpha}_s  C_{A_1} + \varepsilon_c \left( \hat{L}_2 - \hat{L}_5\, \frac{\omega -1}{\omega + 1} \right) + \varepsilon_b \left(  \hat{L}_1 - \hat{L}_4\, \frac{\omega -1}{\omega + 1} \right) , \\ \no
\hat{h}_{A_2} &=&   \hat{\alpha}_s  C_{A_2} + \varepsilon_c \left( \hat{L}_3 + \hat{L}_6 \right) , \\ \no
\hat{h}_{A_3} &=&  1 + \hat{\alpha}_s \left( C_{A_1} + C_{A_3} \right) + \varepsilon_c \left( \hat{L}_2 - \hat{L}_3 + \hat{L}_6 - \hat{L}_5 \right) + \varepsilon_b \left(  \hat{L}_1 - \hat{L}_4 \right)  , \\
\hat{h}_{P} &=&  1 + \hat{\alpha}_s C_P + \varepsilon_c \left[\hat{L}_2 + \hat{L}_3\, (\omega - 1) + \hat{L}_5 - \hat{L}_6\, (\omega +1) \right] + \varepsilon_b \left(  \hat{L}_1 - \hat{L}_4 \right)  , \\ \no
\hat{h}_{T_1} &=&  1 + \hat{\alpha}_s \left[ C_{T_1} + \frac{\omega -1 }{2} \left(C_{T_2} - C_{T_3} \right)  \right] + \varepsilon_c \hat{L}_2 +  \varepsilon_b \hat{L}_1 \, , \\ \no
\hat{h}_{T_2} &=&  \hat{\alpha}_s \, \frac{\omega +1}{2} \left(C_{T_2} + C_{T_3} \right) + \varepsilon_c \hat{L}_5 -  \varepsilon_b \hat{L}_4 \, , \\ \no
\hat{h}_{T_3} &=&   \hat{\alpha}_s C_{T_2} +  \varepsilon_c \left( \hat{L}_6 - \hat{L}_3 \right) ,
\eeqn
for $B \to D^*$. The explicit expressions of the $\omega(q^2)$-dependent factors $\hat L_{1...6}$
and the $\mathcal{O}(\alpha_s)$ corrections $C_i$ can be found in Ref.~\cite{Bernlochner:2017jka}.

The helicity amplitudes of $B_c \to J/\psi$ can be expressed in the same way as for  $B \to D^*$, with the replacements $B \to B_c$ and $D^* \to J/\psi$ and inserting the corresponding form factors. 
The vector and axial-vector form factors have been calculated in the small $q^2$ region, using a perturbative QCD factorization approach, and extrapolated to higher values of $q^2$ with a more model-dependent parametrization
\cite{Watanabe:2017mip,Wen-Fei:2013uea}: 
\beqn
\no
V^c(q^2)  &=&  V^c(0) \;\exp{\left[0.065 \, q^2 + 0.0015\, (q^2)^2 \right]}\, , \\ \no
A_0^c(q^2) &=& A_0^c(0)  \;\exp{\left[0.047\, q^2 + 0.0017 \, (q^2)^2 \right]}\, , \\ 
A_1^c(q^2) &=& A_1^c(0) \; \exp{\left[0.038  \,q^2 + 0.0015 \,(q^2)^2 \right]}\, , \\ \no
A_2^c(q^2) &=& A_2^c(0) \; \exp{\left[0.064\, q^2 + 0.0041\, (q^2)^2 \right]}\, ,  \no
\eeqn
where
$V_c^0 = 0.42 \pm 0.01 \pm 0.01, \, A_0^c(0) =  0.59 \pm 0.02 \pm 0.01, \, A_1^c(0) = 0.46 \pm 0.02 \pm 0.01$ and $A_2^c (0) = 0.64 \pm 0.02 \pm 0.01$ \cite{Wen-Fei:2013uea}. For the tensor form factors the quark-level equations of motion are adopted \cite{Sakaki:2013bfa}:
\beqn
\no
T_1^c(q^2) &=& \frac{m_b+m_c}{m_{B_c} + m_{J/\psi}}\, V^c(q^2)\,  , \\
T_2^c(q^2) &=& \frac{m_b-m_c}{m_{B_c} - m_{J/\psi}}\, A_1^c(q^2)\, , \\
T_3^c(q^2) &=& -  \frac{m_b-m_c}{q^2} \left\{ m_{B_c} \left[ A_2^c(q^2) - A_2^c(q^2) \right] + m_{J/\psi} \left[ A_2^c(q^2) + A_1^c(q^2) - 2 A_0^c (q^2) \right] \right\}. \no
\eeqn

\section{Longitudinal polarization $F_L^{D^*}$}

The $B \to D^* \tau \bar{\nu}_\tau$ differential decay width into longitudinally-polarized ($\lambda_{D^*}=0$) $D^*$ mesons is given by
\label{app:FLDstar}
\begin{eqnarray}
\frac{d\Gamma_{\lambda_{D^*}=0}^{D^*}}{d q^2} &=&
\frac{G_F^2|V_{cb}|^2}{192 \pi^3 m_B^3}\, q^2\sqrt{\lambda_{D^*}(q^2)}\left(1-\frac{m_\tau^2}{q^2}\right)^2 \left\{ 
| 1 + C_{V_L} - C_{V_R}|^2\left[\left(1 + \frac{m_\tau^2}{2q^2}\right)H_{V,0}^2  
+ \frac{3}{2}\,\frac{m_\tau^2}{q^2}\, H_{V,t}^2\right]
\right. \nonumber\\
&&\left.\hskip 1.5cm\mbox{} + 
\frac{3}{2}\, |C_{S_R}-C_{S_L}|^2H_S^2 + 8\, |C_T|^2\left(1+\frac{2m_\tau^2}{q^2}\right)H_{T,0}^2 
\right.\nonumber\\
&&\left.\hskip 1.5cm\mbox{} + 
3\, {\cal R}e [(1+C_{V_L}-C_{V_R})(C_{S_R}^*-C_{S_L}^*)]\,\frac{m_\tau}{\sqrt{q^2}}\, H_S H_{V,t} 
\right.\\
&&\left.\hskip 1.5cm\mbox{} -
12\, {\cal R}e[(1+C_{V_L}-C_{V_R})\, C_T^*]\,\frac{m_\tau}{\sqrt{q^2}}\, H_{T,0}H_{V,0}\right\} \, ,
\nonumber
\end{eqnarray}
where the helicity amplitudes are defined in Appendix~\ref{appendix:helicity}.
\section{UV Lagrangian}
\label{app:UVLagrangian}

Possible new mediators contributing to the effective Hamiltonian of Eq.~(\ref{eq:NPHam}) and their relative effective Lagrangian are summarized in Table~\ref{table:NP}.

%
{\renewcommand{\arraystretch}{2}
\begin{table}[h] 
  \vspace{0.5cm}
  \centering
 \begin{tabular}{| c | c |  c | c  |}
 \hline
 Spin & NP mediator & Contribution & Relevant effective Lagrangian (+ h.c.) \\
 \hline
 \hline
 \multirow{5}{*}{   $0$   } &\multirow{2}{*}{$\displaystyle 
 (\bar{3},1,1/3) \sim \phi$} &  ${\cal O}_{V_L}$ & $\displaystyle 
 \propto \frac{1}{2 M_\phi^2}\, (\overline{d_L}\gamma_\mu u_L)(\overline{\nu_L}\gamma^\mu e_L)$ \\
  			        &								 & $ {\cal O}_{S_L}, \,  {\cal O}_{T}$ & $\displaystyle 
  			        \propto \frac{1}{M_\phi^2}\left[(\overline{u_R}d_L)(\overline{e_R}\nu_L)-\frac{1}{4}(\overline{u_R}\sigma_{\mu \nu} d_L)(\overline{e_R}\sigma^{\mu \nu}\nu_L)\right]$\\
\cline{2-4}
			        & $
			        (3,2,7/6) \sim \left(\phi_{5/3},\phi_{2/3} \right)$ & ${\cal O}_{S_L} , \, {\cal O}_T$ & $\displaystyle 
			        \propto \frac{1}{M_{\phi_{2/3}}^2}\left[(\overline{e_R}\nu_L)(\overline{u_R}d_L)+\frac{1}{4}(\overline{e_R}\sigma_{\mu \nu}\nu_L)(\overline{u_R}\sigma^{\mu \nu}d_L)\right]$\\
\cline{2-4}
			        & $
			        (\bar{3},3,1/3) \sim (\phi_{4/3}, \phi_{1/3}, \phi_{-2/3})$		& ${\cal O}_{V_L}$ &$\displaystyle 
			        \propto \frac{1}{M_{\phi_{1/3}}^2}\, (\overline{d_L}\gamma_\mu u_L)(\overline{\nu_L}\gamma^\mu e_L)$ \\
\cline{2-4}
                    & \multirow{2}{*}{$
                    (1,2,1/2) \sim (h_2^+, h_2^0)$} & ${\cal O}_{S_R}$ & $\displaystyle 
                    \propto \frac{1}{M_{h_2^+}^2}\, (\overline{u_L} d_R)(\overline{e_R}\nu_L)$\\
                    &                                                           &
                    ${\cal O}_{S_L}$ & $\displaystyle 
                    \propto \frac{1}{M_{h_2^+}^2}\, (\overline{u_R} d_L)(\overline{e_R} \nu_L)$\\
\hline
\hline
\multirow{5}{*}{   $1$   }  & $
        (\bar{3},2,5/6) \sim (\phi_{4/3}^\mu, \phi_{1/3}^\mu)$			& $ {\cal O}_{S_R}$ & $\displaystyle
        \propto \frac{1}{M_{\phi_{-1/3}}^2}\, (\overline{e_R}\nu_L)(\overline{u_L}d_R)$ \\
\cline{2-4}
		& \multirow{2}{*}{$
		(3,1,2/3)\sim \phi^\mu$} 	& ${\cal O}_{V_L}$ & $\displaystyle 
		\propto \frac{1}{M_\phi^2}\, (\overline{u_L}\gamma_\mu d_L)(\overline{e_L}\gamma^\mu \nu_L)$\\
		&										& ${\cal O}_{S_R}$ & $\displaystyle 
	\propto	\frac{1}{M_\phi^2}\, (\overline{u_L}d_R)(\overline{e_R}\nu_L)$ \\
\cline{2-4}
		& $
		(3,3,2/3) \sim (\phi_{5/3}^\mu, \phi_{2/3}^\mu, \phi_{-1/3}^\mu)$					& $ {\cal O}_{V_L}$ & $\displaystyle 
		\propto \frac{1}{M_{\phi_{2/3}}^2}\, (\overline{\nu_L}\gamma_\mu e_L)(\overline{d_L}\gamma^\mu u_L) $ \\
\cline{2-4}
        & $(1,3,0) \sim (W^{\prime \mu}_+,W^{\prime \mu}_0, W^{\prime \mu}_-) $ & ${\cal O}_{V_L}$ & $\displaystyle 
        \frac{1}{M_{W^\prime_+}^2}\, (\overline{e_L} \gamma_\mu \nu_L)(\overline{u_L} \gamma^\mu d_L)$\\
\hline
  \end{tabular}
    \caption{Possible fields contributing to the effective Hamiltonian of Eq.~\eqref{eq:NPHam}, at dimension 6: leptoquarks are denoted by $\phi$ and a second Higgs doublet as $h_2$. Their quantum numbers ($SU(3)$, $SU(2)$, $U(1)_Y$), contribution to the EFT operators and their relevant effective Lagrangian after integrating them out are described for each new field. Their $SU(2)$ decomposition is explicitly shown after the "$\sim$".}
    \label{table:NP}
 \end {table}

 \section{Warsaw basis}
 \label{appendix:Warsaw basis}
 
 The operators describing the SMEFT in the Warsaw basis are given by \cite{Grzadkowski:2010es, Buchmuller:1985jz}, 
 \beqn
  \no
\mathcal{O}_{lq}^{(3)} &=& \left(\bar\ell \gamma_{\mu} \tau^{I} \ell \right) \left(\bar{q}\gamma^{\mu} \tau^{I} q \right) , \\
 \no
\mathcal{O}_{lequ}^{(1)} &=& \left( \bar{\ell}^j e \right) \varepsilon_{jk} \left( \bar{q}^k u \right), \\
\mathcal{O}_{ledq} &=& \left( \bar{\ell}^j e \right)  \left(\bar{d}q^j \right) , \\
 \no
\mathcal{O}_{lequ}^{(3)} &=& \left(\bar{\ell}^j \sigma_{\mu \nu} e \right) \varepsilon_{jk}  \left(\bar{q}^k \sigma^{\mu \nu} u \right) ,
 \eeqn
 where $\tau^{I}$ are the Pauli matrices and $\varepsilon_{jk}$ is the totally antisymmetric tensor with $\varepsilon_{12} = +1$. The fields $q$ and $\ell$ are the quark and lepton $SU(2)_L$ doublets, respectively, and $u,d,e$ are the right-handed $SU(2)_L$ singlets.
Neglecting the small corrections proportional to the CKM factors $V_{ub}$ and $V_{cb}$, the relevant contributions to the $b\to c\tau\nu$ transitions originate in the Wilson coefficients
 $[C_{lq}^{(3)}]_{3323}\equiv \tilde{C}_{V_L}$, $[C_{lequ}^{(1)}]_{3332}\equiv \tilde{C}_{S_R}$, $[C_{ledq}]_{3332}\equiv \tilde{C}_{S_L}$ and $[C_{lequ}^{(3)}]_{3332}\equiv \tilde{C}_{T}$, where $[C_X]_{ijkl}$ denotes the coefficient of the corresponding operator $O_X$ with flavour indices $i,j,k,l$.
 The effective Lagrangian relevant for the description of the $B$ anomalies is therefore given by
 \begin{equation} 
    {\cal L}_\text{SMEFT} \,\supset\, \frac{1}{\Lambda_\text{NP}^2}\left( \tilde{C}_{V_L}\, [\mathcal{O}_{lq}^{(3)}]_{3323} + \tilde{C}_{S_R}\, [\mathcal{O}_{lequ}^{(1)}]_{3332} + \tilde{C}_{S_L}\, [\mathcal{O}_{ledq}]_{3332} + \tilde{C}_T\, [\mathcal{O}_{lequ}^{(3)}]_{3332} \right)\, .
 \end{equation}
Notice that there is a correspondence between the effective operators at the SMEFT basis with those at the WET basis, according to:
\beqn
\mathcal{O}_{lq}^{(3)} \,\leftrightarrow\, \mathcal{O}_{V_L}\, , 
\qquad\quad
\mathcal{O}_{lequ}^{(1)} \,\leftrightarrow\, \mathcal{O}_{S_R}\, , 
\qquad\quad
\mathcal{O}_{ledq} \,\leftrightarrow\, \mathcal{O}_{S_L}\, , 
\qquad\quad
\mathcal{O}_{lequ}^{(3)} \,\leftrightarrow\, \mathcal{O}_{T}\, ,
\eeqn
which allow us to use the notation $\tilde{C}_i$ for the Wilson coefficients at the SMEFT basis, with the aim of making the discussion more intuitive for the reader.


\FloatBarrier

\bibliographystyle{utphys}
\bibliography{referencies}

\end{document}